\documentclass[12pt,review,times]{elsarticle}
\journal{Progress in Particle and Nuclear Physics}
\usepackage{amsmath,amssymb,bm}
\usepackage{graphicx}
\usepackage{color}
\usepackage{hyperref,url}

\newcommand{\be}{\begin{equation}}
\newcommand{\ee}{\end{equation}}
\newcommand{\bea}{\begin{eqnarray}}
\newcommand{\eea}{\end{eqnarray}}

\newcommand{\U}{\text{U}}
\renewcommand\d{\partial}
\usepackage{xcolor}

\begin{document}

\begin{frontmatter}

\title{Chiral effects in astrophysics and cosmology}

\author[1]{Kohei Kamada}
\ead{kohei.kamada@resceu.s.u-tokyo.ac.jp}

\author[2]{Naoki Yamamoto\corref{mycorrespondingauthor}}
		\cortext[mycorrespondingauthor]{Corresponding author}
\ead{nyama@rk.phys.keio.ac.jp}

\author[3]{Di-Lun Yang}
\ead{dlyang@gate.sinica.edu.tw}

\affiliation[1]{
organization={Research Center for the Early Universe (RESCEU), Graduate School of Science, The University of Tokyo}, 
addressline={Hongo 7-3-1 Bunkyo-ku}, 
city={Tokyo 113-0033}, 
country={Japan}}
\affiliation[2]{
organization={Department of Physics, Keio University}, 
city={Yokohama 223-8522}, 
country={Japan}}
\affiliation[3]{
organization={Institute of Physics, Academia Sinica}, 
city={Taipei 11529}, 
country={Taiwan}}

\begin{abstract}
The microscopic quantum nature of elementary particles, chirality, leads to macroscopic phenomena like the chiral anomaly, chiral magnetic effect, and chiral plasma instability.
We review recent progress of the studies of these chiral effects in high-energy astrophysics, such as pulsar kicks, magnetars, and core-collapse supernovae, and early Universe cosmology, such as the primordial magnetic field, baryogenesis, and chiral gravitational waves.
We also provide a pedagogical introduction to the chiral effects and low-energy effective theories to describe them in and out of equilibrium---the chiral (magneto)hydrodynamics, chiral kinetic theory, and chiral radiation transport theory for neutrinos.
\end{abstract}

\begin{keyword}
Chiral effects \sep
Neutron stars \sep 
Core-collapse supernovae \sep 
Primordial magnetic field \sep 
Baryogenesis \sep 
Chiral gravitational waves \\
\textit{Preprint number:} RESCEU-12/22
\end{keyword}

\end{frontmatter}

\newpage
\tableofcontents
\newpage

\section{Introduction \label{sec:introduction}}
The notion of chirality emerges in wide areas of natural science from physics and chemistry to biology. Among these, chirality of elementary particles is the most microscopic chirality. One can naturally ask if and how it could be reflected in the properties or evolution of macroscopic systems, including our Universe.

One important consequence of chirality of elementary particles in relativistic quantum field theories (QFT) is the chiral anomaly~\cite{Adler:1969gk,Bell:1969ts}: the violation of an axial symmetry due to the quantum effects. Historically, this phenomenon was found in computing the decay width of the $\pi^0$ meson to two photons. Later, it turned out that the chiral anomaly is rather a universal phenomenon intrinsic to relativistic QFT, and nowadays even it has a connection to the classifications of topological matter in condensed matter physics~\cite{Witten:2015aba}. A remarkable feature of the chiral anomaly is its connection to topology, and as a result, it is independent of the energy scale. This means that the chiral anomaly must manifest itself even at a macroscopic scale.

Another consequence of chirality in QFT that has attracted recent attention is the chiral transport phenomena; see, e.g.,  Ref.~\cite{Landsteiner:2016led} for a review. Examples include the so-called chiral magnetic effect (CME)~\cite{Vilenkin:1980fu,Nielsen:1983rb,Alekseev:1998ds,Fukushima:2008xe}, which is a current along the direction of the magnetic field in relativistic matter with a chirality imbalance. Such unusual transport phenomena are also related to the topological property of a system of chiral fermions~\cite{Son:2012wh}, and can be seen as a macroscopic manifestation of chirality. Because these chiral effects do not depend on the detail of interactions, they may appear in various fields of physics, such as quark-gluon plasmas created in heavy ion collision experiments~\cite{Kharzeev:2015znc} and Weyl/Dirac semimetals~\cite{Armitage:2017cjs,Gorbar2021} where Weyl/Dirac quasi-particles appear emergently close to band touching points.

In this review, we will focus on the chiral effects in high-energy astrophysical systems and early Universe cosmology. The basic idea behind is that  chirality of elementary particles, such as neutrinos, electrons, and quarks, leads to macroscopic chiral effects as above, and consequently, the non-equilibrium evolution of these systems can receive significant modifications. As applications to high-energy astrophysical systems, we will mainly review the relevance of the chiral effects in neutron stars and core-collapse supernovae, and in particular, discuss the theoretical attempts to explain the following long-standing puzzles in astrophysics:
\begin{itemize}
    \item{Pulsar kicks~\cite{Lyne1994,Kaspi1996,Arzoumanian2002,Hobbs2005} (see also Ref.~\cite{Lai_2001} for a review): neutron stars typically have velocities $\sim 100\,{\rm km/s}$ after their formation. How and where do they receive momentum kicks?} 
    \item{Magnetars~\cite{Magnetar} (see also Refs.~\cite{Harding:2006qn,Spruit:2007bt,Enoto:2019vcg} for reviews): a class of neutron stars have gigantic magnetic field $\sim 10^{15}\,{\rm G}$ on the surface. What is the energy source for such a strong magnetic field? Also, what is the mechanism to sustain the strong magnetic field for an astrophysically long time scale?}
    \item{Core-collapse supernova explosions (see, e.g., Refs.~\cite{Kotake:2012nd,Burrows:2012ew,Foglizzo:2015dma,Janka:2016fox,Muller:2016izw,Radice:2017kmj} for reviews): stars with the mass $\gtrsim 10M_{\odot}$ eventually explode with the energy $\sim 10^{51}\,{\rm erg}$ after the gravitational collapse at the end of their evolution. What is the physical mechanism for the explosion with the observed energy?}
\end{itemize}
We will also describe the recent developments of the studies of chiral effects in early Universe cosmology, in particular, to explain the following mysteries in cosmology as well as to establish the tools to explore the dynamical phenomena in the early Universe:
\begin{itemize}
    \item{Primordial magnetic field (see, e.g., Refs.~\cite{Neronov:2010gir,Tavecchio:2010mk,Dolag:2010ni,Finke:2015ona,Fermi-LAT:2018jdy} and also Refs.~\cite{Durrer:2013pga,Subramanian:2015lua} for reviews): } recent blazar observations suggest the existence of the intergalactic magnetic fields. When and how are they generated, if ever? 
    \item{Baryogenesis (see, e.g., Refs.~\cite{Sakharov:1967dj,Kuzmin:1985mm,Fukugita:1986hr,Affleck:1984fy} and also Ref.~\cite{Riotto:1999yt} for review):} the abundance of the antimatter over matter is extremely small in the present Universe. When and how is such an asymmetry generated?
    \item{Chiral gravitational waves (GWs) (see, e.g., Refs.~\cite{Seto:2007tn,Seto:2008sr,Crowder:2012ik,Smith:2016jqs,Domcke:2019zls}):} detection of stochastic GW background is now no longer a dream. What kind of phenomena and physics in the early Universe can we explore if we can detect its circular polarization? 
\end{itemize}

This review is organized as follows. In Section~\ref{sec:chiral_effect}, we provide a pedagogical introduction of the chiral effects, such as the chiral anomaly and chiral transport phenomena. In Section~\ref{sec:theory}, we explain the low-energy effective theories to describe the chiral effects in and out of equilibrium: the chiral (magneto)hydrodynamics and chiral kinetic theory. In Sections~\ref{sec:astro} and \ref{sec:cosmo}, we discuss the chiral effects on astrophysics and cosmology, respectively. Section~\ref{sec:summary} is devoted to summary and outlook.

Throughout this review, we use the natural units $\hbar = c = k_{\rm B} = 1$ and mostly minus metric $\eta_{\mu \nu} = {\rm diag}(1, -1, -1, -1)$ unless otherwise stated. The electron charge is denoted by $e$.

\section{Chiral effects \label{sec:chiral_effect}}
In this section, we review various quantum effects that originate from the chirality of elementary particles: the chiral anomaly, chiral magnetic effect, chiral vortical effect, and chiral plasma instability. We will consider an idealized situation of massless Dirac or chiral fermions in this section. Note that although masses of fermions are much smaller than the typical energy scales in high-energy astrophysical systems and cosmology that we will discuss in Sections~\ref{sec:astro} and \ref{sec:cosmo} below, they can still be important to cause the chirality flipping between right- and left-handed fermions. This effect will be taken into account perturbatively later in each section. 

For the recent status of the studies of the chiral effects in heavy ion collisions and Weyl/Dirac semimetals, see Ref.~\cite{Kharzeev:2015znc} and Refs.~\cite{Armitage:2017cjs,Gorbar2021}, respectively, and references therein.

\subsection{\it Chiral anomaly \label{sec:anomaly}}

One of the most prominent consequences of the chirality of fermions 
is the chiral anomaly~\cite{Adler:1969gk,Bell:1969ts}:
the violation of the $\U(1)$ axial symmetry by quantum effects.
Here we review the chiral anomaly based on the spectral flow argument~\cite{Nielsen:1983rb}.

We first discuss the chiral anomaly for massless Dirac fermions with charge $e$ in $1+1$ dimensions. Taking the spatial direction in the $z$ direction, the Lagrangian of the system is given by
\be
{\cal L}_{1+1{\rm D}} = \psi_{\rm \rm R}^{\dag} {\rm i} (D_t + D_z) \psi_{\rm \rm R} + \psi_{\rm \rm L}^{\dag} {\rm i} (D_t - D_z) \psi_{\rm \rm L},
\ee
where $\psi_{\rm R, L}$ denote the right- and left-handed fermions (or right and left movers), respectively, and $D_{\mu} = \d_{\mu} + {\rm i} e A_{\mu}$ is the covariant derivative. The dispersion relations for right- and left-handed fermions are given by $\varepsilon = \chi p_z$, where $\varepsilon$ is the energy, $p_z$ is the momentum in the $z$ direction, and $\chi = \pm 1$ corresponds to the chirality in $1+1$ dimensions. 

\begin{figure}[t]
\begin{center}
\scalebox{0.45}{{\includegraphics{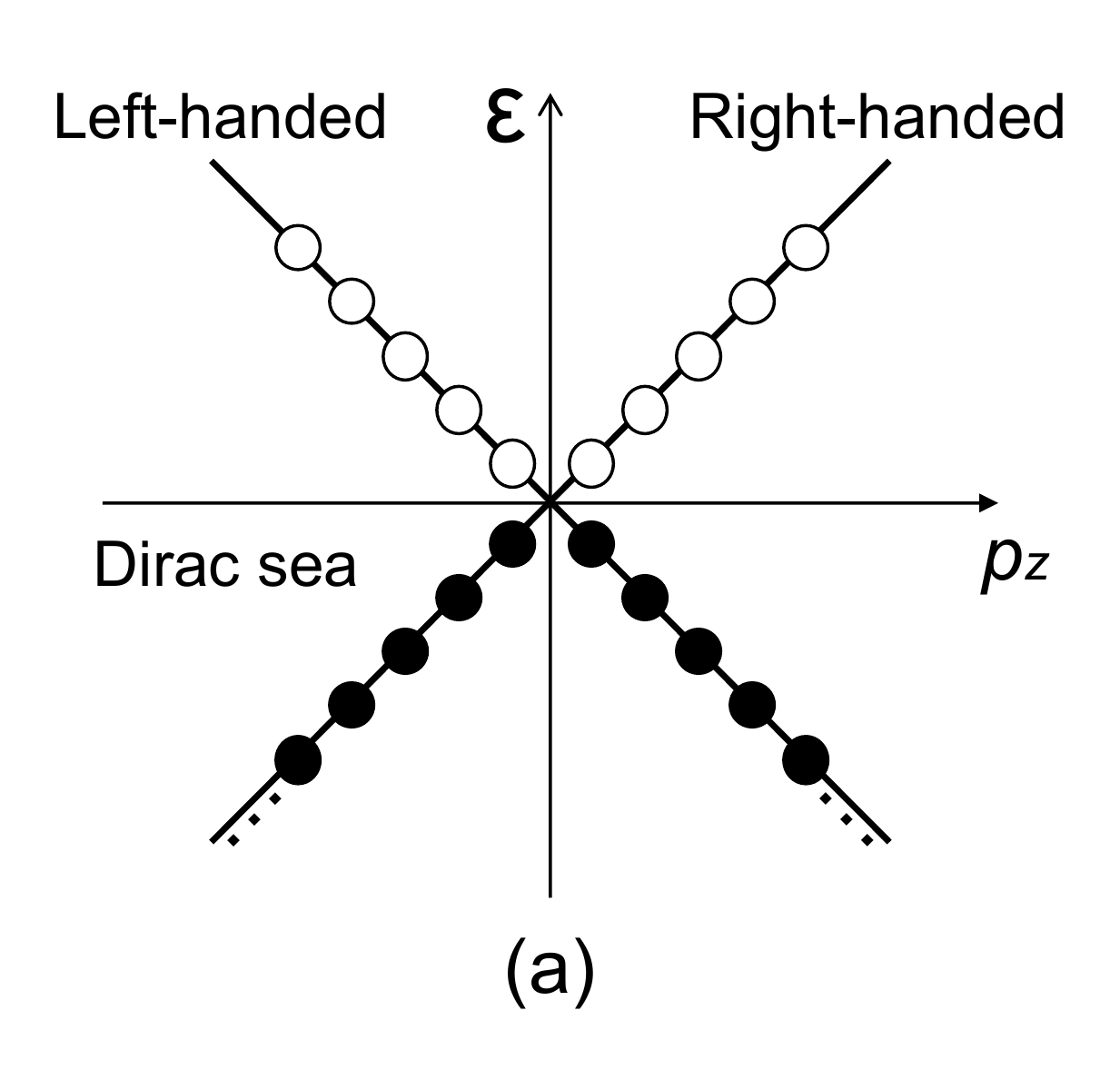}}}
\scalebox{0.45}{{\includegraphics{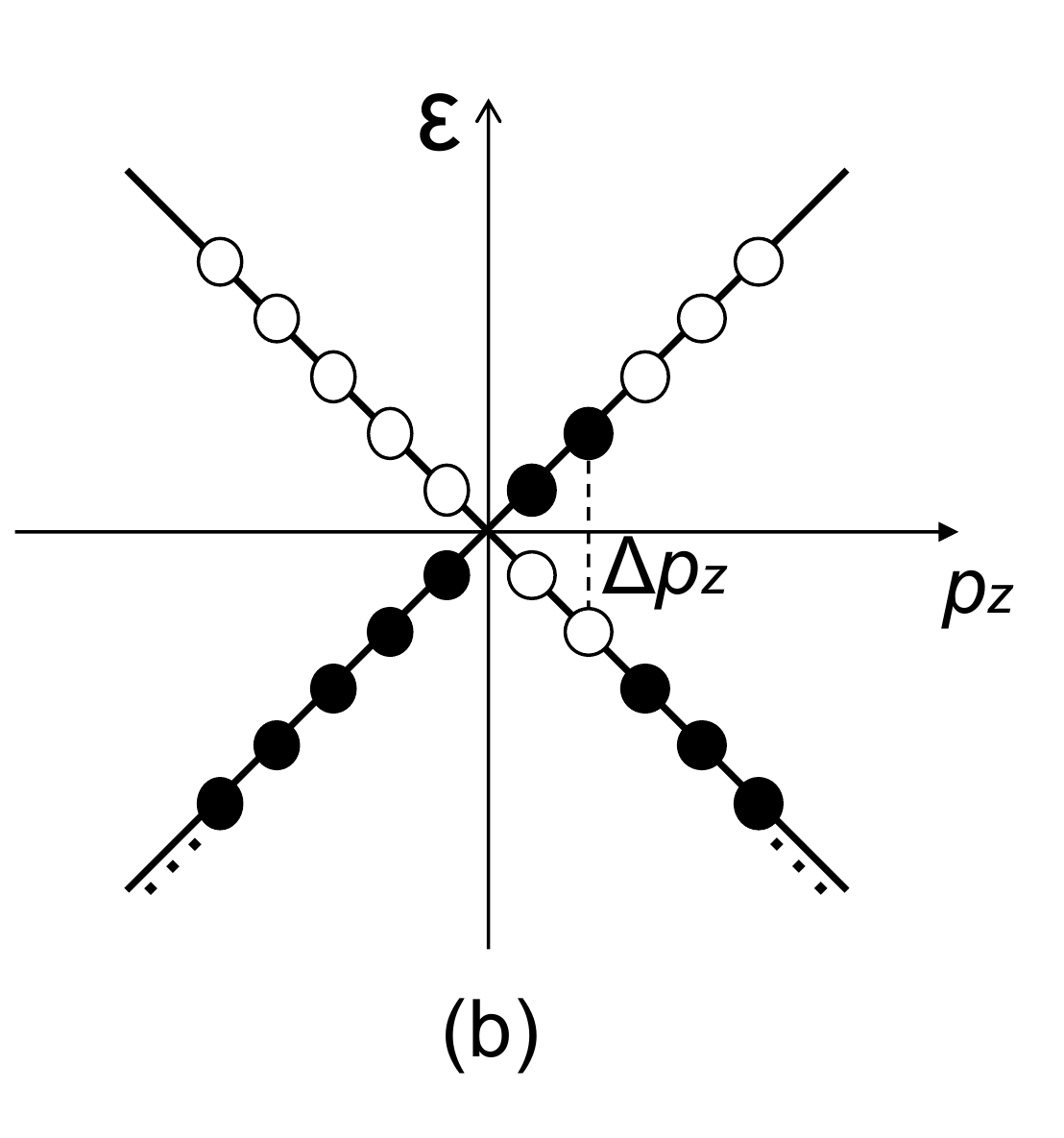}}}
\scalebox{0.45}{{\includegraphics{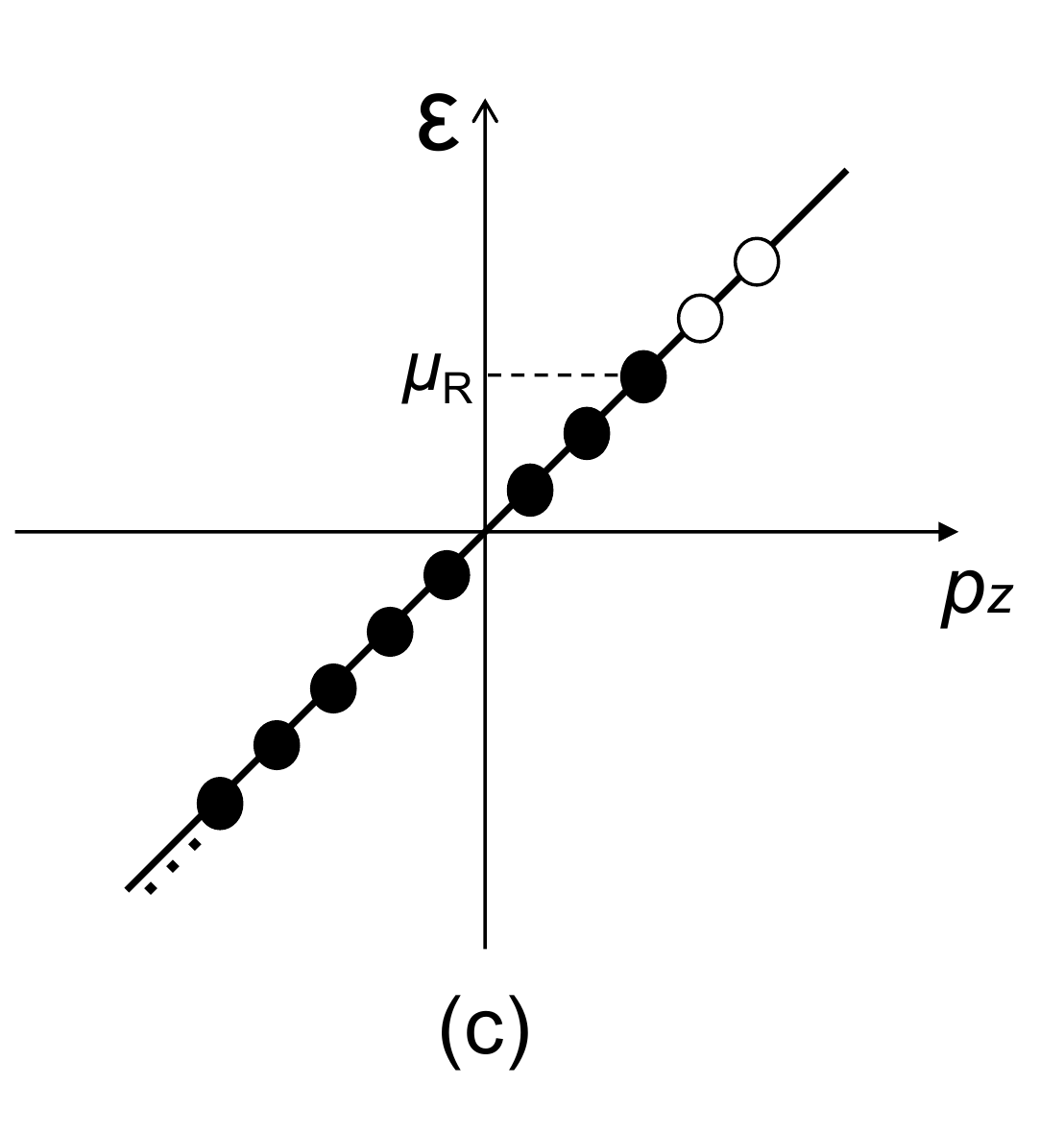}}}
\caption{(a) Filled states of massless Dirac fermions without any external field; (b) Filled states of massless Dirac fermions after turning on an electric field adiabatically; (c) Filled states of massless right-handed fermions at finite chemical potential $\mu_{\rm R}$. All these figures assume $1+1$ dimensions or LLL in $3+1$ dimensions at zero temperature.}
\label{fig1}
\end{center}
\end{figure}

We consider the system in a box with size $L$ with the periodic boundary conditions. As shown in Fig.~\ref{fig1}(a), the vacuum of this system is the Dirac sea where negative energy states are fully occupied. At the classical level, the theory has the $\U(1)_{\rm R} \times \U(1)_{\rm L}$ symmetry, and the numbers of right- and left-handed fermions, $Q_{\rm R}$ and $Q_{\rm L}$, are both conserved.

Let us adiabatically turn on an electric field $E(t)$ in the $z$ direction during a finite time interval $\Delta t$. As the classical equation of motion for charged particles is $\dot {p}_z = e E(t)$, each particle acquires a momentum
\be
\label{p}
\Delta p_z = e \int E(t) {\rm d} t.
\ee
Consequently, the occupied states of right- and left-handed fermions in the quantum theory are shifted as shown in Fig.~\ref{fig1}(b), and so the variations of $Q_{\rm R}$ and $Q_{\rm L}$ satisfy the relation $\Delta Q_{\rm R} = - \Delta Q_{\rm L} \propto \Delta p_z$.
Then, the total number of particles (or vector charge) $Q \equiv Q_{\rm R} + Q_{\rm L}$ is conserved, $\Delta Q = 0$. On the other hand, the axial charge $Q_5 \equiv Q_{\rm R} - Q_{\rm L}$ is not conserved, $\Delta Q_5 \neq 0$.
Recalling that the momentum is quantized as $p_n = 2 n \pi/L$ ($n \in {\mathbb Z}$) in the box, $\Delta Q_5$ can be explicitly evaluated as
\be
\label{Q5_1+1D}
\Delta Q_5 = 2 \times \frac{\Delta p_z}{2\pi/L} = \frac{e}{\pi} \int E {\rm d}t {\rm d}z\,.
\ee
This shows that the conservation of the axial charge is violated by the quantum effect of the particle-antiparticle pair production in the electric field. This is the chiral anomaly in $1+1$ dimensions.

Let us now move on to the chiral anomaly for massless Dirac fermions $\psi$ in $3+1$ dimensions. The Lagrangian is given by
\be
\label{L_3+1D}
{\cal L}_{3+1{\rm D}} = \psi^{\dag}_{\rm R} {\rm i} \sigma^{\mu} D_{\mu} \psi_{\rm R}
+ \psi^{\dag}_{\rm L} {\rm i} \bar \sigma^{\mu} D_{\mu} \psi_{\rm L},
\ee
where $\sigma^{\mu} = (1, {\bm \sigma})$ and $\bar \sigma^{\mu} = (1, -{\bm \sigma})$ with $\sigma^i$ ($i=1,2,3$) being the Pauli matrices. In this case, we first turn on a homogeneous magnetic field in the $z$ direction ${\bm B} = B \hat {\bm z}$, where $\hat {\bm z}$ is the unit vector in the $z$ direction. Then, the energy dispersion is quantized in the $x$-$y$ plane, which is known as the Landau quantization. The explicit expression of the dispersion reads
\be
\varepsilon_n^2 = p_z^2 + (2n+1) eB - 2 e B S_z \quad (n = 0,1,2,\cdots),
\ee
where the integer $n$ characterizes the quantized energy levels in the $x$-$y$ plane, and $S_z=\pm\frac{1}{2}$ is the spin in the $z$ direction. The last term above corresponds to the Zeeman energy in the presence of the magnetic field.

What is special among these energy levels is the lowest Landau level (LLL), corresponding to $n=0$ and $S_z = \frac{1}{2}$: the dispersion relation in  the LLL is $\varepsilon_0^2 = p_z^2$ (or $\varepsilon_0 = \pm p_z$), which is nothing but the dispersion relation for massless chiral fermions in $1+1$ dimensions above. In fact, as $S_z = \frac{1}{2}$, the signs of the dispersion relations $\varepsilon_0 = \pm p_z$ correspond to right- and left-handed fermions, respectively. (This fact can also be checked by deriving the explicit solution to the equation of motion for $\psi_{\rm R,L}$, i.e., the Weyl equation in the magnetic field~\cite{Landsteiner:2016led,Nielsen:1983rb}.) By utilizing the result (\ref{Q5_1+1D}) in $1+1$ dimensions, we get 
\be
\label{Q5_3+1D}
\Delta Q_5 = \int \frac{eE}{\pi} {\rm d} t {\rm d}z \int \frac{eB}{2 \pi} {\rm d}x {\rm d}y = \frac{e^2}{2\pi^2} \int {\rm d}^4 x {\bm E} \cdot {\bm B}\,.
\ee
Here, the factor $eB/(2\pi)$ is the Landau degeneracy per unit area in the $x$-$y$ plane perpendicular to the magnetic field. Note that there is no contribution of higher Landau levels (HLL) to $\Delta Q_5$.

Equation (\ref{Q5_3+1D}) is a global relation in the whole spacetime, but more generically, one can obtain its local covariant version for the axial current $j^{\mu 5} \equiv j^{\mu}_{\rm R} - j^{\mu}_{\rm L}$ with $j^{\mu}_{\rm R} = \psi^{\dag}_{\rm R} \sigma^{\mu} \psi_{\rm R}$ and $j^{\mu}_{\rm L} = \psi^{\dag}_{\rm L} \bar \sigma^{\mu} \psi_{\rm L}$ in the presence of generic electromagnetic fields:
\be
\label{anomaly}
\d_{\mu} j^{\mu 5} = \frac{e^2}{2\pi^2} {\bm E} \cdot {\bm B} = -\frac{e^2}{8\pi^2} F^{\mu \nu} \tilde F_{\mu \nu}\,,
\ee
where $\tilde F_{\mu \nu} = \frac{1}{2} \epsilon_{\mu \nu \alpha \beta} F^{\alpha \beta}$. This is the chiral anomaly in $3+1$ dimensions. Like the argument in $1+1$ dimensions, the vector current $j^{\mu} \equiv j^{\mu}_{\rm R} + j^{\mu}_{\rm L}$ is conserved, $\d_{\mu} j^{\mu} = 0$. In terms of the chiral basis, these relations are equivalent to 
\be
\label{anomaly_RL}
\d_{\mu} j^{\mu}_{\chi}= \chi \frac{e^2}{4\pi^2} {\bm E} \cdot {\bm B}
= - \chi \frac{e^2}{16\pi^2} F^{\mu \nu} \tilde F_{\mu \nu}\,,
\ee
where $\chi = \pm 1$ stands for the chirality in $3+1$ dimensions.
Although we have assumed non-interacting fermions, Eqs.~(\ref{anomaly}) and (\ref{anomaly_RL}) are exact non-perturbatively even in the presence of interactions. This is related to a topological nature of chiral/Dirac fermions, as we will discuss in Section~\ref{sec:Berry}.

While it is often said that the quantum anomaly is the violation of a classical conservation law by quantum effects, the chiral anomaly relation (\ref{anomaly}) can also be understood as the total helicity conservation law. To see this, we perform spatial integration
of Eq.~(\ref{anomaly}) to obtain
\be
\label{helicity}
\frac{{\rm d}}{{\rm d}t} \left(Q_5 + {\cal H}_{\rm mag} \right) = 0\,, \qquad 
{\cal H}_{\rm mag} \equiv \frac{e^2}{4\pi^2} \int {\rm d}^3 {\bm x} {\bm A} \cdot {\bm B}\,,
\ee
where ${\bm A}$ is the vector potential satisfying ${\bm B} = {\bm \nabla} \times {\bm A}$. While the combination ${\bm A} \cdot {\bm B}$ itself is not locally gauge invariant, the quantity ${\cal H}_{\rm mag}$ is gauge invariant under proper boundary conditions (e.g., ${\bm B} \rightarrow {\bm 0}$ at infinity). 
This ${\cal H}_{\rm mag}$ is called the magnetic helicity and corresponds to the helicity of electromagnetic fields. Geometrically, ${\cal H}_{\rm mag}$ has a meaning of the linking number of magnetic fluxes~\cite{Moffatt1969,Davidson2001}.
Therefore, Eq.~(\ref{helicity}) represents the total helicity conservation:
only the summation of the helicity in the fermion sector and the helicity in the gauge sector is conserved, but not separately when they are coupled to each other. 
When the system also possesses the motion of the fluids in the hydrodynamic regime, the total helicity conservation is extended to include the contributions of the fluid helicity ${\bm v} \cdot {\bm \omega}$ and cross helicity ${\bm v} \cdot {\bm B}$, where ${\bm v}$ is the fluid velocity and ${\bm \omega} = \frac{1}{2} {\bm \nabla} \times {\bm v}$ is the vorticity~\cite{Avdoshkin:2014gpa,Yamamoto:2015gzz}.

\subsection{\it Chiral magnetic effect \label{sec:CME}}

The chirality of fermions also leads to anomalous transport phenomena that do not appear in the usual parity-invariant matter. A typical example is the chiral magnetic effect (CME)~\cite{Vilenkin:1980fu,Nielsen:1983rb,Alekseev:1998ds,Fukushima:2008xe}: a current along the direction of the magnetic field $\bm B$ in the presence of a chirality imbalance characterized by the chiral chemical potential $\mu_5 \equiv (\mu_{\rm R} - \mu_{\rm L})/2$.

For simplicity, we first consider a right-handed fermion with charge $e$ at finite chemical potential $\mu_{\rm R}$ at zero temperature $T=0$.
In the presence of a homogeneous magnetic field ${\bm B} = B \hat {\bm z}$, the energy dispersion is quantized as shown in Fig.~\ref{fig1}(c). Then, the contribution of the LLL to the right-handed current in the $z$ direction can be computed as
\be
\label{CME_R}
j_{\rm R}^z = \frac{eB}{2\pi} \int_0^{\mu_{\rm R}} \frac{{\rm d}p_z}{2\pi} = \frac{e \mu_{\rm R}}{4\pi^2}B\,,
\ee
where the factor $eB/(2\pi)$ is again the Landau degeneracy per unit area in the $x$-$y$ plane. Again, the HLL do not contribute to the current in a way similar to the derivation of the chiral anomaly above. In this way, we have the current parallel to the magnetic field at non-zero $\mu_{\rm R}$. This is the CME for right-handed fermions. 

The extension of the CME to finite temperature is straightforward. In this case, since not only particles, but also antiparticles also contribute to the current, the total contribution is given by
\be
\label{CME_T}
j_{\rm R}^z = \frac{e B}{2\pi} \int_0^{\infty} \frac{{\rm d}p_z}{2\pi} \left[n_{p_z} (\mu_{\rm R}) - n_{p_z}(-\mu_{\rm R}) \right]\,,
\ee
where $n_p(\mu) = 1/[{\rm e}^{(p-\mu)/T} + 1]$ is the equilibrium distribution function for massless relativistic particles with momentum $p$. To evaluate this kind of integral, it is convenient to use the identity~\cite{Loganayagam:2012pz}
\be
\label{identity}
\int_0^{\infty} \frac{{\rm d}p}{2\pi} \left(\frac{p}{2\pi} \right)^{\! n} \left[n_p (\mu) - (-1)^n n_p(-\mu) \right] 
= \frac{({\rm i} T)^{n+1}}{n+1} B_{n+1} \left(\frac{1}{2} + \frac{\mu}{2\pi {\rm i} T} \right)\,, \quad (n=0,1,2,\cdots),
\ee
where $B_n(x)$ are Bernoulli polynomials defined by
\be
\frac{t {\rm e}^{xt}}{{\rm e}^t - 1} = \sum_{n=0}^{\infty} \frac{t^n}{n!}B_n (x)\,.
\ee
For example, $B_0(x)=1$, $B_1(x)=x-\frac{1}{2}$, $B_2(x)=x^2-x+\frac{1}{6}$, and so on.
Using this identity (\ref{identity}) for $n=0$, we can compute Eq.~(\ref{CME_T}) to obtain the same result as Eq.~(\ref{CME_R}) even at finite $T$. Therefore, the CME does not receive thermal correction.

Repeating a similar argument for left-handed fermions with chemical potential $\mu_{\rm L}$ at finite $T$, we have
\be
\label{CME_L}
j_{\rm L}^z = -\frac{e \mu_{\rm L}}{4\pi^2}B\,.
\ee
As will be discussed in Section~\ref{sec:CKT}, the transport coefficients of Eqs.~(\ref{CME_R}) and (\ref{CME_L}) are topologically quantized by the chirality $\chi=\pm 1$ and are the same as the chiral anomaly coefficient in Eq.~(\ref{anomaly_RL})~\cite{Son:2012wh}. 

By adding and subtracting Eqs.~(\ref{CME_R}) and (\ref{CME_L}), we obtain the vector and axial currents, ${\bm j} \equiv {\bm j}_{\rm R} + {\bm j}_{\rm L}$ and ${\bm j}_5 \equiv {\bm j}_{\rm R} - {\bm j}_{\rm L}$, for Dirac fermions at finite (vector-like) chemical potential $\mu \equiv (\mu_{\rm R} + \mu_{\rm L})/2$  and chiral chemical potential $\mu_5 \equiv (\mu_{\rm R} - \mu_{\rm L})/2$ as
\be
\label{CME}
{\bm j} = \frac{e \mu_5}{2\pi^2} {\bm B}\,, \qquad 
{\bm j}_5 = \frac{e \mu}{2\pi^2} {\bm B}\,,
\ee
respectively. The former is the CME for Dirac fermions, while the latter is called the chiral separation effect (CSE)~\cite{Son:2004tq,Metlitski:2005pr}. (The electric current due to the CME is given by $e {\bm j}$.) Note again that the transport coefficients of the CME and CSE in Eq.~(\ref{CME}) are the same as the chiral anomaly coefficient in Eq.~(\ref{anomaly}). While the transport coefficient of the CSE can receive corrections in the presence of dynamical gauge fields~\cite{Gorbar:2013upa,Jensen:2013vta}, that of the CME does not and is exact~\cite{Hattori:2017usa}.%
\footnote{The exactness of the transport coefficient of CME is for the DC conductivity in the zero-frequency limit (or long-time limit). Beyond this limit, time-dependent (and inhomogeneous) magnetic fields result in non-equilibrium corrections to the AC conductivity \cite{Satow:2014lva,Kharzeev:2016sut} or viscous corrections \cite{Hidaka:2018ekt} of CME, which can be obtained from the chiral kinetic theory discussed below.} This should be contrasted with the usual Ohmic current of the form ${\bm j} = \sigma {\bm E}$, for which the electrical conductivity $\sigma$ depends on $T$ and the details of the systems.

Note also that unlike the original claim~\cite{Vilenkin:1980fu}, the CME should not be regarded as an equilibrium current. In fact, any electric current vanishes in the ground state (or in global equilibrium) in the thermodynamic limit according to the generalization of the Bloch theorem~\cite{Bohm1949} that can be formulated solely based on the gauge symmetry~\cite{Yamamoto:2015fxa}. The CME should be rather regarded as a non-equilibrium current induced by the ``voltage" $\mu_{\rm R} - \mu_{\rm L}$. On the other hand, the CSE is truly a ground-state current, which can be understood as the Pauli paramagnetism of relativistic matter. It does not contradict the generalized Bloch theorem due to the absence of the corresponding axial gauge symmetry~\cite{Yamamoto:2015fxa}. 

So far, we have focused on the particle number (or electric) current, but the magnetic field also induces the energy current $T^{0z}$ in the direction of the magnetic field, where $T^{\mu \nu}$ is the energy-momentum tensor. For a single right- and left-handed fermions, the contribution reads
\be
\label{T0z}
T_{\rm R, L}^{0z} = \pm \frac{e B}{2\pi} \int_0^{\infty} \frac{{\rm d}p_z}{2\pi} p_z \left[n_{p_z} (\mu) + n_{p_z}(-\mu) \right]
= \pm \left(\frac{\mu^2}{8 \pi^2} + \frac{T^2}{24} \right) e B\,,
\ee
with $\mu = \mu_{\rm R, L}$, respectively, where we used the identity (\ref{identity}) for $n=1$ to obtain the final expression. For Dirac fermions, by adding the contribution of right- and left-handed sector, we obtain
\be
T^{0i} = \frac{e\mu \mu_5}{2\pi^2} B^i\,.
\ee

\subsection{\it Chiral vortical effect \label{sec:CVE}}

There is another chiral transport phenomenon induced by a vorticity 
${\bm \omega} \equiv \frac{1}{2} {\bm \nabla} \times {\bm v}$ with ${\bm v}$ a local fluid velocity or a global rotation ${\bm \Omega}$. 
This is the chiral vortical effect (CVE)~\cite{Vilenkin:1979ui, Son:2009tf,Landsteiner:2011cp}.%
\footnote{It is interesting to note that long after Ref.~\cite{Vilenkin:1979ui}, the CVE was rediscovered by the application of the gauge-gravity duality~\cite{Maldacena:1997re} to study the hydrodynamic properties corresponding to charged black holes in Refs.~\cite{Erdmenger:2008rm,Banerjee:2008th}.}

Physically, one may understand the CVE from the analogy between the magnetic field ${\bm B}$ and the rotation ${\bm \Omega}$. In classical mechanics, such an analogy is formulated as Larmor theorem stating that for a non-relativistic particle with mass $m$, they are related to each other via the correspondence: ${\bm B} \leftrightarrow 2m{\bm \Omega}$.
In fact, the Lorentz force acting on the particle with velocity ${\bm v}$ in a magnetic field ${\bm B}$ is given by ${\bm F} = {\bm v} \times {\bm B}$, while the Coriolis force acting on the particle with velocity ${\bm v}$ in rotating frame with angular velocity ${\bm \Omega}$ is 
${\bm F} = 2m {\bm v} \times {\bm \Omega}$, and so this correspondence is satisfied up to $O(\Omega^1)$ ignoring the centrifugal force of the order $\Omega^2$. For a relativistic particle, one may replace the mass $m$ by the energy $\varepsilon= p$, and one should have the correspondence ${\bm B} \leftrightarrow 2p{\bm \Omega}$. For a more rigorous justification of this replacement, see Ref.~\cite{Liu:2018xip}.

Let us consider the CVE for a right-handed fermion at finite $\mu_{\rm R}$ and at finite $T$ under a global rotation ${\bm \Omega} = \Omega \hat {\bm z}$. By applying the above correspondence to Eqs.~(\ref{CME_T}) and (\ref{T0z}) for CME, we obtain a particle number current and energy current induced by ${\bm \Omega}$ as
\bea
\label{CVE_T}
j_{\rm R}^z = \frac{\Omega}{2\pi} \int_0^{\infty} \frac{{\rm d}p}{2\pi} 2p \left[n_p (\mu_{\rm R}) + n_p(-\mu_{\rm R}) \right]
=\left(\frac{\mu_{\rm R}^2}{4\pi^2} + \frac{T^2}{12} \right)\Omega\,, \\
T^{0z} = \frac{\Omega}{2\pi} \int_0^{\infty} \frac{{\rm d}p}{2\pi} 2p^2 \left[n_p (\mu_{\rm R}) - n_p(-\mu_{\rm R}) \right]
=\left(\frac{\mu_{\rm R}^3}{6\pi^2} + \frac{\mu_{\rm R}T^2}{6} \right)\Omega\,,
\eea
where we used the identity (\ref{identity}) for $n=1$ and $n=2$, respectively. The expressions for left-handed fermions can be obtained by flipping the overall sign with the replacement $\mu_{\rm R} \rightarrow \mu_{\rm L}$. It is suggested that while the coefficient of the $T$-dependent part in Eq.~(\ref{CVE_T}) is related to the gravitational anomaly~\cite{Landsteiner:2011cp} and global anomaly~\cite{Golkar:2015oxw}, it receives renormalization in the presence of dynamical gauge fields~\cite{Golkar:2012kb,Hou:2012xg}.

For Dirac fermions, similarly to the discussion of the CME, we have
\bea
\label{CVE}
{\bm j} = \frac{\mu \mu_5}{\pi^2} {\bm \Omega}\,, \qquad 
{\bm j}_5 = \left(\frac{\mu^2 + \mu_5^2}{2\pi^2} + \frac{T^2}{6} \right) {\bm \Omega}\,, \qquad
T^{0i} =\left(\frac{3\mu^2 + \mu_5^2}{3\pi^2} + \frac{T^2}{3} \right) \mu_5 \Omega^i\,.
\eea

\subsection{\it Chiral plasma instability \label{sec:CPI}}

So far, we have considered chiral matter with a \emph{background} magnetic field or vorticity. When the electromagnetic fields are \emph{dynamical}, charged chiral matter with finite $\mu_5$ is unstable at certain length scale. This is the chiral plasma instability (CPI)~\cite{Joyce:1997uy,Akamatsu:2013pjd}. As reviewed shortly below, the CPI appears in a simple setup of Maxwell's equations including the contributions of the chiral anomaly and CME (sometimes called the anomalous Maxwell's equations)~\cite{Joyce:1997uy}. The CPI can also be described within a kinetic framework~\cite{Akamatsu:2013pjd}. Generally, plasma instabilities, including the Weibel instability~\cite{Weibel1959} and CPI, can be described as unstable collective modes of a system by a kinetic theory. Although usual plasma instabilities appear (or do not appear) whenever the distribution function is anisotropic (or isotropic)~\cite{Arnold:2003rq}, this criterion is only true in parity-invariant systems. In this sense, the CPI can be seen as an extension of the plasma instabilities to parity-violating systems, where the collective modes can become unstable even when the distribution function (except for the magnetic moment contribution) is isotropic~\cite{Akamatsu:2013pjd}. The CPI in anisotropic plasmas are studied in Refs.~\cite{Kumar:2016xuh,Carrington:2021bnk} and the CPI for non-Abelian plasmas are discussed in Refs.~\cite{Akamatsu:2013pjd,Akamatsu:2014yza}.

As the simplest setup of the Abelian CPI, we assume that the slow variables of the system are the electromagnetic fields, ${\bm E}$ and ${\bm B}$, and chiral charge $n_5$. (The coupling to the fluid velocity field will be discussed in Section~\ref{sec:ChMHD}.) 
In this case, the dynamical evolution of the system is described by the anomalous Maxwell's equations,
\begin{gather}
\label{Faraday}
\d_t {\bm B} = - {\bm \nabla} \times {\bm E}\,, \\
\label{Ampere}
{\bm \nabla} \times {\bm B} = \sigma {\bm E} + \frac{e^2 \mu_5}{2\pi^2} {\bm B}\,.
\end{gather}
In Eq.~(\ref{Ampere}), we have ignored the contribution of the displacement current $\d_t {\bm E}$, as can be verified a posteriori below Eq.~(\ref{k_CPI}). By solving Eq.~(\ref{Ampere}) in terms of ${\bm E}$ and inserting it to Eq.~(\ref{Faraday}), we obtain a closed equation in terms of ${\bm B}$ as
\be
\label{B}
\d_t {\bm B} = \frac{1}{\sigma} {\bm \nabla}^2 {\bm B} + \frac{e^2 \mu_5}{2 \pi^2 \sigma} {\bm \nabla} \times {\bm B}\,,
\ee
where we used ${\bm \nabla} \cdot {\bm B} = 0$.

To proceed further, it is convenient to use the helicity decomposition of the gauge field, ${\bm A}_{\pm} = (\hat {\bm x} \pm {\rm i}\hat {\bm y}){\rm e}^{-{\rm i} \omega t + {\rm i} k z}$ corresponding to right- and left-handed photons, where $\hat {\bm x}$ and $\hat {\bm y}$ are the unit vectors in the ${\bm x}$ and ${\bm y}$ directions, respectively. It then follows that ${\bm B}_{\pm} = \pm k {\bm A}_{\pm}$ and ${\bm \nabla} \times {\bm B}_{\pm} = \pm k {\bm B}_{\pm}$. By inserting this expression to Eq.~(\ref{B}), we obtain the dispersion relation for the modes ${\bm A}_{\pm}$,
\be
\omega = {\rm i} \frac{k}{\sigma} \left(\pm \frac{e^2 \mu_5}{2\pi^2} - k \right)\,,
\ee 
respectively. The frequency $\omega$ has a positive imaginary part 
for the wavenumber 
\be
0 < k < \frac{e^2 |\mu_5|}{2\pi^2}\,,
\ee
when $\mu_5 > 0$ for ${\bm A}_+$ and $\mu_5 < 0$ for ${\bm A}_-$. This signals the existence of the instability that the magnetic field  grows exponentially in the linear regime:
\be
{\bm B}_{\bm k}(t) = {\bm B}_{\bm k}(0) {\rm exp} \left[ {\frac{k}{\sigma} \left(\frac{e^2 |\mu_5|}{2\pi^2} - k \right) t} \right]\,.
\ee
The growth rate defined by $\Gamma \equiv {\rm Im}(\omega)$ becomes maximal at $k=k_{\rm CPI}$, where
\be
\label{k_CPI}
k_{\rm CPI} = \frac{e^2 |\mu_5|}{4\pi^2}, \qquad 
\Gamma_{\rm CPI} = \frac{1}{\sigma} \left( \frac{e^2 \mu_5}{4\pi^2} \right)^2\,.
\ee

If we were to include the contribution of the displacement current in Eq.~(\ref{B}), its contribution would be $|(1/\sigma){\bm \nabla} \times \d_t {\bm E}| = |(1/\sigma) \d_t^2 {\bm B}| \sim |(\Gamma_{\rm CPI}^2/\sigma) {\bm B}|$. When $\mu_5 \ll T$, this is much smaller than $|\d_t {\bm B}| \sim |\Gamma_{\rm CPI}{\bm B}|$ on the left-hand side of Eq.~(\ref{B}), where we used the parametric dependence of the conductivity, $\sigma \sim T/e^2$, in QED plasmas~\cite{Arnold:2000dr}.

To understand the fate of the CPI, let us focus on the unstable mode ${\bm A}_+({\bm k}, t)$ with the typical wavenumber $k = k_{\rm CPI}$. In this case, using ${\bm B}_+ = k_{\rm CPI} {\bm A}_+$ and ${\bm E}_+ = - (\Gamma_{\rm CPI}/k_{\rm CPI}) {\bm B}_+$, we obtain the average magnetic helicity $\bar {\cal H}_{\rm mag} \equiv {\cal H}_{\rm mag}/V$ and the time variation of the average chiral charge $\bar n_5 \equiv Q_5/V$ (with $V$ the volume of the system) for the unstable mode as
\begin{gather}
\bar {\cal H}_{\rm mag} = \frac{|{\bm B}_+({\bm k}, t)|^2}{|\mu_5|} > 0\,,
\\
\frac{{\rm d} \bar n_5}{{\rm d} t} = - \frac{2 |\mu_5|}{\sigma} \left( \frac{e^2}{4\pi^2} \right)^2 |{\bm B}_+({\bm k}, t)|^2 < 0\,,
\end{gather}
initially at this linear regime of the CPI.
Therefore, the chiral charge is reduced by the nonlinear effects of the chiral anomaly, and as a consequence, the CPI is weakened. At the same time, the chiral charge is converted to the magnetic helicity such that the total helicity is conserved according to Eq.~(\ref{helicity}).

It is shown in Ref.~\cite{Joyce:1997uy} that the magnetic field in this theory (without including the fluid velocity) exhibits the inverse cascade and the magnetic energy is transferred from small scales to larger scales. It is also revealed that such an inverse cascading feature is self-similar and their time evolution in the nonlinear regime at sufficiently large time $t$ is given by~\cite{Hirono:2015rla}
\be
k_{\rm CPI}(t) \propto \mu_5(t) \propto t^{-1/2}, \qquad g(k,t) \sim t \tilde g(t^{1/2}k)
\ee
where $g(k,t)$ is defined as
\be
{\cal H}_{\rm mag} = \int_0^{\infty} \frac{{\rm d}k}{\pi} k g(k, t)\,
\ee
and $\tilde g(k,t)$ is a scaling function which depends only on the particular combination $t^{1/2}k$. This was extended to the chiral MHD with fluid velocity~\cite{Yamamoto:2016xtu}, where the self-similar inverse cascades of the magnetic and kinetic energies and the scaling laws, e.g., for the correlation length of the magnetic field, $\xi_{B} \propto t^{1/2}$, are found. These scaling laws are also confirmed in Ref.~\cite{Pavlovic:2016gac}.

The classical statistical lattice simulations of the Abelian CPI are performed in Refs.~\cite{Buividovich:2015jfa,Mace:2019cqo}, both of which observe the inverse cascade of the helical magnetic field. Moreover, Ref.~\cite{Mace:2019cqo} observes a self-similar scaling
in the turbulent regime. The non-Abelian CPI is also studied numerically on the lattice~\cite{Akamatsu:2015kau}.

\section{Theoretical tools \label{sec:theory}}

To theoretically describe the dynamical evolution (including transport phenomena) of a given system, it is typically convenient to use low-energy effective theories that are valid for certain energy scales, instead of the microscopic theory itself. Two especially useful theoretical descriptions are hydrodynamics and kinetic theory. The former is formulated based on the local thermal equilibrium while the latter is applicable both in and out of equilibrium. Since the chiral anomaly and chiral transport phenomena are topological effects independent of the energy scale, they must also be incorporated in these effective theories. Below we will review these two theories including the chiral effects: chiral (magneto)hydrodynamics and chiral kinetic theory.

\subsection{\it Chiral (magneto)hydrodynamics \label{sec:Ch(M)HD}}
Generally, hydrodynamics is a low-energy effective field theory (EFT) for slow variables, called hydrodynamic variables, at long distance and long time scale much larger than the mean free path and mean free time of a system. These hydrodynamic variables are typically conserved quantities (such as the energy, momentum, and charge density) and Nambu-Goldstone modes when some continuous global symmetries are spontaneously broken. Here we consider the case without any spontaneous breaking of continuous symmetries and without any Nambu-Goldstone modes in the system.%
\footnote{When there are Nambu-Goldstone modes associated with spontaneous breaking of continuous global symmetries in a system, the chiral effects may also be accounted for by these modes, see, e.g.,  Refs.~\cite{Son:2004tq,Son:2007ny,Fukushima:2012fg,Huang:2017pqe,Manes:2019fyw}}

\subsubsection{\it Chiral hydrodynamics with background electromagnetic fields \label{sec:ChHD_background}}
Let us first consider a relativistic hydrodynamics in the presence of \emph{background} electromagnetic fields with the chiral effects above, which was first formulated in Ref.~\cite{Son:2009tf}. For simplicity, we here consider a system with a single chiral fermion charged under a $\U(1)$ symmetry, but its extension to multiple charges is straightforward~\cite{Son:2009tf,Sadofyev:2010pr,Neiman:2010zi}. The hydrodynamic equations are given by the conservation laws for the energy-momentum tensor $T^{\mu \nu}$ and particle number current $j^{\mu}$: 
\begin{align}
\label{EM_conservation}
\d_{\mu} T^{\mu \nu} &= e F^{\nu \lambda} j_{\lambda}, 
\\
\label{j_anomaly}
\d_{\mu} j^{\mu} &= - C E^{\mu} B_{\mu},
\end{align}
where the electromagnetic fields are defined in the local rest frame as $E^{\mu} = F^{\mu \nu} u_{\nu}$ and $B^{\mu} = \frac{1}{2} \epsilon^{\mu \nu \alpha \beta} u_{\nu} F_{\alpha \beta}$ with $u^{\mu}$ the local fluid velocity. The second equation above is the chiral anomaly relation (\ref{anomaly_RL}) with the coefficient $C=\chi e^2/(4\pi^2)$.

These hydrodynamic equations need to be supplemented by the constitutive equations expressing $T^{\mu \nu}$ and $j^{\mu}$ in terms of hydrodynamic variables order by order. At zeroth order in derivatives, $T^{\mu \nu}$ and $j^{\mu}$ are those of perfect fluids~\cite{Landau}:
\be
T^{\mu \nu}_{(0)} = (\epsilon + P) u^{\mu} u^{\nu} - P ^{\mu \nu},
\ee
\be
j^{\mu}_{(0)} = n u^{\mu},
\ee
where $\epsilon$ is the energy density, $P$ is the pressure, and $n$ is the number density.
For a moment, we will ignore the effects of dissipation (viscosity and conductivity), which we will add later. In this case, we have the corrections of the CME and CVE at first order in derivatives, 
\be
\label{T_s}
T^{\mu \nu}_{s(1)} = D_B (u^{\mu} B^{\nu} + B^{\mu} u^{\nu}) 
+ D_{\omega} (u^{\mu} \omega^{\nu} + \omega^{\mu} u^{\nu}),
\ee
\be
\label{j_s}
j^{\mu}_{s(1)} = C_B B^{\mu} + C_{\omega} \omega^{\mu},
\ee
where $\omega^{\mu}=\frac{1}{2}\epsilon^{\mu\nu\alpha\beta}u_{\nu}\d_{\alpha}u_{\beta}$ is the vorticity. The meaning of the subscript ``$s$" will be clarified shortly. From the results of the CME and CVE in the local rest frame (see Sections~\ref{sec:CME} and \ref{sec:CVE}), these transport coefficients read
\be
D_B = \chi e \left(\frac{\mu^2}{8\pi^2} + \frac{T^2}{24} \right)\,, \quad 
D_{\omega} = \chi \left(\frac{\mu^3}{6\pi^2} + \frac{\mu T^2}{6} \right)\,, \quad 
C_B = \chi \frac{e \mu}{4\pi^2}\,, \quad 
C_{\omega} = \chi \left(\frac{\mu^2}{4\pi^2} + \frac{T^2}{12} \right)\,.
\ee
Notice the relation $D_B = \frac{e}{2} C_{\omega}$, which may be understood as the Onsager relation~\cite{Landsteiner:2012kd}.

We pause here to make remarks on ambiguity of the choice of the local rest frame. In relativistic hydrodynamics, it is known that the choice of the local rest frame is not unique: two typical choices are the Landau frame where the fluid velocity is associated with the energy current $u_{\rm L}^{\mu} \propto T^{\mu \nu} (u_{\rm L})_{\nu}$~\cite{Landau} and the Eckart frame where the fluid velocity is associated with the particle number current $u^{\mu}_{\rm E} \propto j^{\mu}$~\cite{Eckart:1940te}.
As we have the anomalous corrections both in $T^{\mu \nu}_{(1)}$ and $j^{\mu}_{(1)}$, we are neither in the Landau frame nor Eckart frame above. Still, there is no anomalous correction in the entropy current $s^{\mu}$, and so the frame we implicitly adopted above corresponds to the ``entropy frame," where the fluid velocity is associated with the entropy current $u_{s}^{\mu} \propto s^{\mu}$.

We can also go to the Landau frame or Eckart frame from the entropy frame by a proper redefinition of the fluid velocity. For example, in order to go to the Landau frame, we can redefine the fluid velocity as 
\be
u_{\rm L}^{\mu} = u_{s}^{\mu}
+ \frac{1}{\epsilon+P}(D_B B^{\mu}+D_{\omega}\omega^{\mu})\,,
\ee
such that the anomalous corrections to the energy-momentum tensor goes away, $T^{\mu \nu}_{\rm L(1)} = 0$, at first order in derivatives. However, as a consequence of this redefinition, the expression of the anomalous current in Eq.~(\ref{j_s}) is modified as
\be
\label{j_L} 
j^{\mu}_{\rm L(1)} = \xi_B B^{\mu} + \xi_{\omega} \omega^{\mu}\,, 
\ee
where
\be
\label{xi_B}
\xi_B = C_B-\frac{n D_B}{\epsilon+P}\,, \quad
\xi_{\omega} = C_{\omega}-\frac{n D_\omega}{\epsilon+P} \,.
\ee
In this Landau frame, the explicit form of $\d_{\mu} T^{\mu \nu}$ (i.e., the left-hand side of the energy-momentum conservation~(\ref{EM_conservation})) remains the same as the conventional hydrodynamics, but the chiral corrections appear only in the number current $j^{\mu}$ as given by Eq.~(\ref{j_L}). Note that other than the Landau frame, such as the entropy frame, the chiral corrections also appear in $\d_{\mu} T^{\mu \nu}$.

Interestingly, even without the explicit computations of the transport coefficients $C_B$, $C_{\omega}$, $D_B$, and $D_{\omega}$, one can show that the conservation laws and chiral anomaly relation (\ref{j_anomaly}) together with the second law of thermodynamics completely fix their $\mu$-dependent parts by the chiral anomaly coefficient $C$ as~\cite{Son:2009tf,Neiman:2010zi}
\be
e C_B = C \mu, \qquad e D_B(\mu)=\frac{\mu^2}{2}C, \qquad 
e^2 C_{\omega}(\mu) = {\mu^2}C, \qquad e^2 D_{\omega}(\mu)=\frac{2}{3} \mu^3 C\,.
\ee
As the anomaly coefficient $C$ is exact non-perturbatively, this means that these $\mu$-dependent parts of the transport coefficients are also exact independently of interactions. On the other hand, their $T$-dependent parts are not constrained by the second law of thermodynamics~\cite{Neiman:2010zi}.

Finally, we should also add the effects of dissipation at first order in derivatives, which in the Landau frame read~\cite{Landau}
\be
T^{\mu \nu}_{{\rm diss}(1)} = \kappa \left[\d^{\mu}_{\perp}u^{\nu} + \d^{\nu}_{\perp}u^{\mu} - \frac{2}{3}P^{\mu\nu}(\d_{\perp}\cdot u) \right] + \zeta P^{\mu\nu}(\d_{\perp}\cdot u)\,,
\ee
\be
j^{\mu}_{{\rm diss}(1)} = \sigma T \d^{\mu}_{\perp} \left(\frac{\mu}{T} \right) + \sigma E^{\mu}\,,
\ee
where $\kappa$ is the shear viscosity, $\zeta$ is the bulk viscosity, $\sigma$ is the conductivity, $P^{\mu\nu}=g^{\mu\nu}-u^{\mu}u^{\nu}$ is the projection operator with respect to $u^{\mu}$, and $\d^{\mu}_{\perp} = P^{\mu \nu} \d_{\nu}$. We can derive the relativistic version of the Navier-Stokes equation for the evolution of $u^{\mu}$ by taking the transverse projection $P_{\nu}^{\rho}$ of Eq.~(\ref{EM_conservation}). In addition, the energy relation for the evolution of $\epsilon$ is obtained by multiplying Eq.~(\ref{EM_conservation}) by $u^{\nu}$.

The chiral hydrodynamics with external electromagnetic fields is extended to second order in derivatives in Ref.~\cite{Kharzeev:2011ds}.

\subsubsection{\it Chiral magnetohydrodynamics \label{sec:ChMHD}}
We then turn to chiral magnetohydrodynamics (MHD) coupled to 
\emph{dynamical} electromagnetic fields~\cite{Giovannini:2013oga,Boyarsky:2015faa,Yamamoto:2015gzz,Yamamoto:2016xtu,Hattori:2017usa}.
In this case, we also have to consider the equation of motion for the gauge field, i.e., Maxwell's equations, 
\be
\label{Maxwell}
\d_{\nu} F^{\nu \mu} = j^{\mu},
\ee
where we redefine $e j^{\mu}$ (with $j^{\mu}$ the particle number current previously) as the electric current $j^{\mu}$ for notational simplicity. 
Note however that the chiral MHD for a single chiral fermion is not well defined due to the gauge anomaly; the chiral anomaly relation (\ref{anomaly_RL}) for the single chiral fermion would be inconsistent with Eq.~(\ref{Maxwell}), since the latter leads to $\d_{\mu} j^{\mu} = \d_{\mu} \d_{\nu}F^{\nu \mu} = 0$. Hence, the chiral MHD is well defined only when such possible gauge anomalies are cancelled out. 

As a simple example, we consider a chiral MHD for a massless charged Dirac fermion at finite chemical potential $\mu$ and chiral chemical potential $\mu_5$, coupled to a heavy charged fermion. (An example of such a system is relativistic electrons coupled to non-relativistic protons.)
In this system, the total electric current $j^{\mu}$ is conserved while the axial current for the Dirac fermion, $j^{\mu 5} \equiv j^{\mu}_{\rm R} - j^{\mu}_{\rm L}$, is anomalous as we have seen in Section~\ref{sec:anomaly}.
Together with the energy-momentum conservation, the hydrodynamic equations read
\bea
\d_{\mu} T^{\mu \nu} &=& F^{\nu \lambda} j_{\lambda}, \\
\label{j_conservation}
\d_{\mu} j^{\mu} &=& 0, \\
\label{j5_anomaly}
\d_{\mu} j^{\mu 5} &=& - \frac{e^2}{2\pi^2} E^{\mu} B_{\mu}.
\eea
Here, the conservation of $n_5$ is violated by the chiral anomaly, but as long as the variation of $n_5$ is sufficiently slow, it can be regarded as a slow variable.

Note that the electric field $E^{\mu}$ is gapped and is decoupled from the low-energy dynamics at long time scale $t \gg \eta$, where $\eta \equiv 1/\sigma$ is the resistivity. In fact, by inserting the Ohm's law ${\bm j} = \sigma {\bm E}$ into Eq.~(\ref{j_conservation}) and by using the Gauss's law ${\bm \nabla} \cdot {\bm E} = n$ with $n$ the total charge density, we obtain $\d_t n + \sigma n = 0$. This equation shows that the charge density and the electric field are relaxation modes with the gap $\sigma$. In particular, the medium can be regarded as locally charge neutral, $n \approx 0$, in the long time scale $t \gg \eta$. For this time scale, we also have $|\d_t {\bm E}| = \eta |\d_t {\bm j}| \ll |{\bm j}|$, and so the displacement current in the Amp\`ere's law can be ignored.

As a demonstration to see the differences from the conventional MHD, let us now write down a part of hydrodynamic equations in the Landau frame explicitly for $|{\bm v}| \ll 1$ (which is satisfied both in the astrophysical systems and the early Universe to be discussed below). We will suppress the subscript ``L" for the Landau frame for notational simplicity. We also assume $\mu_5 \ll T$.
In this case, the spatial components of $j^{\mu}$ and $j^{\mu5}$ are given by
\begin{align}
\label{j_Dirac}
{\bm j} &= \sigma ({\bm E} + {\bm v} \times {\bm B}) + \xi_B {\bm B} + \xi_{\omega} {\bm \omega}, \\
{\bm j}_5 &= n_5 {\bm v} + \sigma_5 ({\bm E} + {\bm v} \times {\bm B}) + \xi_{B5} {\bm B} + \xi_{\omega 5} {\bm \omega},
\end{align}
respectively. Here,  
\be
\label{xi_B2}
\xi_B = \frac{e^2\mu_5}{2\pi^2}\,, \quad 
\xi_{\omega} = \frac{e\mu \mu_5}{2\pi^2}\,, \quad
\quad\xi_{B5} = \frac{e\mu}{2\pi^2}\,, \quad
\quad\xi_{\omega 5} = \frac{T^2}{12}\,,
\ee
and $\sigma_5 \propto \mu_5$ depends the detail of the system similarly to the conductivity $\sigma$. The axial current proportional to the electric field, $j^{\mu}_5 \propto E^{\mu}$, is called the chiral electric separation effect~\cite{Huang:2013iia}. Note that the corrections proportional to $n/(\epsilon + P)$ in Eq.~(\ref{xi_B}) do not appear when going to the Landau frame in Eq.~(\ref{xi_B2}) for the system with local charge neutrality $n=0$ (see, e.g., Ref.~\cite{Matsumoto:2022lyb}).

Combining Eq.~(\ref{j_Dirac}) and Amp\`ere's law, one can solve for $\bm E$ as
\be
{\bm E} = -{\bm v} \times {\bm B} + \eta {\bm \nabla} \times {\bm B} - \eta (\xi_B {\bm B} + \xi_{\omega} {\bm \omega}).
\ee
By inserting it to Faraday's law $\d_t {\bm B} = - {\bm \nabla} \times {\bm E}$ and using ${\bm \nabla} \times ({\bm \nabla} \times {\bm B}) = - {\bm \nabla}^2 {\bm B}$, one derives the induction equation for the magnetic field:
\be
\d_t {\bm B} = {\bm \nabla} \times ({\bm v} \times {\bm B}) + \eta {\bm \nabla}^2 {\bm B} + \eta {\bm \nabla} \times (\xi_B {\bm B} + \xi_{\omega} {\bm \omega}).
\ee
In the chiral MHD, the zeroth component of $j^{\mu 5}$ also acquires the additional terms proportional to $B^{0} \approx {\bm v} \cdot {\bm B}$ and $\omega^{0} \approx {\bm v} \cdot {\bm \omega}$, which can be understood as the contributions of cross helicity and fluid helicity, respectively~\cite{Avdoshkin:2014gpa,Yamamoto:2015gzz}. 

The construction of the chiral MHD above may look somewhat phenomenological because these equations of motion are obtained by adding the chiral effects to the usual MHD by hand. From the viewpoint that hydrodynamics is an EFT, it would be more desirable to obtain the chiral MHD equations based on a symmetry principle together with a systematic derivative expansion in a way similar to the construction of the conventional MHD as an EFT~\cite{Grozdanov:2016tdf,Glorioso:2018kcp}. The latter became possible thanks to the recent development of the notion of generalized global symmetries or higher-form symmetries~\cite{Gaiotto:2014kfa}. One systematic derivation is given in Ref.~\cite{Hattori:2017usa}, which assumes not only the derivative expansion, but also the expansion in terms of the anomaly coefficient, and so the domain of validity is limited. Alternative attempts are made in Refs.~\cite{Iqbal:2020lrt,Das:2022auy} using holography. 


\subsection{\it Chiral kinetic theory \label{sec:CKT}}
To study the dynamical evolution of a relativistic system out of equilibrium, one may resort to a relativistic kinetic theory which describes the evolution of the distribution function of quasi-particles in phase space.
However, the classical kinetic theory, such as the Boltzmann equation, does not capture quantum effects like the chiral anomaly and chiral transport phenomena above. In recent years, there have been intensive studies on the developments and applications of such a quantum kinetic theory for massless fermions, known as the chiral kinetic theory (CKT)~\cite{Son:2012wh,Stephanov:2012ki}. See Ref.~\cite{Hidaka:2022dmn}
for a review of the CKT and its massive extensions and Ref.~\cite{Gorbar2021} for formulations and applications of the CKT in the context of Weyl/Dirac semimetals.

\subsubsection{\it Chiral kinetic theory from Berry curvature \label{sec:Berry}}
The quantum effects responsible for the chiral anomaly and chiral transport phenomena can be incorporated into the kinetic theory by introducing a notion widely applied in condensed matter physics: the Berry phase~\cite{Berry}. Generically, when a Hamiltonian depends on a set of parameters ${\bm n} = (n_1, n_2, \cdots)$ and the Hilbert space of the wave function $u_{\bm n}$ is constrained on a hypersurface ${\cal S}$ in the ${\bm n}$ space (which corresponds to the so-called adiabatic approximation), one can define a Berry connection and Berry curvature on ${\cal S}$. For simplicity and concreteness, let us assume that the parameter space is three-dimensional  ${\bm n} = (n_1, n_2, n_3)$. In this case, the Berry connection and Berry curvature are given by
\be
\bm{\mathcal A}_{\bm n} \equiv {\rm i}u_{\bm n}^{\dag} {\bm \nabla}_{\bm n} u_{\bm n}, \qquad {\bm \Omega}_{\bm n} \equiv {\bm \nabla}_{\bm n} \times \bm{\mathcal A}_{\bm n},
\ee
respectively. One can also define a ``Wilson loop" of $\bm{\mathcal A}_{\bm n}$ for a closed loop $C$ in the ${\bm n}$ space as
\be
\theta({\cal C}) = \int_{\cal C} {\bm{\mathcal A}}_{\bm n} \cdot {\rm d}{\bm n} = \int_{{\cal S}_{\cal C}} {\bm \Omega}_{\bm n} \cdot {\rm d}{\bm S}_{\bm n},
\ee
where we used the Stokes theorem. Here, ${\cal S}_{\cal C}$ is a two-dimensional surface whose boundary is ${\cal C}$.
This $\theta(\cal{C})$ is called the Berry phase.

Since the spin of the chiral fermion is aligned with its momentum direction depending on the chirality, a Berry phase is generated when the particle propagates under the adiabatic approximation. For the Hamiltonian of chiral fermions, $H = \chi {\bm \sigma} \cdot {\bm p}$, the Berry curvature in momentum space is given by
\be
{\bm \Omega}_{\bm p}=\chi \frac{\bm p}{2|\bm p|^3}\,,
\ee
which takes the form of the ``magnetic field" of a magnetic monopole at ${\bm p} = {\bm 0}$ in momentum space.%
\footnote{In order for the Berry curvature to be well defined, particle motion must be sufficiently away from the origin in momentum space. This corresponds to the condition of the adiabatic approximation that the two energy levels $\pm |{\bm p}|$ do not cross with each other.}
In fact, the monopole charge $K$, given by the integral over a surface surrounding the origin, is equal to the chirality:
\be
\label{monopole}
K \equiv \frac{1}{2\pi} \int_S {\bm \Omega}_{\bm p} \cdot {\rm d}{\bm p} = \chi\,.
\ee

The semi-classical action of a chiral fermion with charge $e$ is given by~\cite{Son:2012wh,Stephanov:2012ki,Son:2012zy}
\be
S = \int \left[({\bm p} + e {\bm A}) \cdot {\rm d}{\bm x} - {\bm{\mathcal A}}_{\bm p} \cdot {\rm d} {\bm p} - (\varepsilon_{\bm p} + e\phi){\rm d}t \right]
\ee
where ${\bm A}$ and $\phi$ are the vector potential and the scalar potential in position space, respectively, ${\bm{\mathcal A}}_{\bm p}$ is the Berry connection in momentum space, and $\varepsilon_{\bm p}$ is the energy of the chiral fermion. Note here that $\varepsilon_{\bm p}$ generally depends on $\bm x$, as will be explained below. The equations of motion derived from this action read
\be
\label{EOM_Berry0}
\dot{\bm x}={\bm v}+\dot{\bm p}\times {\bm \Omega}_{\bm p}, \qquad
\dot{\bm p}=e\big(\tilde{\bm E}+\dot{\bm x}\times\bm B\big),
\ee
where ${\bm v} \equiv {\bm \nabla}_{\bm p} \varepsilon_{\bm p}$ and $e \tilde {\bm E} \equiv e {\bm E} - {\bm \nabla}_{\bm x} \varepsilon_{\bm p}$. The term $\dot{\bm p}\times {\bm \Omega}_{\bm p}$ in the first equation is the ``Lorentz force" in momentum space, which is dual of the Lorentz force $\dot{\bm x}\times {\bm B}({\bm x})$ in position space. Solving these simultaneous equations for $\dot{\bm x}$ and $\dot{\bm p}$, we get
\bea
\label{EOM_Berry}
\dot{\bm x}&=&\frac{1}{\sqrt{G}}\left[{\bm v}+ e\tilde{\bm E}\times {\bm \Omega}_{\bm p}+ (\bm v\cdot {\bm \Omega}_{\bm p})e\bm B\right]\,,
\nonumber \\
\dot{\bm p}&=&\frac{1}{\sqrt{G}}\left[e(\tilde{\bm E}+{\bm v}\times \bm B)+ e^2(\tilde{\bm E}\cdot {\bm B}){\bm \Omega}_{\bm p}\right]\,,
\eea
where $\sqrt{G}=1+e{\bm B}\cdot {\bm \Omega}_{\bm p}$.

Generally, the kinetic equation for the distribution function of on-shell chiral particles, $f_{\chi}(t, \bm x, \bm p)$, is given by
\bea
\label{Liouville_th}
\frac{{\rm d}f_{\chi}}{{\rm d}t}= \d_t f_{\chi}+\dot{\bm x}\cdot {\bm \nabla}_{\bm x} f_{\chi}+\dot{\bm p}\cdot {\bm \nabla}_{\bm p}f_{\chi}=C[f_{\chi}]\,,
\eea
where $C[f_{\chi}]$ is the collision term. 
For chiral fermions, due to the Berry curvature corrections in the equations of motion above, the kinetic equation is also modified. By inserting Eq.~(\ref{EOM_Berry}) into Eq.~(\ref{Liouville_th}), we arrive at the CKT~\cite{Son:2012zy}:%
\footnote{We here focus on the CKT for particles, but the inclusion of antiparticles in the CKT is straightforward~\cite{Manuel:2013zaa}.}
\be
\label{CKT}
\d_t f_{\chi} + \frac{1}{\sqrt{G}} 
\left[\bm v + e \tilde {\bm E} \times {\bm \Omega}_{\bm p} 
+ (\bm v \cdot {\bm \Omega}_{\bm p})e{\bm B} \right] \cdot 
{\bm \nabla}_{\bm x} f_{\chi}
+ \frac{1}{\sqrt{G}} \left[e \tilde {\bm E} + \bm v \times e{\bm B} 
+ e^2(\tilde {\bm E} \cdot {\bm B}) {\bm \Omega}_{\bm p} \right] \cdot {\bm \nabla}_{\bm p} f_{\chi}
= C[f_{\chi}]\,.
\ee
From the Lorentz invariance of the system, one can also show that the dispersion of chiral fermions has a correction related to the Berry curvature as~\cite{Son:2012zy,Chen:2014cla}
\be
\label{dispersion}
\varepsilon_{\bm p}=|\bm p|(1- e\bm B\cdot{\bm \Omega}_{\bm p}).
\ee
which may physically be understood as the magnetic moment coupling (and hence, $\varepsilon_{\bm p}$ generally depends on $\bm x$ through the magnetic field).

One can define the number density and current of chiral fermions from Eq.~(\ref{EOM_Berry}) as~\cite{Son:2012wh,Stephanov:2012ki,Son:2012zy}
\bea
\label{n_CKT}
&&n_{\chi} = \int \frac{{\rm d}^3{\bm p}}{(2\pi)^3} \sqrt{G} f_{\chi}\,,
\\
\label{j_CKT}
&&{\bm j}_{\chi} = \int \frac{{\rm d}^3{\bm p}}{(2\pi)^3} \sqrt{G} \dot {\bm x} f_{\chi}
= \int \frac{{\rm d}^3{\bm p}}{(2\pi)^3} \left[{\bm v} + e \tilde {\bm E} \times {\bm \Omega}_{\bm p} 
+ (\bm v \cdot {\bm \Omega}_{\bm p})e{\bm B} \right]f_{\chi}\,,
\eea
where the phase-space modifications by the factor $\sqrt{G}$ originate from the Berry curvature corrections in the equations of motion~(\ref{EOM_Berry})~\cite{Xiao:2005qw,Duval:2005}.
By multiplying Eq.~(\ref{CKT}) by $\sqrt{G}$ and integrating over momentum, we find~\cite{Son:2012wh,Stephanov:2012ki,Son:2012zy}
\be
\d_t n_{\chi} + {\bm \nabla} \cdot {\bm j}_{\chi} = - \int \frac{{\rm d}^3{\bm p}}{(2\pi)^3}
\left({\bm \Omega}_{\bm p} \cdot {\bm \nabla}_{\bm p} f_{\chi} \right)
e^2 {\bm E} \cdot {\bm B}
= \chi \frac{e^2}{4\pi^2} {\bm E} \cdot {\bm B}\,,
\ee
which reproduces the chiral anomaly relation (\ref{anomaly_RL}) above.
Here we assumed that the collision term satisfies
\be
\int \frac{{\rm d}^3{\bm p}}{(2\pi)^3} \sqrt{G}C[f_{\chi}] = 0\,.
\ee
This derivation shows that the coefficient of the chiral anomaly is quantized by the monopole charge $K$ (or chirality $\chi$) in Eq.~(\ref{monopole}), independent of the details of the interactions. Also, in local thermal equilibrium without any flow velocity, only the third term in Eq.~(\ref{j_CKT}) survives to yield
\be
{\bm j}_{\chi} = \chi \frac{e \mu}{4\pi^2} {\bm B}\,,
\ee
reproducing the CME. This shows the quantization of the coefficient of the CME by the monopole charge $K$ in Eq.~(\ref{monopole}). Analogously, one can also derive the CVE from CKT~\cite{Stephanov:2012ki,Chen:2014cla,Chen:2015gta,Liu:2018xip}.

\subsubsection{\it Chiral kinetic theory from Wigner functions}
A drawback of the formulation of CKT in terms of the Berry curvature is that it lacks the manifest Lorentz invariance, which is necessary especially to extend CKT in curved spacetime. One can attempt to construct a manifestly Lorentz-covariant CKT based on the finite Lorentz transformation for a spin tensor (see Eq.~(\ref{S}) below)~\cite{Chen:2015gta}. In this approach, however, the form of the collision term may be guessed, but cannot be systematically determined. These issues can be resolved by constructing CKT from quantum field theories based on the Wigner function. 

The derivation of quantum kinetic theories from Wigner functions can be traced back to the studies in Refs.~\cite{Elze:1986qd,Vasak:1987um,Elze:1989un} for Dirac fermions (see, e.g., Ref.~\cite{Blaizot:2001nr} for a review), while its extension to CKT for Weyl/Dirac fermions was made by Refs.~\cite{Son:2012zy,Chen:2012ca,Manuel:2013zaa,Manuel:2014dza,Hidaka:2016yjf,Hidaka:2017auj,Huang:2018wdl} in Minkowski spacetime (see also Ref.~\cite{Hidaka:2022dmn} for a more recent review). However, it is more useful to further generalize the formalism to curved spacetime for astrophysical applications~\cite{Liu:2018xip,Yamamoto:2020zrs}. In curved spacetime, the Wigner functions are defined as the quantum expectation values of correlation functions of chiral fermions in phase space on a tangent bundle as
\bea
\label{def_WF_c}
\grave{W}_{\chi}^{\lessgtr}(q,x)\equiv\int \frac{{\rm d}^4y}{\sqrt{-g(x)}} {\rm e}^{-{\rm i}q\cdot y} W_{\chi}^{\lessgtr}(x,y)\,,
\eea
where $W_{\chi}^<(x,y) \equiv \langle \psi^{\dagger}_{\chi}(x+y/2)\psi_{\chi}(x-y/2)\rangle$ and $W_{\chi}^>(x,y) \equiv \langle\psi_{\chi}(x-y/2){\psi}^{\dagger}_{\chi}(x+y/2)\rangle$ are the lesser and greater propagators for the chiral fermion $\psi_{\chi}$ with chirality $\chi = \pm 1$ and $g(x)$ denotes the determinant of the metric. 
Also, we introduce $\psi_{\chi}(x,y)={\rm e}^{y\cdot\tilde{D}}\psi_{\chi}(x)$ and $\psi^{\dagger}_{\chi}(x,y)=\psi^{\dagger}_{\chi}(x){\rm e}^{y\cdot\overleftarrow{\tilde{D}}}$, where $\tilde{D}_{\mu}=\nabla_{\mu}+{\rm i}e A_{\mu}+\Gamma^{\lambda}_{\mu\nu}y_{\lambda}\partial_y^{\nu}$ is the horizontal lift and $\nabla_{\mu}$ is the covariant derivative with respect to $x^{\mu}$. The lesser and greater propagators of right- and left-handed fermions can be parameterized as 
$\grave{W}_{+}^{\lessgtr}(q,x)=\bar{\sigma}^{\mu} {\cal W}^{\lessgtr}_{+\mu}(q,x)$ and $\grave{W}_{-}^{\lessgtr}(q,x)=\sigma^{\mu}{\cal W}_{-\mu}^{\lessgtr}(q,x)$, respectively.

The dynamics of Wigner functions in phase space are dictated by the so-called Kadanoff-Baym equations. Based on the approach in Refs.~\cite{Hidaka:2016yjf,Hidaka:2017auj,Liu:2018xip}, we obtain the following master equations up to $O(\d)$ from the Kadanoff-Baym equations with collisions for chiral fermions:%
\footnote{We here ignored the one-particle potential denoted by $\Sigma^{\delta}$ in Ref.~\cite{Hidaka:2016yjf}, which may modify the dispersion relation of the fermions.}
\bea
\label{m_eq1}
\mathcal{D}_{\mu} {\cal W}_{\chi}^{< \mu}&=&0,
\\\label{m_eq2}
q_{\mu} {\cal W}_{\chi}^{< \mu}&=&0,
\\\label{m_eq3}
\mathcal{D}_{\mu}{\cal W}^{<}_{\chi,\nu}-\mathcal{D}_{\nu}{\cal W}^{<}_{\chi,\mu}&=&2\chi \epsilon_{\mu\nu\rho\sigma}q^{\rho}{\cal W}_{\chi}^{< \sigma},
\eea
where
\begin{eqnarray}
\mathcal{D}_{\mu}\mathcal{W}_{\chi,\nu}^< \equiv \Delta_{\mu}\mathcal{W}^<_{\chi,\nu}-\Sigma^<_{\chi,\mu}\mathcal{W}^>_{\chi,\nu}+\Sigma^>_{\chi,\mu}\mathcal{W}^<_{\chi,\nu}
\end{eqnarray}
and $\Delta_{\mu}\mathcal{W}_{\chi,\nu} \equiv \big(D_{\mu}+e F_{\lambda\mu}\partial^{\lambda}_q\big)\mathcal{W}_{\chi,\nu}$ with $D_{\mu} \equiv \nabla_{\mu}-\Gamma^{\lambda}_{\mu\nu}q^{\nu}\partial_{q\lambda}$ being the dual operator of the horizontal lift $\tilde{D}_{\mu}$ in the tangent space $(x^{\mu},q^{\mu})$ in the absence of gauge fields. An important property of this dual operator is that $D_{\mu}q^{\nu}=0$. In addition, $\Sigma^{<}_{\chi,\mu}$ and $\Sigma^{>}_{\chi,\mu}$ are the lesser and greater self-energies for chiral fermions, which depend on details of interactions in a given system. 

In order to solve these equations, one has to further perform the gradient expansion. Note that this expansion is equivalent to the $\hbar$ expansion and is generally different from the gradient expansion in the hydrodynamics. 
The solutions of Eqs.~(\ref{m_eq2}) and (\ref{m_eq3}) up to $O(\d)$ take the form~\cite{Hidaka:2016yjf,Liu:2018xip,Yamamoto:2020zrs}%
\footnote{We here ignored the contribution of antiparticles, which can be included by multiplying the right-hand side of Eq.~(\ref{WF_L_full}) by the sign of $q\cdot n$.}
\begin{eqnarray}\label{WF_L_full}
\mathcal{W}^{\lessgtr \mu}_{\chi}=2\pi\Big[\delta(q^2)\big(q^{\mu}+ S^{\mu\nu}_{\chi(n)}\mathcal{D}_{\nu}\big) + \chi e \tilde{F}^{\mu\nu}q_{\nu}\delta'(q^2)\Big]f_{\chi}^{\lessgtr}\,,
\end{eqnarray} 
where $\mathcal{D}_{\mu}f_{\chi} \equiv \Delta_{\mu}f_{\chi}-\mathcal{C}_{\mu}[f_{\chi}]$ and $\mathcal{C}_{\mu} [f_{\chi}^{\lessgtr}] \equiv\Sigma_{\chi,\mu}^{\lessgtr}f^{\gtrless}_{\chi}-\Sigma_{\chi,\mu}^{\gtrless}f^{\lessgtr}_{\chi}$, with $f^{<}_{\chi}=f_{\chi}$ and $f^{>}_{\chi}=1-f_{\chi}$ the distribution functions of incoming and outgoing fermions, respectively. Here 
\begin{eqnarray}
\label{S}
S^{\mu\nu}_{\chi(n)}=\chi \frac{\epsilon^{\mu\nu\alpha\beta}q_{\alpha}n_{\beta}}{2q\cdot n}
\end{eqnarray}
denotes the spin tensor, which depends on a timelike frame vector $n^{\mu}(x)$ satisfying $n^2=1$ and $\delta'(q^2)=\partial\delta(q^2)/\partial q^2$. The frame vector $n^{\mu}(x)$ appears due to the ambiguity to define the spin for massless particles. 
Note that $\mathcal{W}^{\lessgtr \mu}_{\chi}$ themselves are independent of the choice of $n^{\mu}$. Also, the $O(\d)$ corrections proportional to $q^{\mu}\delta(q^2)$, which are the trivial solutions for Eqs.~(\ref{m_eq2}) and (\ref{m_eq3}), can be absorbed into $f_{\chi}$~\cite{Hidaka:2016yjf}.
	
Equation~(\ref{WF_L_full}) incorporates two terms related to the quantum corrections at $O(\d)$. One is the $\delta'(q^2)$ term, which leads to the modification on the on-shell condition due to the magnetic-moment coupling in the presence of background electromagnetic fields (see Eq.~(\ref{dispersion}))~\cite{Son:2012zy,Chen:2014cla}. 
The other is the term associated with the spin tensor $S^{\mu\nu}_{\chi(n)}$, called the side-jump term, which exists even without background electromagnetic fields. Physically, this term originates from the spin-momentum locking of chiral fermions under the angular-momentum conservation~\cite{Chen:2014cla,Chen:2015gta,Yang:2018lew}.
Due to the side-jump term, the distribution functions $f_{\chi}$ are now frame dependent and they follow the modified Lorentz transformation between different frames. Such a modified frame transformation ensures the Lorentz covariance of Wigner functions and CKT~\cite{Chen:2014cla}.

By inserting Eq.~(\ref{WF_L_full}) into Eq.~(\ref{m_eq1}), we derive the CKT with collisions in curved spacetime up to $O(\d)$~\cite{Yamamoto:2020zrs} (see Refs.~\cite{Hidaka:2016yjf, Hidaka:2017auj} for the case in Minkowski spacetime):
\begin{align}
\label{eq:CKT_full}
\left[(q + \ell) \cdot \Delta + S^{\mu\nu}_{\chi(n)}\left(e \nabla_{\mu}F^{\lambda}_{\,\,\,\nu}-q^{\rho}R^{\lambda}_{\rho\mu\nu}\right)\partial_{q\lambda}
\right]f_{\chi} \nonumber 
=(q + \ell) \cdot \mathcal{C} [f_{\chi}]+S^{\alpha\beta}_{\chi(n)}\big[(1-f_{\chi})\Delta^>_{\chi,\alpha}\Sigma^{<}_{\chi,\beta}-f_{\chi }\Delta^<_{\chi,\alpha}\Sigma^{>}_{\chi,\beta}\big]
\,,\\
\end{align}
where 
\be
\ell^{\nu} = \frac{e S^{\mu\nu}_{\chi(n)}F_{\mu\rho}n^{\rho}}{q\cdot n}+D_{\mu}S^{\mu\nu}_{\chi(n)}
\ee
and $\Delta^{\gtrless}_{\chi, \mu} \equiv \Delta_{\mu}+\Sigma_{\chi, \mu}^{\gtrless}$, supplemented by the on-shell condition 
\be
\label{on-shell}
q^2+ e F_{\alpha\beta}S^{\alpha\beta}_{\chi(n)}=0.
\ee
The on-shell condition (\ref{on-shell}) matches the dispersion relation (\ref{dispersion}) above. Equation~(\ref{eq:CKT_full}) also reproduces the CKT with the Berry corrections in Eq.~(\ref{CKT})~\cite{Son:2012zy}. Here, collision terms can be expressed in the language of quantum field theory. 
	
Solving the CKT and obtaining $f_{\chi}$, one can then insert $f_{\chi}$ into the Wigner functions~(\ref{WF_L_full}) to obtain physical observables. For example, based on the definition in field theory, the particle number current and symmetric energy-momentum tensor for chiral fermions can be derived from the lesser propagators as
\bea
\label{j_and_T}
j^{\mu}_{\chi}=2\int_q\mathcal{W}^{<\mu}_{\chi},
\qquad
T^{\mu\nu}_{\chi}=\int_q \big(\mathcal{W}^{<\mu}_{\chi}q^{\nu}+\mathcal{W}^{<\nu}_{\chi}q^{\mu}\big),
\eea
where $\int_q \equiv \int{\rm d}^4q\sqrt{-g}/(2\pi)^{4}$.

\section{Chiral effects in astrophysics}
\label{sec:astro}
In this section, we review the relevance of chiral effects in astrophysical systems. One realization of chiral matter in astrophysical systems is the lepton matter in core-collapse supernovae. The chirality imbalance of leptons is generated via the electron capture process, ${\rm p} + {\rm e}_{\rm L}^- \rightarrow {\rm n} + \nu^{\rm e}_{\rm L}$, which involves only left-handed leptons. In particular, because neutrinos are only left-handed in the Standard Model (SM) of particle physics, their chiral effects may be relevant to pulsar kicks, magnetars, and core-collapse supernova explosions, as we will describe in Sections~\ref{sec:kick}, \ref{sec:magnetar}, and \ref{sec:CCSN}, respectively. For this purpose, we will first review the non-equilibrium effective theories for neutrinos, such as the chiral radiation transport theory and chiral hydrodynamics for neutrinos in Section~\ref{sec:EFT_neutrino}.
One may also attempt to study the possible impacts of the chirality imbalance of electrons on neutron star and supernova physics, although it is suppressed by the chirality flipping, as will be touched upon in Sections~\ref{sec:kick} and \ref{sec:magnetar}. The other applications in astrophysical systems will be discussed in Section~\ref{sec:other_astro}.

\subsection{\it Non-equilibrium effective theories for neutrinos \label{sec:EFT_neutrino}}

In high-energy astrophysical environments, such as supernovae and neutron stars, the typical energy scale is much smaller than the $W$ or $Z$ boson masses $\sim 100\,{\rm GeV}$, and we can use the low-energy effective theory of the Weinberg-Salam electroweak theory---the four-Fermi theory of the weak interaction. Below we summarize the four-Fermi interactions that will be used in the later subsections.

The neutrino absorption on nucleons
$\nu^{\rm e}_{\rm L}+{\rm n} \rightleftarrows {\rm e}_{\rm L}+{\rm p}$
is mediated by the charged $W$ boson. By integrating out heavy $W$ at low energy, this is described by charged current interactions,
\begin{eqnarray}
\label{4-Fermi_cc}
\mathcal{L}^{\rm cc}_{\rm int}=\frac{{G}_{\rm F}}{\sqrt{2}}(J_{\ell}^{-})_{\mu} (J_{\rm N}^{+})^{\mu}+{\rm h.c.}\,,
\end{eqnarray}
where $G_{\rm F}$ is the Fermi constant and $(J_{\ell}^{-})_{\mu} = \bar{\psi}_{\rm e} \gamma_{\mu}(1-\gamma^5)\psi_{\nu}$ and $(J_{\rm N}^{+})^{\mu} = \bar{\psi}_{{\rm p}}\gamma^{\mu}(g_{\rm V}-g_{\rm A}\gamma^5)\psi_{\rm n}$ are the lepton and nucleon charged currents, respectively, with $g_{\rm V}=1$ and $g_{\rm A} \approx 1.27$; for recent calculations of $g_{\rm A}$ in lattice QCD, see Ref.~\cite{Chang:2018uxx} and references therein.

On the other hand, the elastic neutrino-nucleon scattering
$\nu^{\ell}_{\rm L}+{\rm N} \rightleftarrows \nu^{\ell}_{\rm L}+{\rm N}$
is mediated by the charge neutral $Z$ boson. After integrating out $Z$, this scattering is described by the neutral current interactions,
\begin{eqnarray}
\label{4-Fermi_nc}
\mathcal{L}^{\rm nc}_{\rm int}=\frac{G_{\rm F}}{\sqrt{2}}(J_{\nu})_{\mu} (J_{\rm N})^{\mu}+{\rm h.c.}\,,
\end{eqnarray}
where $(J_{\nu})_{\mu} = \bar{\psi}_{\nu}\gamma_{\mu}(1-\gamma^5)\psi_{\nu}$ 
and $(J_{\rm N})^{\mu} = \frac{1}{2}\bar{\psi}_{\rm N}\gamma^{\mu}(c_{\rm V}-c_{\rm A}\gamma^5)\psi_{\rm N}$ are the lepton and nucleon neutral currents, respectively.
Here $c_{\rm V}= -1$ and $c_{\rm A} = - g_{\rm A}$ for ${\rm N} = {\rm n}$, 
and $c_{\rm V}= 1-4 \sin^2 \theta_{\rm w}$ and $c_{\rm A} = g_{\rm A}$ for ${\rm N} = {\rm p}$, where $\theta_{\rm w}$ is the weak mixing angle.

To describe the non-equilibrium evolution of astrophysical systems, however, this microscopic theory itself is not convenient, as we need to consider non-equilibrium many-body problems of nucleons, electrons, and neutrinos.

\subsubsection{\it Chiral radiation transport theory for neutrinos}\label{sec:ChRTT}

In core-collapse supernovae, most of the gravitational binding energy of a massive star is released in the form of neutrinos. It is for this reason that proper treatment of neutrino transport is indispensable to unravel the core-collapse supernova explosions and subsequent evolution. The theoretical formulations and numerical simulations is based on the neutrino radiation transport theory~\cite{Lindquist1966,Castor1972,Bruenn:1985en,Mihalas,Sumiyoshi:2012za}, where neutrinos treated as radiation are coupled to the matter sector described by hydrodynamics. However, the conventional radiation transport theory misses the important feature of left-handedness of neutrinos.

In order to describe such chiral effects of neutrinos, we have to construct a chiral radiation transport theory for left-handed neutrinos including quantum corrections on the interaction with matter, which has been derived in Ref.~\cite{Yamamoto:2020zrs}. For our purpose, it is convenient to work in the inertial frame with a proper choice of the frame vector, $n^{\mu}=(1,{\bm 0})$, such that $q\cdot n=E_{\rm i}$.%
\footnote{For construction of the radiation transport equation in core-collapse supernovae from the Einstein-Vlasov equation, one could alternatively work in the comoving frame. See, e.g., Ref.~\cite{Mihalas} for detailed comparisons and discussions of the transport equations in two frames.}
A choice of the frame vector is for technical simplifications but the physical results do not depend on such a choice.   

From Eq.~(\ref{eq:CKT_full}) for left-handed neutrinos in the absence of electromagnetic fields, the chiral radiation transport equation can be written as
\begin{gather}
	\label{CKT_inertial_con}
	\Box_{\rm i}f_{q}^{(\nu)}
	=\frac{1}{{E_{\rm i}}}\left[(1-f^{(\nu)}_{q})\Gamma^{<}_{q}-f_{q}^{(\nu)}\Gamma^{>}_{q} \right]\,, \\
	\Gamma^{\lessgtr}_{q} = (q^{\nu} - S^{\mu\nu}_{q} D_{\mu}) \Sigma^{\lessgtr}_{\nu}\,,
\end{gather}
where $\Box_{\rm i}\equiv q\cdot D/{E_{\rm i}}$ is given by~\cite{Sumiyoshi:2012za}
\begin{eqnarray}
	\Box_{\rm i} = \partial_{t_{\rm i}}+ \frac{\mu_{\rm i}}{r^2}\partial_{r}r^2+\frac{\sqrt{1-\mu_{\rm i}^2}}{r}\left(\frac{\cos\bar{\phi}_{\rm i}}{\sin\theta}\partial_{\theta}\sin\theta+\frac{\sin\bar{\phi}_{\rm i}}{\sin\theta}\partial_{\phi}\right)
	+\frac{1}{r}\partial_{\mu_{\rm i}}(1-\mu_{\rm i}^2)
	-\frac{\sqrt{1-\mu_{\rm i}^2}}{r}\cot\theta\partial_{\bar{\phi}_{\rm i}}\sin\bar{\phi}_{\rm i}\,
\end{eqnarray}
in the spherical coordinate system $(r, \theta, \phi)$ for the position and $(E_{\rm i}, \bar \theta_{\rm i}, \bar \phi_{\rm i})$ for the momentum of the neutrino and the subscripts ``${\rm i}$" stand for the quantities in the inertial frame. We also defined $\mu_{\rm i} \equiv \cos \bar \theta_{\rm i}$ and $S^{\mu\nu}_{q} \equiv \epsilon^{\mu\nu\alpha\beta}q_{\alpha}n_{\beta}/ (2q\cdot n)$. Since we have fixed the frame vector and chirality (left-handedness of neutrinos), we eliminate the frame dependence and $\chi$ dependence of the spin tensor and instead use the subscript ``$q$" to specify its momentum dependence hereafter. We will also use similar subscripts to indicate the momenta of distribution functions. The emission and absorption rates are given by $R_{\rm emis} = \Gamma^{<}_q/E_{\rm i}$ and $R_{\rm abs} = \Gamma^{>}_q/E_{\rm i}$, respectively. Note that the corrections to the conventional transport theory involve the spin tensor $S^{\mu \nu}_q$ (and hence, the antisymmetric tensor $\epsilon^{\mu \nu \alpha \beta}$), which explicitly breaks the spherical symmetry and axisymmetry of the system. Physically, this is because chirality involving the spin degree of freedom can only be defined in 3D.

In addition to taking neutrinos as chiral fermions, we will assume that electrons are ultra-relativistic and treat them as approximate chiral fermions. Then, the self-energies  $\Sigma^{\lessgtr}_{\nu}$ incorporate the implicit quantum corrections from the Wigner functions of scattered electrons or neutrinos. On the contrary, nucleons are regarded as massive Dirac fermions. 

Based on the four-Fermi theory of the weak interaction in Eq.~(\ref{4-Fermi_cc}) for the neutrino absorption $\nu_{\rm L}(q)+{\rm n}(k) \rightleftarrows {\rm e}_{\rm L}(q')+{\rm p}(k')$, with $q^{\mu}$ and $q'^{\mu}$ $(k^{\mu}$ and $k'^{\mu}$) the four-momenta of incoming or outgoing leptons (nucleons), one finds the self-energies,
\begin{eqnarray}\label{Sigma_ab}
	\Sigma^{(\text{ab})\lessgtr}_{q\mu}=\int_p {\Pi}^{({\rm np})\lessgtr}_{p,\mu\nu}\mathcal{L}^{({\rm e})\lessgtr\nu}_{q-p}\,.
\end{eqnarray} 
Here, $\mathcal{L}^{({\rm e})\lessgtr\nu}_{q-p}$ represent Wigner functions of left-handed electrons given by $\mathcal{W}^{\lessgtr \nu}_{-}(q-p,x)$ in Eq.~(\ref{WF_L_full}) and
\begin{eqnarray}
	{\Pi}^{({\rm np})\lessgtr}_{p,\mu\nu} = 8{G}_{\rm F}^2\int_k \left(g_{+}^2k_{\mu}k'_{\nu}+g_{-}^2 k'_{\mu}k_{\nu}-g_{+}g_{-}m_{\rm n}m_{\rm p}\eta_{\mu\nu}\right)
	(2\pi)^2\delta(k^2-m_{\rm n}^2)\delta(k'^2-m_{\rm p}^2)f^{({\rm n})\gtrless}_{k}f^{({\rm p})\lessgtr}_{k'}\bigg|_{k'=p+k}\,,
\end{eqnarray}
with $p^{\mu} = (k'-k)^{\mu} = (q-q')^{\mu}$ being the four-momentum transfer in scattering, $m_{\rm n,p}$ the masses of neutrons/protons, $f^{({\rm N})<}$ and $f^{({\rm N})>}$ (${\rm N}={\rm n,p}$) the distribution functions of incoming and outgoing nucleons, respectively, and $g_{\pm} \equiv g_{\rm V}\pm g_{\rm A}$. Up to $\mathcal{O}(\d)$ and the leading order in coupling, one finds
\begin{eqnarray} \nonumber
	\Gamma^{(\text{ab})\lessgtr}_{q}=
	\int_p q^{\mu} {\Pi}^{({\rm np})\lessgtr}_{p,\mu\nu}
	\Bigg\{ \bigg[\delta(q'^2) \Big(q'^{\nu}- S^{\nu\rho}_{q'}\Delta_{q' \rho}\Big)
	-e\tilde{F}^{\nu\rho}q'_{\rho}\delta'(q'^2) \bigg]f_{{\rm L}q'}^{({\rm e})\lessgtr}
	+ S^{\mu\rho}_{q}D_{q\rho}\int_p {\Pi}^{({\rm np})\lessgtr}_{p,\mu \nu} \delta(q'^2)q'^{\nu}
	f_{{\rm L}q'}^{({\rm e})\lessgtr}\Bigg\}_{\! q'=q-p} \,.\\
	\label{qsigma_ab}
\end{eqnarray}
Note that $\Gamma^{(\text{ab})\lessgtr}_{q}$ involve electromagnetic fields due to the quantum corrections on electron Wigner functions. 

Similarly, for the elastic neutrino-nucleon scattering $\nu_{\rm L}(q)+{\rm N}(k) \rightleftarrows \nu_{\rm L}(q')+{\rm N}(k')$ described by Eq.~(\ref{4-Fermi_nc}), one obtains
\begin{eqnarray}
	\Gamma^{(\text{el})\lessgtr}_{q}
	=\int_p q^{\mu} {\Pi}^{({\rm NN})\lessgtr}_{p, \mu \lambda}\bigg[\delta(q'^2)\Big(q'^{\lambda}- S^{\lambda\rho}_{q'}D_{q'\rho}\Big)f_{{\rm L}q'}^{(\nu)\lessgtr}+ S^{\mu\rho}_{q}D_{q\rho}\int_p {\Pi}^{({\rm NN})\lessgtr}_{p,\mu \lambda} \delta(q'^2) q'^{\lambda} 
	f_{{\rm L}q'}^{(\nu)\lessgtr}\bigg]_{q'=q-p}\,,
\end{eqnarray}
where 
\begin{eqnarray}
	{\Pi}^{({\rm NN})\lessgtr}_{p,\mu\nu}
	=8G_{\rm F}^2\int_k\left(c_{+}^2k_{\mu}k'_{\nu} +c_{-}^2k'_{\mu}k_{\nu}-c_{+}c_{-}m_{\rm N}^2\eta_{\mu\nu}\right)
	(2\pi)^2\delta(k^2-m_{\rm N}^2)\delta(k'^2-m_{\rm N}^2)f^{({\rm N})\gtrless}_{k}f^{({\rm N})\lessgtr}_{k'}\bigg|_{k'=p+k}\,,
\end{eqnarray}
with $c_{\pm} \equiv (c_{\rm V}\pm c_{\rm A})/2$.

In the following, we will focus on the neutrino absorption process. In core-collapse supernovae, the mean free paths of electrons and nucleons are much smaller than the system size, so that they reach local thermal equilibrium. Then, in the presence of magnetic fields $B^{\mu}$ and fluid vorticity $\omega^{\mu}$ with a sufficiently small fluid velocity $u^{\mu}=\gamma(1,{\bm v})\approx (1,{\bm 0})$, the Wigner functions of electrons read~\cite{Hidaka:2017auj}
\begin{eqnarray}
	\mathcal{\bar L}^{({\rm e})\lessgtr\mu}_q=2\pi\bigg[\delta(q^2)\Big(q^{\mu}+\frac{1}{2}\big(\omega^{\mu}q\cdot u-u^{\mu}q\cdot \omega\big)\partial_{q\cdot u}\Big)
	-e\big(B^{\mu}q\cdot u-u^{\mu}q\cdot B\big)\delta'(q^2)\bigg]f^{({\rm e})\lessgtr}_{0,q}, 
\end{eqnarray}
where we ignored the viscous corrections and the gradients of the temperature and chemical potentials.
The nucleons obey the Fermi-Dirac distribution, $f^{(\rm N)\lessgtr}=f^{(\rm N)\lessgtr}_{0,q}$,
where
\begin{eqnarray}
	f^{(i)<}_{0,q}=1-f^{(i)>}_{0,q}=f^{(i)}_{0,q}\equiv \frac{1}{{\rm e}^{\beta (q\cdot u-\mu_i)}+1}\,, \quad (i={\rm n}, {\rm p}, {\rm e})
\end{eqnarray}
represent the equilibrium distribution functions with $\beta=1/T$ and $\mu_i$ being chemical potentials for $i={\rm n}, {\rm p}, {\rm e}$, respectively. Here and below, $\mathcal{\bar O}$ stands for a quantity $\mathcal{O}$ in local thermal equilibrium. Schematically, for the neutrino absorption, we may decompose $\bar \Gamma^{\lessgtr}_{q}$ as
\begin{eqnarray}
	\label{ab_rad_rates}
	\bar \Gamma^{({\rm ab})\lessgtr}_{q}\approx \bar \Gamma^{(0)\lessgtr}_{q}
	+ \bar \Gamma^{(B)\lessgtr}_{q} (q\cdot B)
	+ \bar \Gamma^{(\omega)\lessgtr}_{q} (q\cdot \omega).
\end{eqnarray} 
Note that the combinations $q\cdot \omega$ and $q\cdot B$ break the parity symmetry. The emergence of this type of collision term is a consequence of the parity-violating nature of the weak interaction.

Using the non-relativistic approximation for nucleons with setting $m_{\rm n}\approx m_{\rm p}\approx m_{\rm N}$ and the ``quasi-isoenergetic" approximation that allows for the energy transfer up to $\mathcal{O}(1/m_{\rm N})$, the classical terms $\bar \Gamma^{(0)\lessgtr}_{q}$ in Eq.~(\ref{ab_rad_rates}) become~\cite{Reddy:1997yr}
\begin{eqnarray}
	\nonumber
	\bar \Gamma^{(0)>}_{q}&\approx& \frac{1}{\pi} \big(g_{\rm V}^2+3g_{\rm A}^2\big){G}_{\rm F}^2 (q\cdot u)^3(1-f^{({\rm e})}_{0,q})\left(1-\frac{3q\cdot u}{m_{\rm N}}\right)	\frac{n_{{\rm n}}-n_{\rm p}}{1-{\rm e}^{\beta(\mu_{\rm p}-\mu_{{\rm n}})}}\,,
	\label{Gamma_0}
	\\
	\bar \Gamma^{(0)<}_{q}&\approx& \frac{1}{\pi} \big(g_{\rm V}^2+3g_{\rm A}^2\big){G}_{\rm F}^2 (q\cdot u)^3f^{({\rm e})}_{0,q}\left(1-\frac{3q\cdot u}{m_{\rm N}}\right)	\frac{n_{{\rm p}}-n_{\rm n}}{1-{\rm e}^{\beta(\mu_{\rm n}-\mu_{{\rm p}})}}\,,
\end{eqnarray}
while the quantum corrections $\bar \Gamma^{(B)\lessgtr}_{q}$ and $\bar \Gamma^{(\omega)\lessgtr}_{q}$ take the form~\cite{Yamamoto:2020zrs}
\begin{eqnarray}
	\nonumber
	\bar \Gamma^{(B)>}_{q} &\approx& \frac{e}{2\pi m_{\rm N}} \big(g_{\rm V}^2+3g_{\rm A}^2\big){G}_{\rm F}^2 (q\cdot u)(1-f^{({\rm e})}_{0,q})\left(1-\frac{8q\cdot u}{3 m_{\rm N}}\right)	\frac{n_{{\rm n}}-n_{\rm p}}{1-{\rm e}^{\beta(\mu_{\rm p}-\mu_{{\rm n}})}} \,,
	\\
	\bar \Gamma^{(B)<}_{q}&\approx& \frac{e}{2\pi m_{\rm N}} \big(g_{\rm V}^2+3g_{\rm A}^2\big){G}_{\rm F}^2 (q\cdot u)f^{({\rm e})}_{0,q}\left(1-\frac{8q\cdot u}{3 m_{\rm N}}\right)	\frac{n_{{\rm p}}-n_{\rm n}}{1-{\rm e}^{\beta(\mu_{\rm n}-\mu_{{\rm p}})}} \,,
	\label{Gamma_B}
\end{eqnarray}
and
\begin{eqnarray}
	\nonumber
	\bar \Gamma^{(\omega)>}_{q}&\approx & \frac{1}{2\pi} \big(g_{\rm V}^2+3g_{\rm A}^2\big){G}_{\rm F}^2 (q\cdot u)^2(1-f^{({\rm e})}_{0,q})\left(\frac{2}{E_{\rm i}}+\beta  f^{({\rm e})}_{0,q}\right)
	\frac{n_{{\rm n}}-n_{\rm p}}{1-{\rm e}^{\beta(\mu_{\rm p}-\mu_{{\rm n}})}}\,,
	\\
	\bar \Gamma^{(\omega)<}_{q}&\approx & \frac{1}{2\pi} \big(g_{\rm V}^2+3g_{\rm A}^2\big){G}_{\rm F}^2 (q\cdot u)^2f^{({\rm e})}_{0,q}\left(\frac{2}{E_{\rm  i}}-\beta (1-f^{({\rm e})}_{0,q})\right)
	\frac{n_{{\rm p}}-n_{\rm n}}{1-{\rm e}^{\beta(\mu_{\rm n}-\mu_{{\rm p}})}}\,,
	\label{Gamma_omega}
\end{eqnarray}
where $n_{{\rm n}/{\rm p}} = (1/(2\pi)^{3})\int {\rm d}^3 {\bm k} f^{({\rm n}/{\rm p})}_{0,k}$ are neutron/proton densities.

For a given $f^{(\nu)}_{q}$, the neutrino number current and energy-momentum tensor are given by
\begin{align}
\label{j_nu}
j^{\mu}_{(\nu)}& =\int_q 4\pi\delta(q^2)\Big(q^{\mu} f^{(\nu)}_{q}-S_q^{\mu \rho}\mathcal{D}_{\rho}f^{(\nu)}_{q}\Big)\,,
\\
\label{T_nu}
T^{\mu\nu}_{(\nu)}&=\int_q 4\pi\delta(q^2)\Big(q^{\mu}q^{\nu}f^{(\nu)}_{q}- q^{\{\mu}S_q^{\nu\}\rho}\mathcal{D}_{\rho}f^{(\nu)}_{q}\Big)\,.
\end{align}
Note that these are not just the moments of $f^{(\nu)}_{q}$, but there are additional contributions of the spin tensor due to the chirality of neutrinos. 

The energy-momentum transfer from the neutrino sector to the matter sector is dictated by the energy-momentum conservation law, 
\begin{eqnarray}
\label{cons_EM0}
\nabla_{\mu}T^{\mu\nu}_{\text{mat}} = -\nabla_{\mu}T^{\mu\nu}_{(\nu)}\,,
\end{eqnarray}
where $T^{\mu\nu}_{\text{mat}}$ is the energy-momentum tensor of the matter sector composed of nucleons and electrons. In the presence of the electromagnetic fields, the energy-momentum conservation law is given by
\begin{eqnarray}
\label{cons_EM}
\nabla_{\mu}T^{\mu\nu}_{\text{mat}}=e F^{\nu\mu}\big(j_{({\rm p})\mu}-j_{({\rm e})\mu}\big)-\nabla_{\mu}T^{\mu\nu}_{(\nu)}\,,
\end{eqnarray}
where $j_{(\rm p)}^{\mu}$ and $j_{(\rm e)}^{\mu}$ are the number currents of protons and electrons, respectively.

\subsubsection{\it Chiral hydrodynamics and non-equilibrium corrections for neutrinos \label{sec:hydro_neutrino}}

Since neutrinos are only weakly interacting, they are typically out of equilibrium, but the exception is realized at the core of supernovae. There, the matter density there is so high that even neutrinos reach local thermal equilibrium, where hydrodynamics is applicable to neutrinos.

To see this, let us estimate the neutrino mean free path $l_{\rm mfp}$. The most efficient scattering involving neutrinos is the coherent scattering with nuclei, for which the mean free path is given by $l_{\rm mfp} = (\sigma_A n_A)^{-1}$, where $\sigma_A$ is the cross section and $n_A$ is the number density of nuclei. By using  $\sigma_A \sim G_{\rm F}^2 E_{\nu}^2 A^2$~\cite{Tubbs:1975jx} and $n_A = \rho/(A m_{\rm N})$, where $A$ is the atomic mass number, $E_{\nu} \simeq \mu_{\rm e} = (3\pi^2 \rho Y_{\rm e}/m_{\rm N})^{1/3}$ is the neutrino energy, $\rho$ is the mass density of nuclear matter, and $Y_{\rm e}$ is the electron fraction, we have (see, e.g., Ref.~\cite{Kotake:2005zn})
\be
\label{mfp}
l_{\rm mfp} \sim 10^7\,{\rm cm} \left(\frac{\rho}{10^{10}\,{\rm g}/{\rm cm}^3} \right)^{-\frac{5}{3}} \left(\frac{A}{56} \right)^{-1} \left(\frac{Y_{\rm e}}{26/56} \right)^{-\frac{2}{3}}\,.
\ee
For the typical magnitudes of the quantities at the core, $\rho \gtrsim 10^{13}\,{\rm g}/{\rm cm}^3$, $A \simeq 56$, and $Y_{\rm e} \sim 0.1$, we obtain $l_{\rm mfp} \lesssim 1\,{\rm m}$. This is much smaller than the typical length scale of the supernova core of order $10$-$100\,{\rm km}$. Therefore, the local thermal equilibrium is achieved for neutrinos, and the neutrino chemical potential $\mu_{\nu}$ is well defined. 

From the $\beta$ equilibrium condition $\mu_{\rm p} + \mu_{\rm e} \simeq \mu_{\rm n} + \mu_{\nu}$, charge neutrality condition $n_{\rm p} = n_{\rm e}$, and the typical lepton fraction $Y_{\ell} \equiv (n_{\rm e} + n_{\nu})/(n_{\rm p} + n_{\rm n}) \sim 0.1$, it follows that $\mu_{\nu} \sim \Lambda_{\rm QCD}$ at the core, where $\Lambda_{\rm QCD} \simeq 200\,{\rm MeV}$ is the QCD scale. Since the temperature of the supernova is at most $\sim 10\,{\rm MeV}$, neutrinos there are mildly Fermi degenerate. 
As neutrinos are only left-handed within SM, the degenerate neutrino matter possesses a non-trivial Berry curvature expressed by Eq.~(\ref{monopole}) with the monopole charge $K=-1$, and so it is topological matter in nature (despite being a mixture with electron and nuclear matter)~\cite{Yamamoto:2015gzz}.

In this regime, the neutrino sector can be described by chiral hydrodynamics. Using the results in Section~\ref{sec:ChHD_background}, the equilibrium neutrino energy-momentum tensor $\bar T^{\mu \nu}_{(\nu)}$ and number current $\bar j^{\mu}_{(\nu)}$ in the entropy frame can be written by using the neutrino temperature $T$, chemical potential $\mu_{\nu}$, and four velocity $u^{\mu}$ as~\cite{Yamamoto:2015gzz,Yamamoto:2021hjs}
\begin{align}
\bar T^{\mu \nu}_{(\nu)} &= (\epsilon_{\nu} + P_{\nu}) u^{\mu} u^{\nu}-P_{\nu} g^{\mu \nu} + D_{\omega}^{(\nu)}(u^{\mu} \omega^{\nu} + \omega^{\mu} u^{\nu}),
\\
\bar j^{\mu}_{(\nu)} &=n_{\nu}u^{\nu}+ C_{\omega}^{(\nu)} \omega^{\mu},
\end{align}
where 
\begin{gather}
\epsilon_{\nu}=3P_{\nu} = \frac{\mu_{\nu}^4}{8\pi^2}+\frac{\mu_{\nu}^2 T^2}{4}+\frac{7\pi^2}{120}T^4\,, 
\quad
D_{\omega}^{(\nu)} = -\bigg(\frac{\mu_{\nu}^3}{6\pi^2}+\frac{\mu_{\nu}T^2}{6}\bigg)\,.
\\
n_{\nu} = \frac{\mu_{\nu}^3}{6\pi^2} + \frac{\mu_{\nu} T^2}{6}\,,
\quad
C_{\omega}^{(\nu)} = -\left(\frac{\mu_{\nu}^2}{4\pi^2}+\frac{T^2}{12}\right)\,.
\end{gather}
Here, we ignored the dissipative corrections on $\bar T^{\mu \nu}_{(\nu)}$ and $\bar j^{\mu}_{(\nu)}$, and the corrections due to interactions on $\epsilon_{\nu}$, $P_{\nu}$, and $n_{\nu}$ for simplicity.

Outside the core of supernovae where the density is not sufficiently high and where neutrinos are not in complete local equilibrium, one generically needs to use the chiral radiation transport theory for neutrinos in Section~\ref{sec:ChRTT} instead. Still, when neutrinos are close to equilibrium, one can systematically derive the non-equilibrium corrections on the neutrino chiral hydrodynamics~\cite{Yamamoto:2021hjs} based on the chiral radiation transport theory.
Focusing on the neutrino absorption on nucleons under this condition, we may rewrite the collision term in the relaxation-time approximation, 
\begin{eqnarray}
	\frac{1}{{E_{\rm i}}}\left[(1-f^{(\nu)}_{q})\bar{\Gamma}^{({\rm ab})<}_{q}-f_{q}^{(\nu)}\bar{\Gamma}^{({\rm ab})>}_{q} \right]\approx -\frac{\delta f^{(\nu)}_{q}}{\tau}\,,
\end{eqnarray}  
where $\delta f^{(\nu)}_{q}\equiv f^{(\nu)}_{q}-\bar{f}^{(\nu)}_{q}$ is the fluctuation of the distribution function and  $\tau={E_{\rm i}}/\big(\bar{\Gamma}^{({\rm ab})>}_{q}+\bar{\Gamma}^{({\rm ab})<}_{q}\big)$ denotes a momentum-dependent relaxation time which describes how long the system returns to the equilibrium state. From the detailed balance such that the collision term vanishes when all scattered particles reach thermal equilibrium, the equilibrium distribution function of neutrinos reads
\begin{eqnarray}
	\bar{f}^{(\nu)}_{q}=\frac{1}{{\rm e}^{h}+1}\,,
\end{eqnarray}
where
\begin{eqnarray}
	h \approx \beta(q\cdot u-\mu_{\nu})+\beta\frac{q\cdot\omega}{2q\cdot u}+O\left(\frac{q\cdot u}{m_{\rm N}}\right)\,.
\end{eqnarray}
The expression for $\bar{f}^{(\nu)}_{q}$ above agrees with the equilibrium distribution function in Refs.~\cite{Chen:2015gta,Hidaka:2017auj}. From Eqs.~(\ref{Gamma_0}), (\ref{Gamma_B}), and (\ref{Gamma_omega}), the relaxation time can be decomposed as $\tau=\tau^{(0)}+ \tau^{(1)}$ via the gradient expansion, where $\tau^{(0)}$ and $\tau^{(1)}$ are zeroth and first order in derivatives and are given by
\begin{align}
	\tau^{(0)}&
	= \frac{\pi(1-{\rm e}^{\beta(\mu_{\rm n}-\mu_{{\rm p}})})}{G_{\rm F}^2 (g_{\rm V}^2 + 3 g_{\rm A}^2) (n_{{\rm p}}-n_{\rm n})}
	\frac{E_{\rm i} (1-f_{0,q}^{(\nu)})}{(q\cdot u)^3 (1-f^{({\rm e})}_{0,q}) }\,, 
	\\\label{eq:tau1}
	\tau^{(1)}&
	=-\tau^{(0)}\left[\left(\frac{2}{E_{\rm  i}}+\beta \big(f^{({\rm e})}_{0,q}-f^{({\nu})}_{0,q}\big)\right)\frac{q\cdot \omega}{2q\cdot u}+\frac{e q\cdot B}{2m_{\rm N}(q\cdot u)^2}\right]\,.
\end{align}
Solving Eq.~(\ref{CKT_inertial_con}) with the simplified collision term, the perturbative solution of $\delta f^{(\nu)}_{q}$ is given by 
\begin{eqnarray}\label{delf_sol}
	\delta f^{(\nu)}_{q}\approx -\tau\Box_{\rm i}\bar{f}_{q}^{(\nu)}=-\tau\frac{q\cdot D}{{E_{\rm i}}}\bar{f}_{q}^{(\nu)}\,.
\end{eqnarray}

Considering neutrinos near equilibrium and again focusing on the neutrino absorption on nucleons, one may insert Eqs.~(\ref{WF_L_full}) and (\ref{delf_sol}) into Eq.~(\ref{j_and_T}) to compute the energy-momentum tensor $T^{\mu\nu}_{(\nu)}$ and number current $j^{\mu}_{(\nu)}$. We decompose them into the equilibrium and non-equilibrium parts as $T^{\mu\nu}_{(\nu)}=\bar T^{\mu\nu}_{(\nu)}+\delta T^{\mu\nu}_{(\nu)}$ and $j^{\mu}_{(\nu)} = \bar j^{\mu}_{(\nu)}+\delta j^{\mu}_{(\nu)}$,
and will focus on just the corrections from magnetic fields. 
It turns out that $\bar T^{\mu\nu}_{(\nu)}$ and $\bar j^{\mu}_{(\nu)}$ match the results obtained from chiral hydrodynamics for neutrinos  above, while the non-equilibrium corrections are given by
\begin{align}
	\label{deltT_decomp}
	\delta T^{(1)\mu\nu}_{(\nu)B}
	&=\delta\epsilon_{B} u^{\mu}u^{\nu}-\delta P_{B_{\rm T}}\Delta^{\mu\nu}_{B}-\delta P_{B_{\rm L}}\hat{B}^{\mu}\hat{B}^{\nu}+2h_{\perp}^{\{\mu}B^{\nu\}}+2u^{\{\mu}V_{\perp}^{\nu\}},
	\\
	\label{deltj_decomp}
	\delta j^{(1)\mu}_{(\nu)B}&=\delta N_{B}u^{\mu}+\sigma_{B}^{\mu\nu}B_{\nu},
\end{align}
where $\Delta^{\mu\nu}_{B}=\Delta^{\mu\nu}+\hat{B}^{\mu}\hat{B}^{\nu}$, $h_{\perp}\cdot B=0$, and $\sigma_{B}^{\mu\nu}u_{\mu}=0$. The explicit forms of these transport coefficients for $\delta T^{(1)\mu\nu}_{(\nu)B}$ and $\delta j^{(1)\mu}_{(\nu)B}$ can be found in Ref.~\cite{Yamamoto:2021hjs}. 
When we further assume that ${\mu}_{\rm n}-{\mu}_{\rm p}={\mu}_{\rm e}-{\mu}_{\nu} \gg T$ and $u^{\mu} \approx (1, {\bm v})$ with $|{\bm v}| \ll 1$, the explicit expressions for the asymmetric momentum current $\delta T^{(1)i0}_{(\nu) B}$ and number current $\delta j^{(1)i}_{(\nu)B}$ driven by the magnetic field can be simplified as
\be
	\label{main_T}
	\delta T^{(1)i 0}_{(\nu) B} \approx \mu_{\nu} \delta j^{(1)i}_{(\nu)B} \approx -\frac{e}{72 \pi m_{\rm N} G_{\rm F}^2 (g_{\rm V}^2 + 3 g_{\rm A}^2)}
	\frac{{\rm e}^{2\beta(\mu_{\rm n}-\mu_{\rm p})}}{n_{\rm n}-n_{{\rm p}}}
	({\bm \nabla} \cdot {\bm v})\mu_{\nu} B^i\,.
\ee
These chiral transport are prohibited in the usual parity-invariant matter by parity symmetry but become possible in the present case again due to the parity-violating nature of the weak interaction.

\subsection{\it Pulsar kicks \label{sec:kick}}
It has long been observed that neutron stars have velocities around hundreds of km/s, which are much greater than the velocities of their progenitors~\cite{Lyne1994,Kaspi1996,Arzoumanian2002,Hobbs2005}. Nevertheless, the physical mechanisms that ``kick" the neutron stars during their formation and cause the net momentum asymmetry remain an open question. There have been several mechanisms proposed in the past, including hydrodynamic instabilities and asymmetric neutrino emissions induced by strong magnetic fields among others, see Ref.~\cite{Lai_2001} for a brief review and, e.g., Refs.~\cite{Kuroda:2020pta,Janka:2021deg} for more recent discussions and references therein. In this review, we will focus on the latter as a possible mechanism of the pulsar kick in the context of the chiral effects of leptons.

The impact of chiral transport for neutrinos with the strong magnetic field on pulsar kick was first considered in Refs.~\cite{Chugai:1984,Vilenkin:1995um}, where the non-equilibrium transport of nearly trapped neutrinos in the proto-neutron star is studied. Similar to the relaxation-time approximation of the chiral radiation transport theory in Section~\ref{sec:hydro_neutrino} (see Eq.~(\ref{eq:tau1})), it is proposed that the mean free time (or the relaxation time) receives the correction from the magnetic field due to the magnetic-moment coupling of the electrons that participate in the neutrino absorption on nucleons as $\tau=\tau^{(0)}(1+\kappa_{\rm m}{{\bm q}\cdot {\bm B}}/E_{\rm i})$, where $\kappa_{\rm m}$ is proportional to the effective electron magnetic moment. The involvement of the magnetic field in the collision term hence leads to the asymmetric flux of neutrinos. A more rigorous derivation of the scattering cross section was presented in the follow-up studies~\cite{Bezchastnov:1996gy,Roulet:1997sw}. Further corrections from the magnetic moments of nucleons and neutrinos have been later investigated~\cite{Lai:1997mm,Horowitz:1997mk,Horowitz:1997fb}, from which the asymmetric neutrino emission is instead due to the elastic neutrino-nucleon scattering. The systematic study of both the neutrino absorption on nucleons and elastic neutrino-nucleon scattering affected by magnetic fields and the associated neutrino flux can be found in Ref.~\cite{Arras:1998mv}. 

More recently, possible mechanisms relating pulsar kicks to the chiral effects, such as CME and CSE, were proposed in Refs.~\cite{Charbonneau:2009ax,Kaminski:2014jda}.
In the scenario of Ref.~\cite{Charbonneau:2009ax}, the electron number current ${\bm j}_{\rm e}= e \mu_{5}{\bm B}/(2\pi^2)$ is induced by the magnetic field in the presence of the electron chiral chemical potential $\mu_{5}$, which is considered to be generated by the weak interaction ignoring the effect of chirality flipping (see, however, the effect of the chirality flipping in Section~\ref{sec:magnetar}). Then, it is argued that the electrons also transfer their momentum into space, thereby kicking the neutron star along the direction of the magnetic field.
In the scenario of Ref.~\cite{Kaminski:2014jda}, the asymmetric neutrino flux is driven by a leptonic axial current, ${\bm j}_{\ell 5}= e\mu_{\ell}{\bm B}/(2\pi^2)$, through the leptonic CSE in Eq.~(\ref{CME}). Nonetheless, the microscopic interaction connecting the neutrinos to magnetic fields and the exact portion of neutrino number current incorporated in the full leptonic axial-charge current lack explanations. See also Refs.~\cite{Arras:1998cv,Goyal:1998nq,Lai:1998sz,Sagert:2007as,Maruyama:2010hu,Adhya:2013ima,Ayala:2018kie,Patino:2018tyq} for other studies of pulsar kicks related to neutrino scattering influenced by magnetic fields and Ref.~\cite{Shaverin:2014xya} for a possible mechanism of pulsar kick induced by rotation.

In all the above studies with the magnetic field, the neutrino momentum current is indirectly estimated from its number current. On the other hand, by adopting the chiral radiation transport theory, one can directly and systematically evaluate the energy-momentum tensor and accordingly derive the momentum current of neutrinos with chiral effects~\cite{Yamamoto:2021hjs}, as we have seen in Section~\ref{sec:hydro_neutrino}. By using Eq.~(\ref{main_T}) and momentum conservation between the matter and neutrino sectors, we may estimate the kick velocity of the core due to the magnetic-field contribution as $v_{\rm kick} \sim \delta T^{i 0}_{(\nu) B}/\rho_{\rm core}\,$, where we assumed the homogeneous core mass density $\rho_{\rm core}$ and $\delta T^{i 0}_{(\nu) B}$ for an order of estimate. 
Taking $n_{\rm n} - n_{\rm p} \sim 0.1\,{\rm fm}^{-3}$, $\mu_{\rm n} - \mu_{\rm p} \sim 100\,{\rm MeV}$, $\mu_{\nu} \sim 100\,{\rm MeV}$, $T \sim 10\,{\rm MeV}$, typical length scale for the variation of the hydrodynamic variables, $L \sim 10\,{\rm km}$, $|{\bm v}| \sim 0.01$, and $\rho_{\rm core} \sim m_{\rm N} (n_{\rm n} + n_{\rm p})$ with $n_{\rm n} + n_{\rm p} \sim 0.1\,{\rm fm}^{-3}$ in the regime where the incompressibility condition ${\bm \nabla}\cdot {\bm v} = 0$ does not apply, we obtain
\begin{equation}
\label{eq:vkick_est}
	v_{\rm kick} \lesssim \left( \frac{B}{10^{13{\text -}14}\,{\rm G}} \right)\,{\rm km/s}\,.
\end{equation}
The required magnetic field at the core to account for the observed pulsar velocity $v_{\rm kick} \sim 10^2\,{\rm km/s}$ only from this contribution is of the order of $10^{15{\text -}16}\,{\rm G}$.
Note that this estimate depends sensitively on the choice of the parameters and that the right-hand side of Eq.~(\ref{eq:vkick_est}) should be regarded as an upper bound for the estimate. 

There are also other neutrino chiral transport induced by the vorticity and gradients of temperature and chemical potentials that have not yet been studied in detail so far. Also, neutrino chiral transport far away from equilibrium is not captured by the adopted relaxation time approximation, whereas such a contribution should play an essential role for neutrinos that can escape from the proto-neutron stars to engender momentum asymmetry~\cite{Arras:1998mv}. Eventually, it would be necessary to perform numerical simulations of the chiral radiation transport theory for neutrinos to study the consequences of fully non-equilibrium chiral effects in the future.

\subsection{\it Magnetars \label{sec:magnetar}}

Magnetars are a type of neutron stars with the most gigantic magnetic field ($\sim 10^{15}\,{\rm G}$ on the surface) in the Universe~\cite{Magnetar}; see Refs.~\cite{Harding:2006qn,Spruit:2007bt,Enoto:2019vcg} for reviews. The origin of this strong magnetic field is one of the mysteries in astrophysics. Among others possible mechanisms proposed so far include the fossil field and dynamo hypothesis~\cite{Harding:2006qn,Spruit:2007bt}. However, the produced magnetic field in these mechanisms is typically messy and unstable and cannot be sustained for a long time scale. In fact, a purely poloidal magnetic field configuration is unstable~\cite{Wright,Markey,Flowers}, as confirmed by numerical MHD simulations~\cite{Braithwaite:2005md}. 

A new mechanism to generate a strong \emph{and} stable magnetic field is proposed in Ref.~\cite{Ohnishi:2014uea} based on the chiral effects of electrons. The chirality imbalance of electrons is produced by the electron capture process ${\rm p} + {\rm e}_{\rm L} \rightarrow {\rm n} + {\nu}^{\rm e}_{\rm L}$ during core collapse of supernovae. Here, only left-handed electrons and neutrinos are involved, as it is described by the four-Fermi theory of the weak interaction in Eq.~(\ref{4-Fermi_cc}). Ignoring the effect of chirality flipping by the finite electron mass, the total (or time-integrated) electron chiral chemical potential $\mu_{5}$ during core collapse is estimated to be the QCD scale, $\mu_{5} \sim \Lambda_{\rm QCD}$. If this is converted to the magnetic field via the CPI in Section~\ref{sec:CPI}, the corresponding magnetic field strength would be $B \sim 10^{18}\,{\rm G}$. An advantage of this mechanism is that it can naturally generate a non-zero magnetic helicity that ensures the stability of the resulting magnetic field; see also the earlier work~\cite{Charbonneau:2009ax} that pointed out a potential importance of the CME to the generation of a magnetic helicity in neutron stars.

However, it was pointed out in Refs.~\cite{Grabowska:2014efa,Kaplan:2016drz} that the generation of the chirality imbalance of electrons can be highly suppressed by the chirality flipping due to the finite electron mass $m_{\rm e}$ although $m_{\rm e} \ll \mu_{\rm e}$. The dominant contribution to the chirality flipping is the Rutherford scattering of electrons off the ambient protons. The other processes, such as the electron-electron scattering and Compton scattering are comparably suppressed. The former is because the electrons are more degenerate than protons. The latter is because the proton density $\sim \mu_{\rm e}^3$ is much larger than the photon density $\sim T^3$ during the core collapse where $T/\mu_{\rm e} \lesssim 10^{-1}$. Based on the Boltzmann equation for degenerate electrons, the chirality-flipping rate by the Rutherford scattering is computed as
\begin{align}
\label{flipping}
    \Gamma_{\rm f}
    &\simeq
    \frac{\alpha^2 m_{\rm e}^2}{3\pi {\mu}_{\rm e}}\left(\ln\frac{4\mu_{\rm e}^2}{q_{\rm D}^2}-1\right)\,,
\end{align}
where $\alpha = e^2/(4\pi)$ and $q_{\rm D}$ is the Debye momentum in the medium. For $\mu_{\rm e} \sim 100\,{\rm MeV}$ and $T \sim 30\,{\rm MeV}$, for example, $\Gamma_{\rm f} \sim 10^{14}\,{\rm s}^{-1}$.
The equation for the evolution of the electron chiral charge $n_5$ is given by
\bea
\d_t n_5 = \Gamma_{\rm w} (n_{\rm e} - n_5) - (\Gamma_{\rm CPI} + \Gamma_{\rm f}) n_5,
\eea 
where $\Gamma_{\rm w} \equiv \dot Y_{\rm e}/Y_{\rm e}$ is the depletion rate of the electron fraction $Y_{\rm e}$ due to the electron capture process and $\Gamma_{\rm CPI}$ is the depletion rate of $n_5$ due to the CPI. By ignoring the contribution of $\Gamma_{\rm CPI}$, the steady-state solution is given by 
\bea
\label{steady}
n_5 = \frac{\Gamma_{\rm w}}{\Gamma_{\rm f}} n_{\rm e}\,.
\eea
If the change $\delta Y_{\rm e} \simeq 0.4$ occurs on a time scale of the free-fall time $\sim 100\,{\rm ms}$ during core collapse, then $\Gamma_{\rm w}\sim 1$-$10\,{\rm s}^{-1}$.
Using the value $\Gamma_{\rm w} \simeq 1\,{\rm s}^{-1}$, it is found that $n_5 \sim 10^{-14} n_{\rm e}$, and correspondingly, $\mu_5 \sim 10^{-7}\,{\rm eV}$~\cite{Grabowska:2014efa}. This seems to be too small to be astrophysically relevant. 

On the other hand, the rate of electron capture, $\Gamma_{\rm w}$, for $\mu_{\rm e} \gg T$ is evaluated in Ref.~\cite{Sigl:2015xva} based on the result~\cite{Lattimer:1991ib} of the microscopic four-Fermi theory of the weak interaction as
\bea 
\Gamma_{\rm w} = \frac{457\pi}{10080}G_{\rm F}^2 (g_{\rm V}^2 + 3 g_{\rm A}^2) \cos^2 \theta_{\rm C} \frac{m_{\rm n} m_{\rm p} T^6}{n_{\rm e}}\,,
\eea
where $\theta_{\rm C} \approx 0.24$ is the Cabibbo angle.
This shows that $\Gamma_{\rm w}$ is highly sensitive to the temperature of proto-neutron stars. Using this result, the steady-state value of $n_5$ in Eq.~(\ref{steady}) can become much larger than the one in Ref.~\cite{Grabowska:2014efa} for higher $T$. The estimate given in Ref.~\cite{Sigl:2015xva} is $\mu_5 \sim {\rm eV}$-${\rm keV}$ and the resulting magnetic field strength due to the CPI is up to $\sim 10^{14}\,{\rm G}$.

In Ref.~\cite{Dvornikov:2014uza}, it is argued that the electron-nucleon weak interaction leads to the term $(G_{\rm F}/\sqrt{2}) n_{\rm n} \bar {\rm e} \gamma^0 \gamma_5 {\rm e}$ in the mean-field approximation, where $n_{\rm n} \equiv \langle \bar {\rm n} \gamma^0 {\rm n} \rangle$ is the neutron density. One might then regard this as an ``effective chiral chemical potential" $V_5 = (G_{\rm F}/2\sqrt{2}) n_{\rm n}$, which then leads to the CPI.%
\footnote{A similar mechanism was proposed earlier in Ref.~\cite{Boyarsky:2012ex} in the context of the early Universe to explain the observed large-scale magnetic fields in galaxies and clusters.}
It was estimated that this mechanism could generate a helical magnetic field with the magnitude $\gtrsim 10^{15}\,{\rm G}$~\cite{Dvornikov:2014uza} or even $\sim 10^{17}\,{\rm G}$ when a seed magnetic field $\sim 10^{12}\,{\rm G}$ (which is the typical magnitude of magnetic fields in pulsars) is present~\cite{Dvornikov:2015lea}. However, the CPI induced by $V_5$ was questioned by Ref.~\cite{Grabowska:2014efa} simply because there is no energy source for the magnetic field there unlike the case of the electron chiral chemical potential.

There are in fact other sources that play the same role of the electron chiral chemical potential. One is the fluid helicity (or kinetic helicity), ${\bm v} \cdot {\bm \omega}$. In the presence of this quantity, the electric current of the form ${\bm j} = C_{\rm HME} ({\bm v} \cdot {\bm \omega}) {\bm B}$ with $C_{\rm HME}$ some constant becomes possible. This is referred to as the helical magnetic effect (HME)~\cite{Yamamoto:2016xtu,Yamamoto:2021gts}. This is similar to the so-called $\alpha$ effect in the usual MHD~\cite{Moffatt1970,Davidson2001}, where a non-zero $\langle \delta {\bm v} \cdot \delta {\bm \omega} \rangle$ may be generated by turbulence. Here, $\delta {\bm v}$ and $\delta {\bm \omega}$ are the turbulent components of $\bm v$ and $\bm \omega$, and the expectation value denotes the ensemble average. What is distinct from the usual $\alpha$ effect is that in the case of the chiral hydrodynamics for neutrinos in Section~\ref{sec:hydro_neutrino}, the fluid helicity can simply be generated via the CVE~\cite{Yamamoto:2016xtu} even without considering turbulent helicity $\langle \delta {\bm v} \cdot \delta {\bm \omega} \rangle$. For purely relativistic Dirac fermions, it is shown that $C_{\rm HME}$ coincides with the anomaly coefficient and $C_{\rm HME} = e^2/(4\pi^2)$~\cite{Yamamoto:2021gts}. The HME may lead to an instability similar to the CPI, which could generate a strong helical magnetic field.

Another source is the kick from non-equilibrium neutrinos~\cite{Matsumoto:2022lyb}. When neutrinos are not in complete equilibrium but close to equilibrium, their momentum density is given by Eq.~(\ref{main_T}). From the momentum conservation law, the matter sector receives momentum kick from neutrinos as $\delta T_{(\rm e)}^{i0} = - \delta T_{(\nu)}^{i0}$, where $T_{(\rm e)}^{i0}$ is the electron momentum density. Here, we ignored the nucleon recoil in the neutrino scattering consistently, which is also used to derive Eq.~(\ref{main_T}) under the expansion of $|{\bm p}|/M$ (with $\bm p$ the momentum transfer). Assuming a relation between the current and momentum density for electrons similar to Eq.~(\ref{main_T}), $\delta T_{(\rm e)}^{i0} = \mu_{\rm e} \delta j_{(\rm e)}^i$, one finds an electric current proportional to the magnetic field:
\be
\label{main_j}
{\bm j} = \xi_{B} {\bm B}, \qquad \xi_B = - \frac{e^2}{72 \pi m_{\rm N} G_{\rm F}^2 (g_{\rm V}^2 + 3 g_{\rm A}^2)}
\frac{{\rm e}^{2\beta(\mu_{\rm n}-\mu_{\rm p})}}{n_{\rm n}-n_{{\rm p}}} ({\bm \nabla} \cdot {\bm v})\frac{\mu_{\nu}}{\mu_{\rm e}}\,.
\ee
Taking the typical values $n_{\rm n} - n_{\rm p} \sim 0.1\,{\rm fm}^{-3}$, $\mu_{\rm n} - \mu_{\rm p} \sim 100\,{\rm MeV}$, 
$\mu_{\nu} \sim \mu _{\rm e} \sim 100\,{\rm MeV}$, $T \sim 10\,{\rm MeV}$, $|{\bm v}| \sim 0.01$, and the typical length scale of the system, $L \sim 10\,{\rm km}$, we have $\xi_{B} \sim 10\,{\rm MeV}$. Note that this value is again sensitive to the choice of the parameters, and it should be regarded as the upper bound for $\xi_B$, similar to Eq.~(\ref{eq:vkick_est}) for pulsar kicks. The maximum magnitude of the magnetic field generated through the CPI due to this chiral current is again $\sim10^{18}\,{\rm G}$. These chiral effects provide possible new mechanisms for the strong and stable magnetic field of magnetars.

\subsection{\it Chiral turbulence in core-collapse supernovae \label{sec:CCSN}}

Another unsolved question in high-energy astrophysics is the physical mechanism of core-collapse supernova explosions. For recent reviews on the theoretical aspects of core-collapse supernovae, see, e.g., Refs.~\cite{Kotake:2012nd,Burrows:2012ew,Foglizzo:2015dma,Janka:2016fox,Muller:2016izw,Radice:2017kmj}. Two crucial components for successful explosions, among others, are the spherically asymmetric flow and turbulence. As we have seen in Section~\ref{sec:ChRTT}, the chiral effects in the radiation transport theory necessarily breaks the spherical symmetry of the system, implying their potential importance in the explosion dynamics. Below we focus on the chiral effects on the turbulence.

Most of the conventional 3D supernova models without the chiral effects exhibit the direct turbulent cascade that transfers energy from large scales to small scales in the post-shock region~\cite{Hanke:2011jf,Takiwaki:2013cqa,Radice:2017kmj}.
The axisymmetric (2D) models, on the other hand, often exhibit the inverse cascade transferring energy from small scales to large scales~\cite{Janka:2016fox}. In this case, due to the large-scale coherent fluid motion, explosion is easier compared with the 3D models. In this way, the turbulent direction plays a key role in the supernova explosion mechanism.

One can ask how the chiral effects neglected so far can modify the conventional turbulent behavior in 3D supernovae, as studied in Refs.~\cite{Masada:2018swb,Matsumoto:2022lyb}. Here, there are several possible sources for the CME-like electric current ${\bm j} = \xi_B {\bm B}$: e.g., the electron chiral chemical potential $\mu_{5}$ (which can be suppressed by chirality flipping), the fluid helicity generated via the CVE of neutrinos, and the kick from non-equilibrium neutrinos, as we have seen in Section~\ref{sec:magnetar}. Here, we do not first specify the origin of $\xi_B$, and simply include it in this form at the proto-neutron stars. The corresponding effective chiral density is denoted by $n_{5, {\rm eff}}$, which is related to $\xi_B$ by
\be
\xi_{B} = \frac{1}{4} \biggl ( \frac{3}{\pi^4} \biggr )^{1/3} [(n_{\rm e} + n_{5, {\rm eff}})^{1/3}-(n_{\rm e} - n_{5, {\rm eff}})^{1/3}]\,.
\ee

The chiral MHD equations for relativistic electrons and non-relativistic protons in the Landau frame are then given by~\cite{Masada:2018swb,Matsumoto:2022lyb}
\begin{align} 
&\d_t \rho + {\bm \nabla} \cdot (\rho {\bm v}) = 0\,, 
\label{eq:mass_cons} \\
&\d_t(\rho {\bm v}) + {\bm \nabla} \cdot \biggl [ \rho {\bm v} {\bm v} - {\bm B} {\bm B}
+ \biggl (P + \frac{B^2}{2} \biggr ) \textbf{I} \biggr ] = {\bm S} \,, 
\label{eq:momentum_cons} \\
&\d_t \biggl ( \frac{1}{2} \rho v^2 + \frac{1}{\Gamma - 1}P + \frac{B^2}{2} \biggr ) 
+ {\bm \nabla} \cdot \biggl [ \biggl ( \frac{1}{2} \rho v^2 + \frac{\Gamma}{\Gamma - 1}P \biggr ) {\bm v} + {\bm E} \times {\bm B} \biggr] 
= {\bm S} \cdot {\bm v} - \xi_B {\bm B} \cdot {\bm E} \;, 
\label{eq:energy_cons} \\
&\d_t {\bm B} = 
{\bm \nabla} \times ({\bm v} \times {\bm B}) + \eta {\bm\nabla}^2 {\bm B} + \eta  \; {\bm \nabla} \times (\xi_B {\bm B}) \;, 
\label{eq:induction} \\
&\d_t n_{5, {\rm eff}} = \frac{e^2}{2 \pi^2} {\bm E} \cdot {\bm B} \;,
\label{eq:anomaly}
\end{align}
corresponding to the mass conservation, momentum conservation, 
energy conservation, induction equation, and anomaly equation (or helicity conservation), respectively.
Here, $\rho = n_{\rm e} (m_{\rm p} + m_{\rm e}) \simeq n_{\rm e} m_{\rm p}$ is the rest-mass density, $\Gamma = 5/3$ is the ratio of specific heats assuming the ideal gas, $\textbf{I}$ is the unit matrix, and ${\bm S} = \rho \nu {\bm\nabla}^2 {\bm v} + ({1}/{3}) \rho \nu {\bm \nabla} ({\bm \nabla} \cdot {\bm v})$ with $\nu = \kappa/\rho$ the kinematic viscosity. We also used a shorthand notation such as $[{\bm \nabla} \cdot (\rho {\bm v} {\bm v})]^j \equiv \d_i (\rho v^i v^j)$ and wrote the above equations in the conservative form for numerical simulations.
The electric field can be eliminated from the above equations by
\be
{\bm E}= \eta({\bm \nabla}\times {\bm B} - \xi_B {\bm B}) -{\bm v} \times {\bm B}.
\ee
Here, CVE, CSE, fluid helicity, and cross helicity, are ignored for simplicity. Also, advection and diffusion are ignored in Eq.~(\ref{eq:anomaly}).

\begin{figure}
\begin{center}
\scalebox{0.13}{{\includegraphics{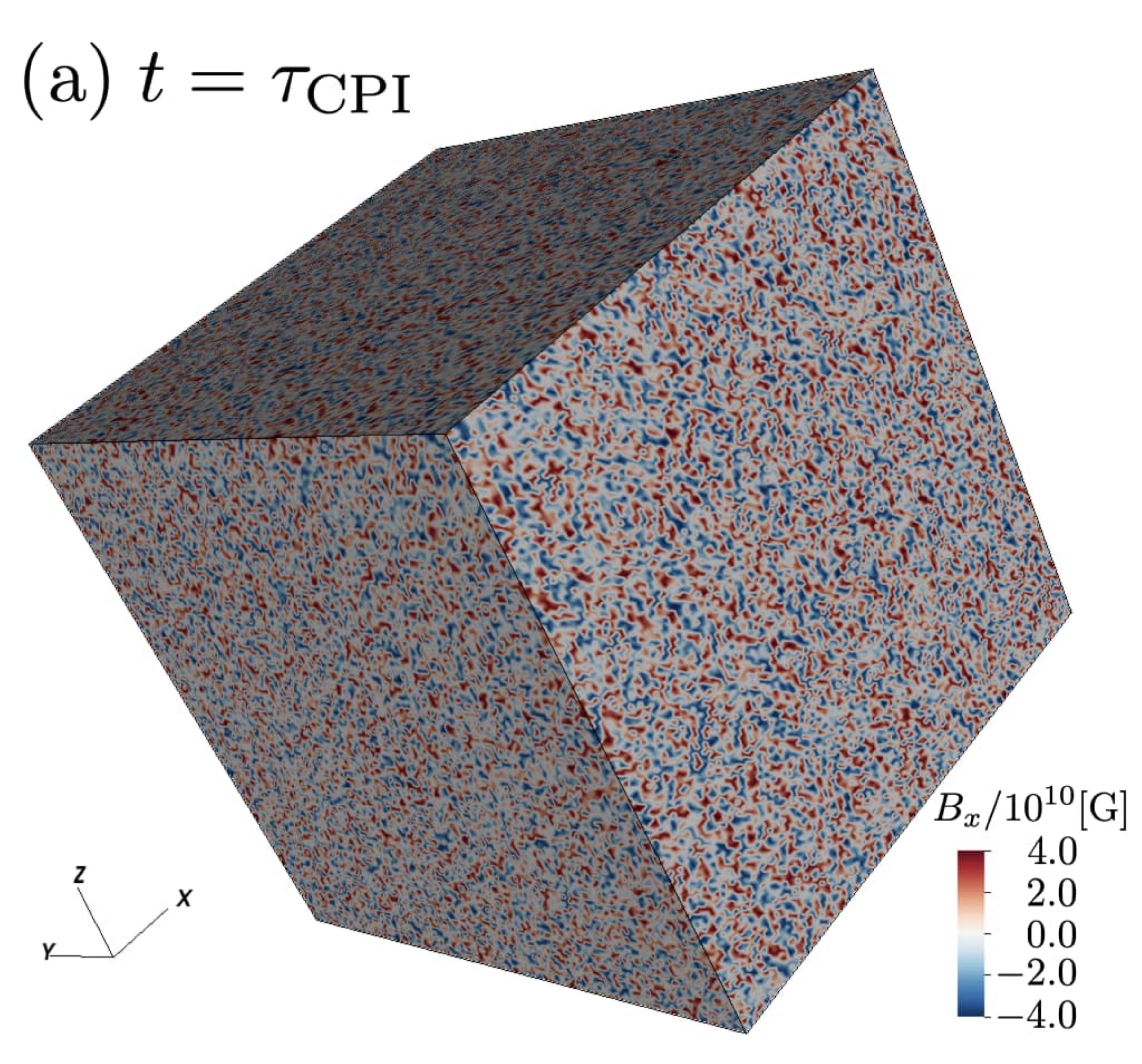}}}
\scalebox{0.13}{{\includegraphics{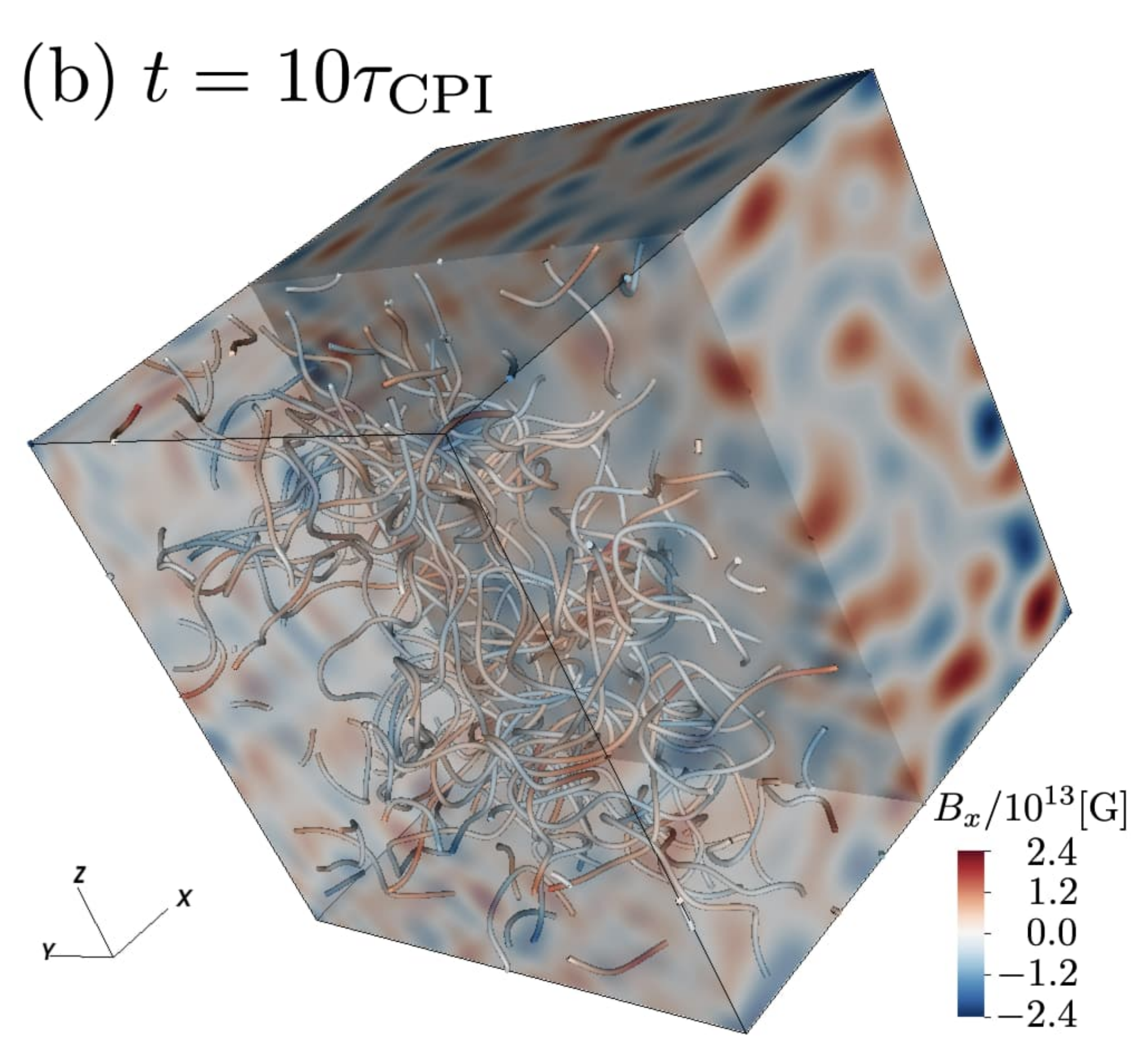}}}
\scalebox{0.13}{{\includegraphics{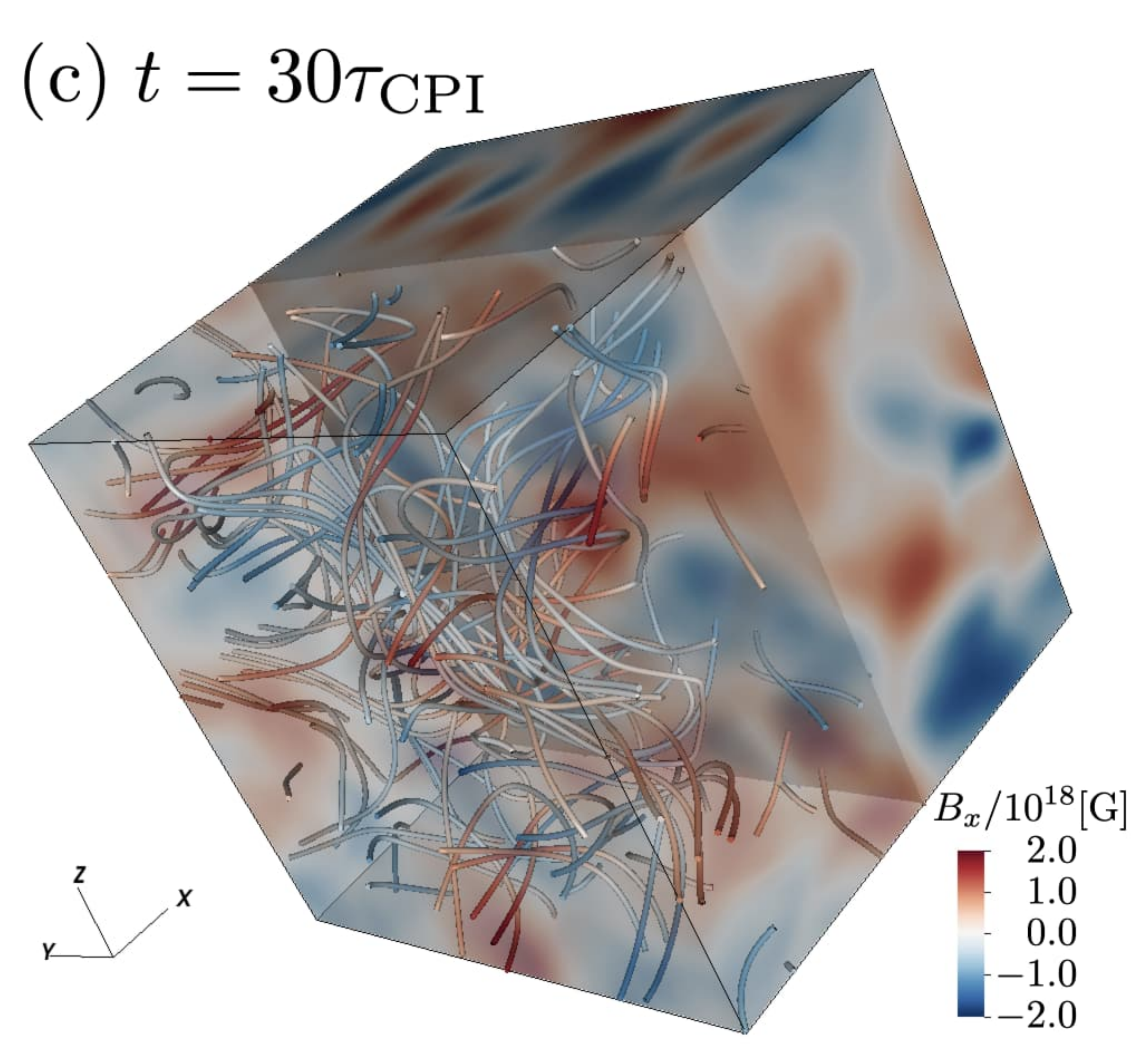}}}\\
\scalebox{0.13}{{\includegraphics{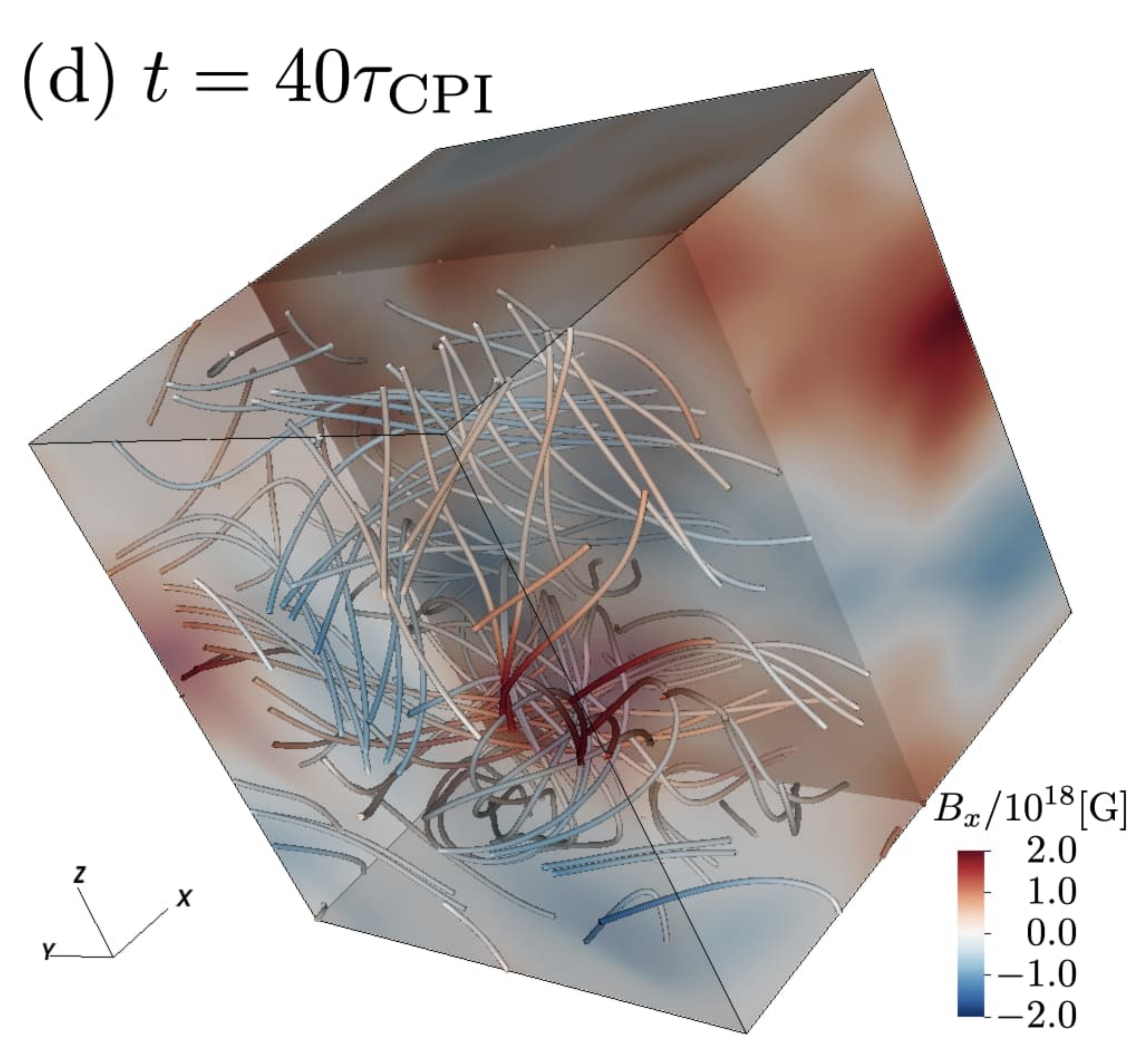}}}
\scalebox{0.13}{{\includegraphics{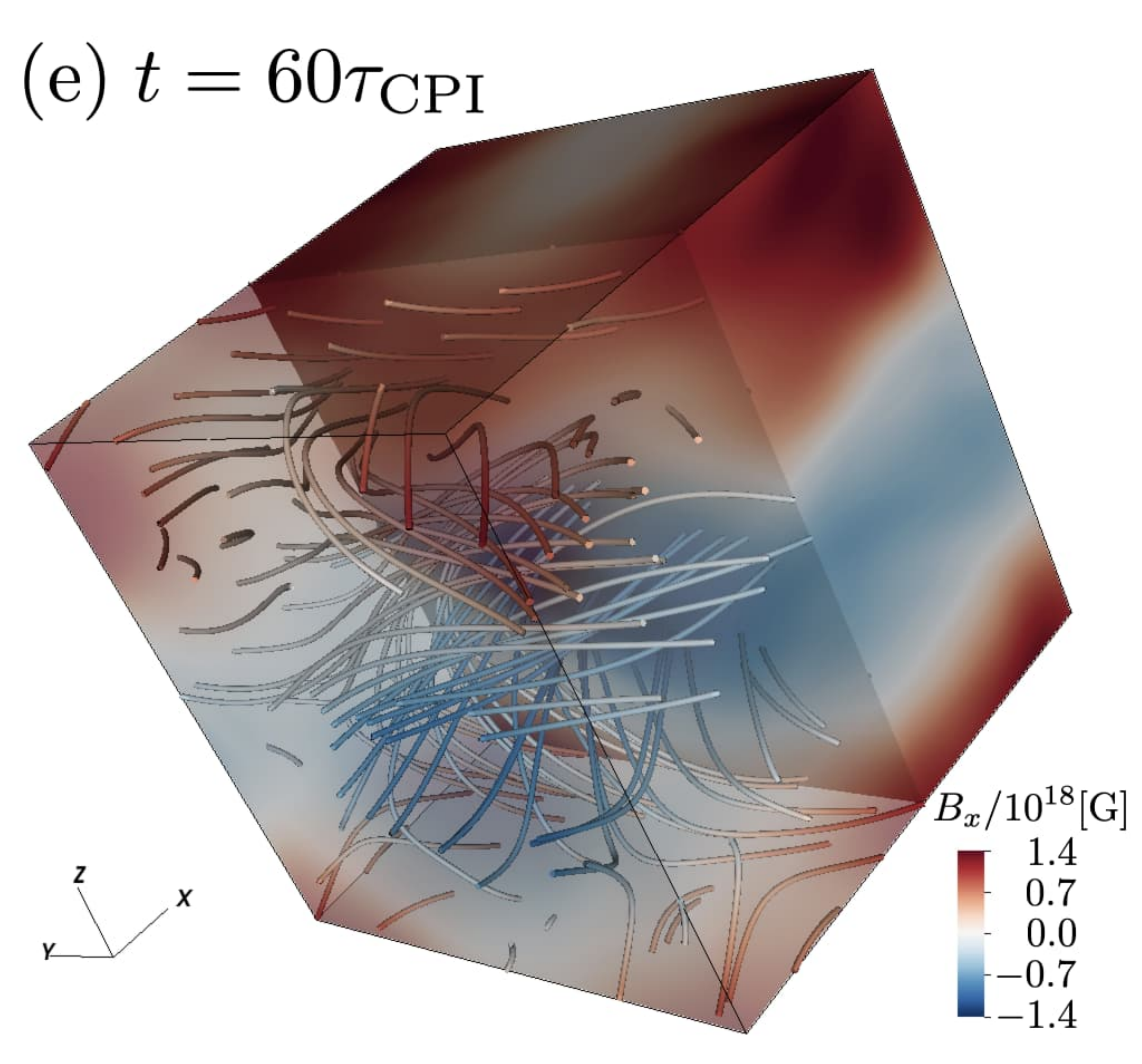}}}
\scalebox{0.13}{{\includegraphics{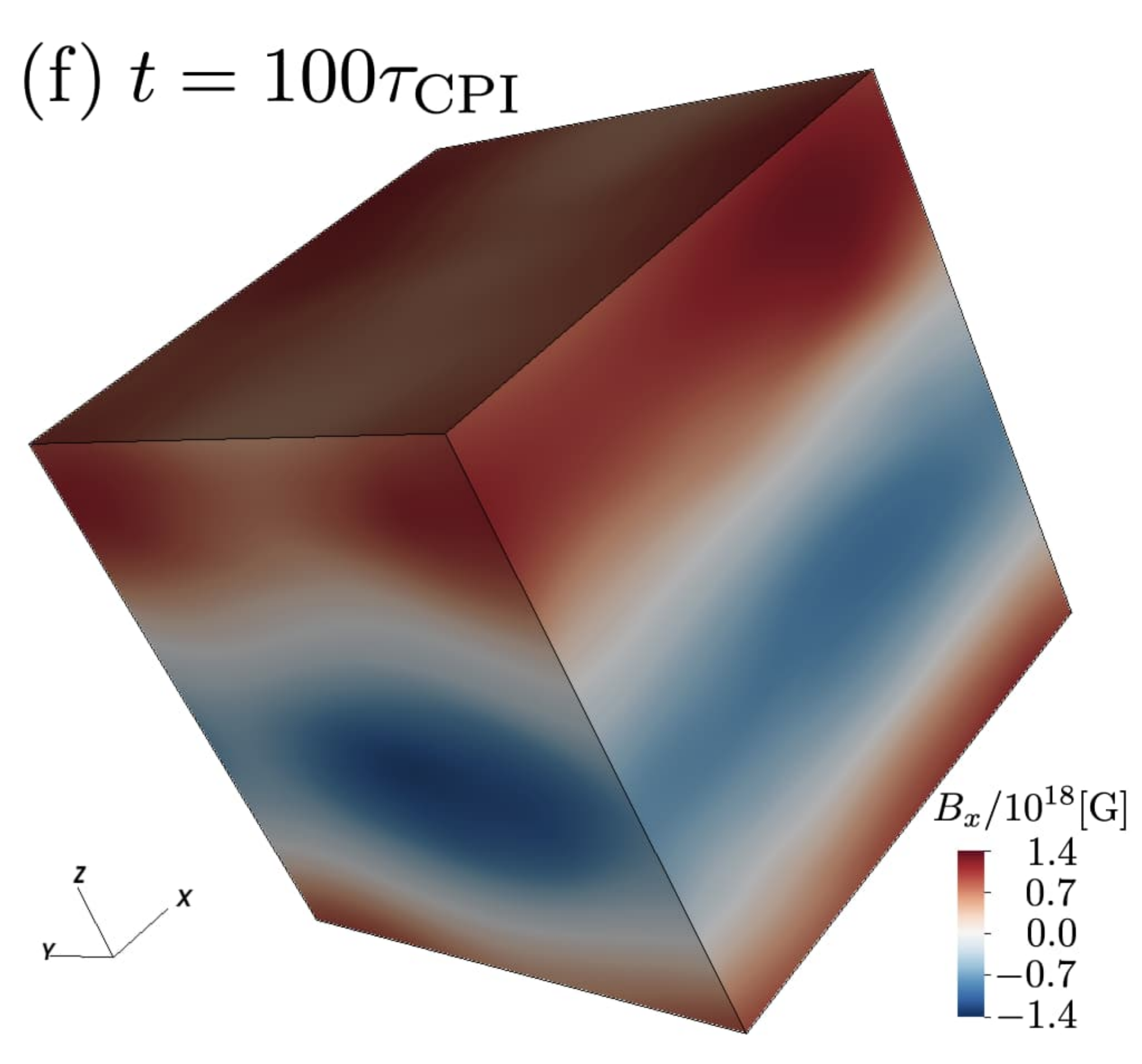}}}
\caption{Time evolution of 3D visualization of $B_x$ and magnetic field lines. Panels (a), (b), (c), (d), (e), and (f) correspond to the time $t/\tau_{\rm CPI}$=$1$, $10$, $30$, $40$, $60$, and $100$, respectively.
Figures are taken from Ref.~\cite{Matsumoto:2022lyb}.}
\label{fig2}
\end{center}
\end{figure}

\begin{figure}
\begin{center}
\scalebox{0.13}{{\includegraphics{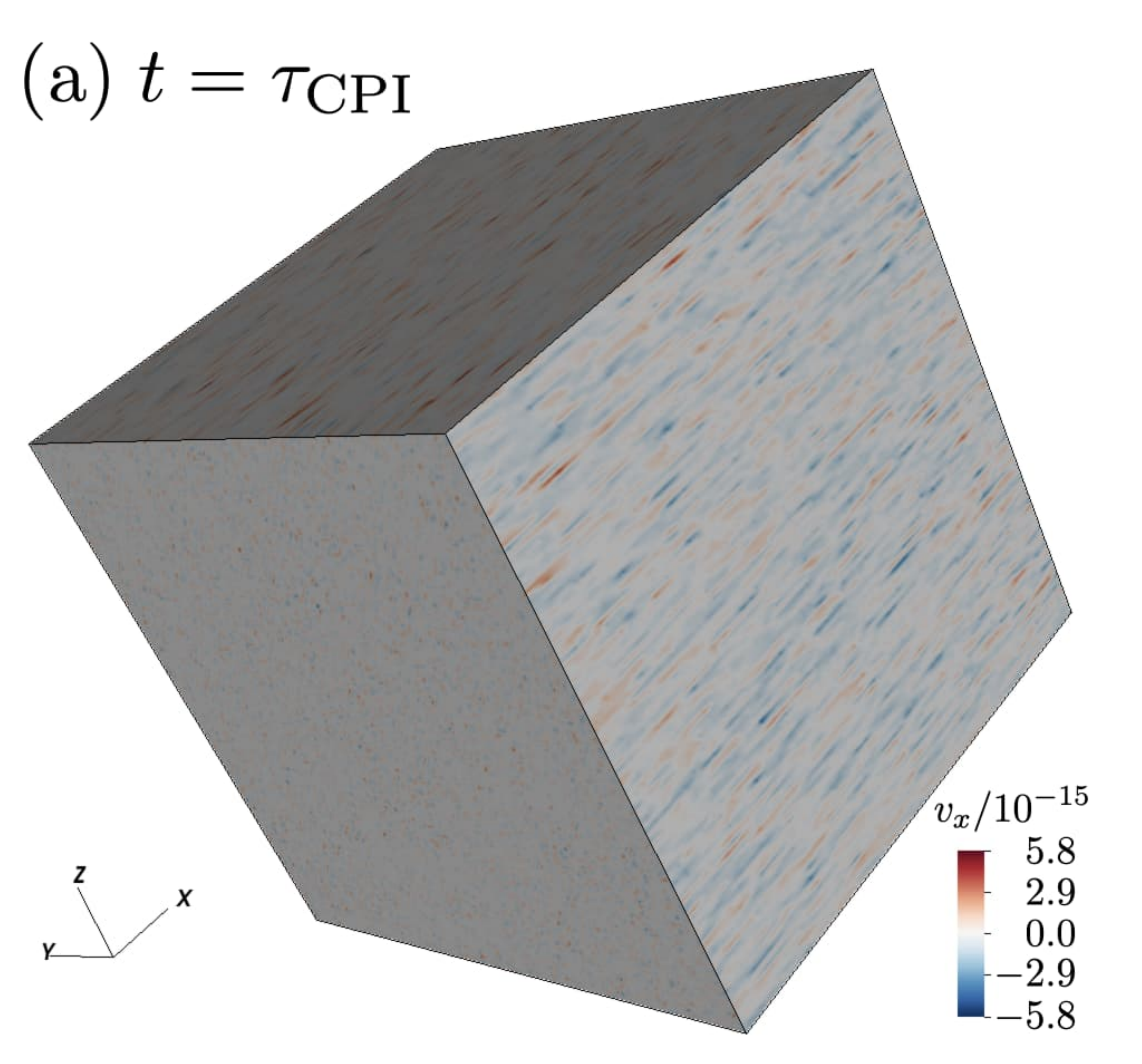}}}
\scalebox{0.13}{{\includegraphics{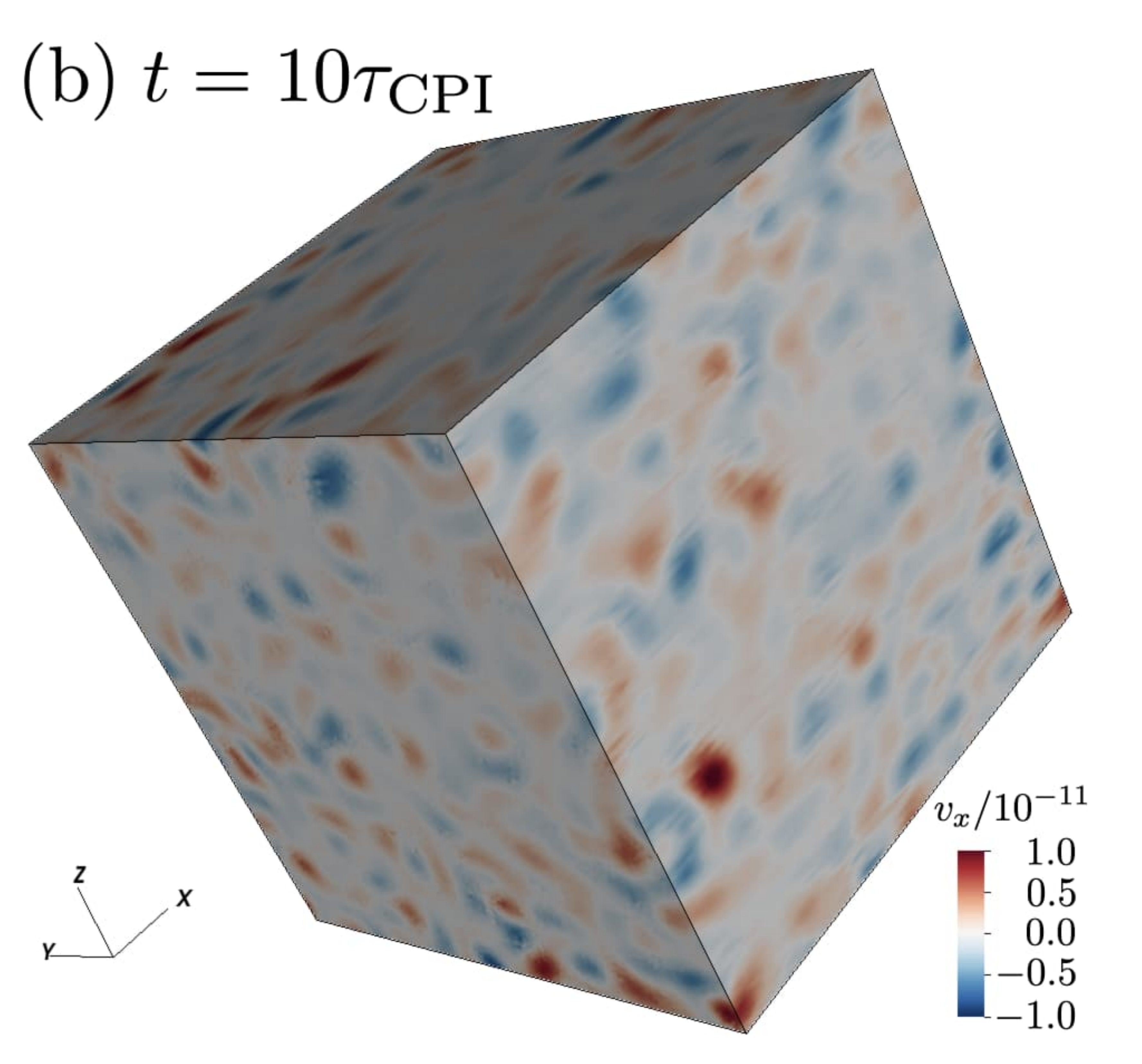}}}
\scalebox{0.13}{{\includegraphics{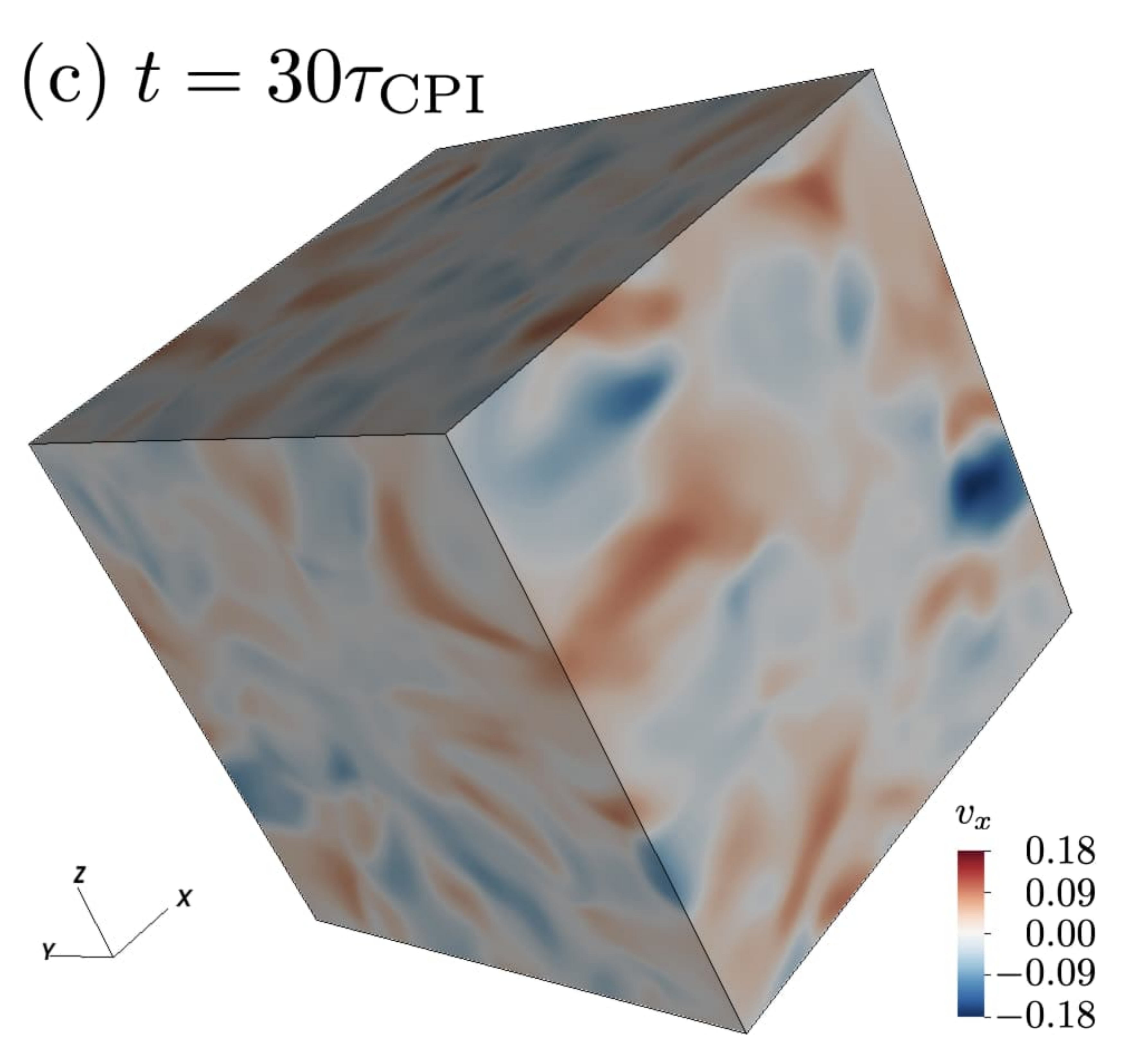}}}\\
\scalebox{0.13}{{\includegraphics{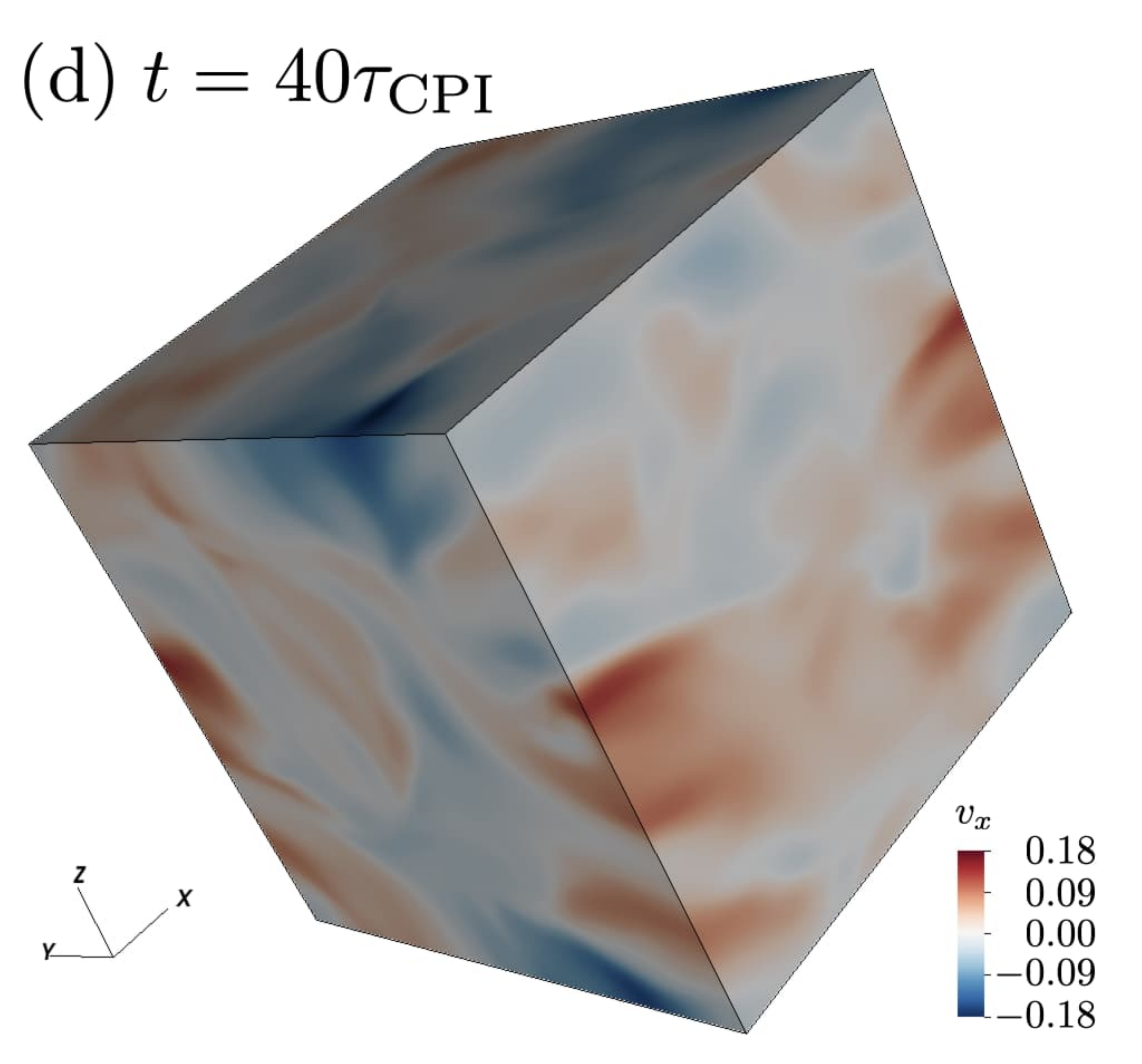}}}
\scalebox{0.13}{{\includegraphics{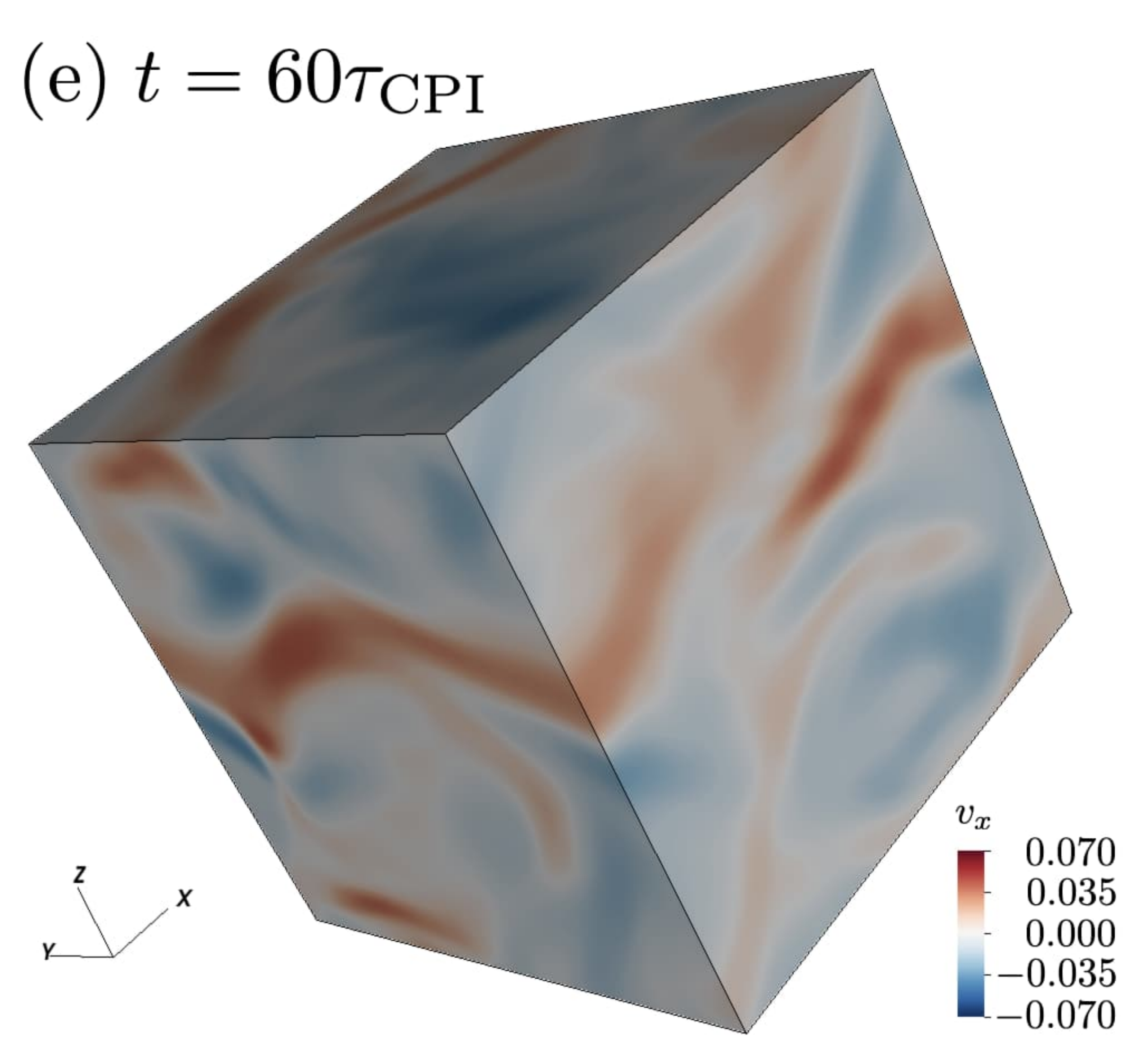}}}
\scalebox{0.13}{{\includegraphics{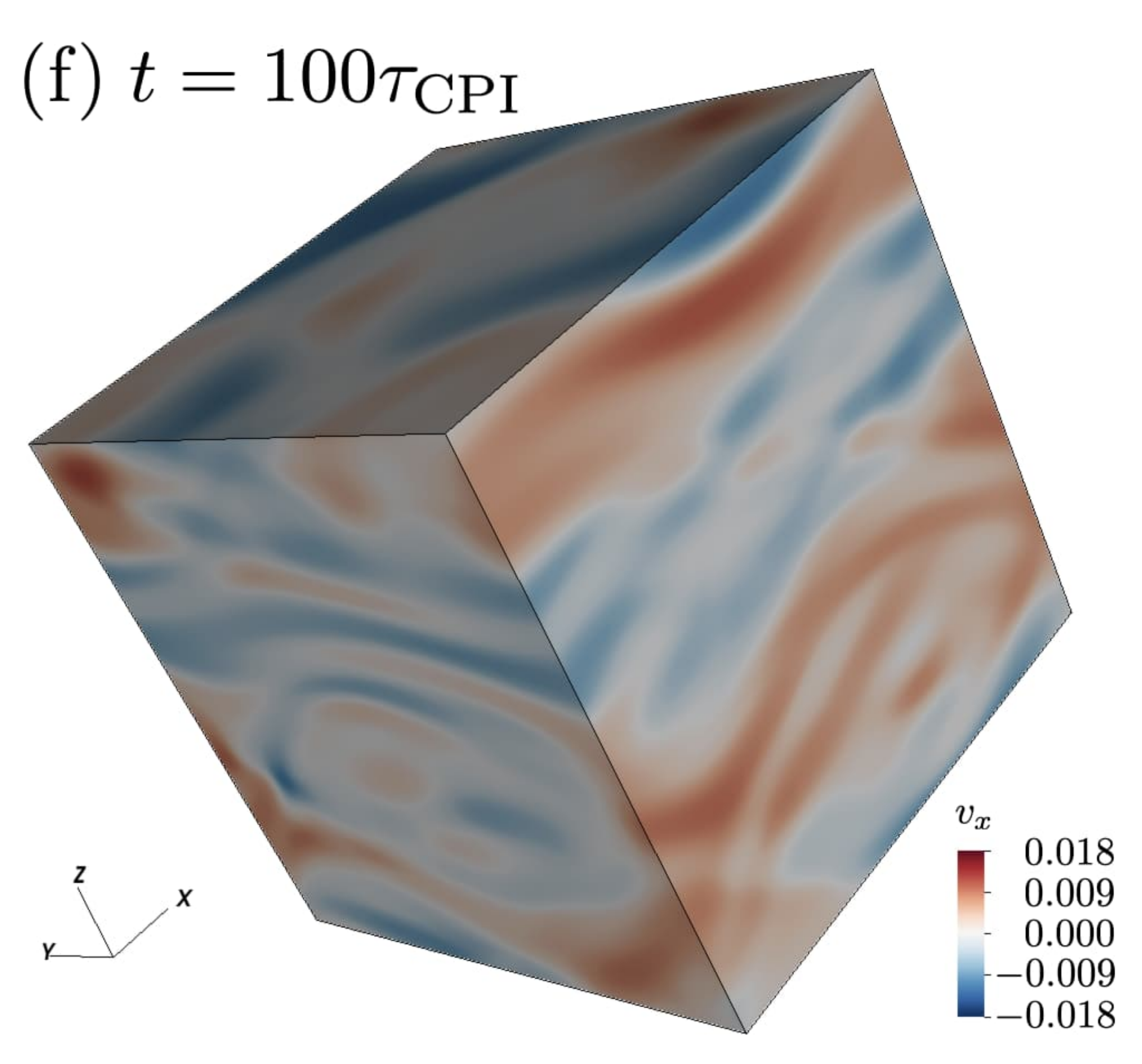}}}
\caption{Temporal evolution of 3D visualization of $v_x$. 
Time in each panel is the same as that in Fig.~\ref{fig2}.
Figures are taken from Ref.~\cite{Matsumoto:2022lyb}.}
\label{fig3}
\end{center}
\end{figure}

The numerical results of local simulations for these chiral MHD equations are shown in Figs.~\ref{fig2} and \ref{fig3}~\cite{Matsumoto:2022lyb}.
Here, $\nu=0.01$, $\eta = 1$, and $\xi_B = 0.1$ corresponding to the maximal value for Eq.~(\ref{main_j}) are taken as an example in the unit of $100\,{\rm MeV} = 1$; for the technical detail of the simulations, see Ref.~\cite{Matsumoto:2022lyb}. The inverse cascade of the magnetic energy and fluid kinetic energy can be observed in Fig.~\ref{fig2} and Fig.~\ref{fig3}, respectively, where $\tau_{\rm CPI} = 1/\Gamma_{\rm CPI}$ (see Eq.~(\ref{k_CPI})) is the typical time scale of the CPI. 
Eventually, the correlation lengths of the magnetic field and fluid velocity become comparable to the box size. These features can also be seen for smaller values of $\xi_B$ up to $\xi_B=10^{-5}$ in Ref.~\cite{Matsumoto:2022lyb}. The spectra of the magnetic energy and kinetic energy, $E_{\rm M}(k)$ and $E_{\rm K}(k)$ as a function of the wavenumber $k$, are studied in Ref.~\cite{Masada:2018swb}, and the scalings $E_{\rm M}(k) \propto k^{-2}$ and $E_{\rm K}(k) \propto k^{-5/3}$ in the small $k$ region are observed; see also the discussion in Section~\ref{sec:cosmo_CPI} in the context of the early Universe.

The qualitative difference between the MHD turbulence with and without the chiral/helical effects may be understood in two ways. 
One is based on the conservation laws: while only the energy is conserved in the conventional 3D hydrodynamic theory, {\it helicity} is also (approximately) conserved in the chiral MHD, which can change the turbulent behavior qualitatively.%
\footnote{Note also that in the usual MHD without the chiral effects, but with a non-zero magnetic helicity (which is a good conserved quantity for turbulent plasmas), inverse cascade has been observed for the maximally helical magnetic fields in the cosmological MHD simulations~\cite{Muller:2000zz,Banerjee:2004df,Kahniashvili:2012uj,Brandenburg:2016odr}.}
A similar situation is known to occur in 2D fluids, in which an additional conserved quantity---the {\it enstrophy} defined by $\int {\rm d}^2{\bm x}\,{\bm \omega}^2$---leads to the inverse energy cascade~\cite{Kraichnan1967}. 
Another way to understand the inverse cascading feature in chiral MHD is based on the properties of the CPI. As reviewed in Section~\ref{sec:CPI}, the CPI tends to reduce $\xi_B$ to weaken the instability, and consequently, the typical length scale of the CPI, $\ell_{\rm CPI} = 2\pi/k_{\rm CPI}$ with $k_{\rm CPI}$ given in Eq.~(\ref{k_CPI}), becomes larger as a function of time. Although this argument is valid when the nonlinear terms in the chiral MHD equations are subdominant, it provides a simple physical reason for the inverse cascade of the magnetic energy.

In summary, the chiral effects can reverse the turbulent direction from direct to inverse cascade in 3D. As the inverse cascade is known to help the supernova explosion in 2D~\cite{Radice:2017kmj}, the chiral effects and resulting inverse cascading tendency may be relevant to the mechanism of supernova explosions in 3D.

\subsection{\it Other applications in astrophysics \label{sec:other_astro}}

Here we summarize other applications of the chiral effects to astrophysical systems that are not covered above.

In most of applications above, we considered situations where chirality imbalance of leptons is produced by the parity-violating weak interaction. Yet, there is another mechanism to generate the chirality imbalance of leptons in astrophysical systems: the parallel electric and magnetic fields in the gap regions of magnetospheres of supermassive black holes and pulsars~\cite{Gorbar:2021tnw}. From the chiral anomaly relation (\ref{anomaly}) for electrons including the effect of chirality flipping,
\be
\d_t n_5 = \frac{e^2}{2\pi^2}{\bm E} \cdot {\bm B} - \Gamma_{\rm f }n_5\,,
\ee
where $\Gamma_{\rm f} \sim \alpha^2 m_{\rm e}^2/T$ is the chirality-flipping rate at high temperature $T \gg m_{\rm e}$ similar to Eq.~(\ref{flipping}), one obtains the steady-state solution for $n_5$,
\be
n_5 = \frac{e^2}{2\pi^2 \Gamma_{\rm f}} {\bm E} \cdot {\bm B}\,.
\ee
It is argued in Ref.~\cite{Gorbar:2021tnw} that while the produced chiral chemical potential is negligibly small near a supermassive black hole, it is much larger in the gap region of magnetars, $\mu_5 \sim 10^{-5}\,{\rm MeV}$-$10^{-3}\,{\rm MeV}$. It is also suggested that this triggers spontaneous generation and emission of helical electromagnetic modes for the frequency window $0 \lesssim \omega \lesssim \alpha |\mu_5|$. 

The chiral effects can also be important even in the absence of the mean chirality imbalance itself. In fact, fluctuations of chiral charges lead to novel collective modes, or chiral waves, such as the so-called chiral magnetic wave (CMW)~\cite{Newman:2005hd,Kharzeev:2010gd}, chiral vortical wave (CVW)~\cite{Jiang:2015cva}, chiral Alfv\'en wave~\cite{Yamamoto:2015ria}, chiral heat wave~\cite{Chernodub:2015gxa}, and chiral shock wave~\cite{Sen:2016jzl}; see also Ref.~\cite{Rybalka:2018uzh} for a related work. 
Originally, the possible observation of CMW was studied in the context of quark-gluon plasmas in heavy ion collision experiments~\cite{Burnier:2011bf}. Recently, this idea was imported to the area of seismology relating oscillations of neutron stars and supernovae to their internal structures, and the frequency and magnitude of CMW/CVW at the core of these compact stars are estimated~\cite{Hanai:2022yfh}. In particular, seismic oscillations due to the CMW and resulting GWs may provide a possible new probe for measuring the magnetic field in neutron stars and supernovae.

As the other application of chiral effects, Ref.~\cite{Flachi:2017vlp} attempted to relate the neutrino flux due to the CVE to collimated jets from rotating astrophysical compact sources. It is also argued in Ref.~\cite{Yamamoto:2017uul} that the photonic analogue of the axial CVE (the second of Eq.~(\ref{CVE})) proportional to $T^2 {\bm \Omega}$ provides a novel contribution to the photon emission from pulsars and rotating massive stars. This and related effects are also studied in Refs.~\cite{Avkhadiev:2017fxj,Zyuzin2017,Chernodub:2018era} theoretically or in a different context. A distinct feature of the photonic CVE is that photons emission has dependence on the circular polarizations in that only right-handed photons are emitted from one of the poles while only left-handed ones from the other. This effect is expected to be most significant in the accretion-powered millisecond pulsars where the temperature and angular velocity are comparably large.

The framework of the chiral radiation transport theory for neutrinos in Section~\ref{sec:ChRTT} may also be extended to circularly polarized photons, which should be applicable to other astrophysical systems involving photon radiation. The chiral kinetic theory for photons is derived based on the notion of Berry curvature in Ref.~\cite{Yamamoto:2017uul} and from the Wigner function in Refs.~\cite{Huang:2020kik,Hattori:2020gqh,Lin:2021mvw}, and an extension to curved spacetime is made in Ref.~\cite{Mameda:2022ojk}.

\section{Chiral effects in cosmology}
\label{sec:cosmo}
In this section, we review the recent developments of the studies on the early Universe cosmology where the interplay between the cosmological magnetic fields and the chirality carried by the fermions plays important roles. 
One may wonder that the chiral symmetry is violated by the Yukawa interaction in the SM and hence any interesting chiral phenomena cannot occur since the chirality would be washed out quickly even if it is produced in the early Universe. 
However, when the temperature of the Universe is sufficiently high so that the electron Yukawa interaction is weak compared to the cosmic expansion, the right-handed electron number is a well-conserved quantity at the temperature $T>T_{h_\mathrm{e}} \simeq10^2\,{\rm TeV}$, where $T_{h_\mathrm{e}}$ is the equilibration temperature of the electron Yukawa interaction~\cite{Campbell:1992jd,Bodeker:2019ajh}.
This makes the net chirality under the hyper U(1) gauge interaction of the SM also a well-conserved quantity, which opens up the possibility that the chiral effects cause interesting phenomena. 
In the following, we examine how the chiral effects appear in the models of the early Universe such as primordial (hyper)magnetic field generation (magnetogenesis), baryogenesis from the hypermagnetic  helicity decay, and production of the chiral GWs.  

We first discuss the generation of magnetic fields with helicity in the early Universe in Section~\ref{sec:primordial} with several examples. The first example, pseudoscalar inflation, is not directly related to the topics discussed in the previous sections, but it uses the chiral anomaly to induce the Chern-Simons term essential for the mechanism of magnetogenesis. On the other hand, the second example is a direct application of the CME and CPI discussed in the previous sections to the early Universe cosmology. 
Then we discuss the baryogenesis from the primordial magnetic fields in Section~\ref{sec:baryogenesis}. This mechanism is not only a direct consequence of the chiral anomaly but also can be seen as the backreaction of the CME or CPI. Finally, we discuss the generation of chiral gravitational waves from the helical magnetic fields in Section~\ref{sec:GW}, which will be a possible probe of the chiral phenomena in the early Universe. In this way, the early Universe cosmology in this section shares the similar underlying physics related to the chiral effects.

\subsection{\it Primordial magnetic field generation \label{sec:primordial}}

Honestly speaking, for the standard cosmology, the primordial magnetic fields that spread over the entire very early Universe with relatively large coherence length are not necessary ingredients. 
However, there are several advantages to imagine that cosmological magnetic fields are generated in the early Universe. 
In particular, in this decade, the existence of the intergalactic magnetic fields (IGMFs) suggested by the blazar observations~\cite{Neronov:2010gir,Tavecchio:2010mk,Dolag:2010ni,Finke:2015ona,Fermi-LAT:2018jdy,Durrer:2013pga,Subramanian:2015lua} has inspired the interest for the exploration of the primordial magnetic fields, which evolve according to the MHD and remain until today. 
The latest constraints from the Fermi-LAT collaboration~\cite{Fermi-LAT:2018jdy} gives the constraint on the present IGMF strength $B_0$ as $B_0 \gtrsim 10^{-16}\,{\rm G}$ at the coherence length $\lambda > 10\,{\rm kpc}$ and stronger at smaller scales. 
Note that it has not been clear if the IGMFs are really responsible for the observations (see, e.g., Refs.~\cite{2012ApJ...752...22B,2012ApJ...758..102S,Broderick:2018nqf}) and are the remnants from the very early Universe, if ever. 
Nevertheless we hereafter discuss the several phenomena associated with the primordial (hyper)magnetic fields with these observations in mind. 

An important property of the gauge fields is the magnetic helicity, as has been discussed in the previous sections. 
Helical primordial magnetic fields are, indeed, of interest, since once they enter the MHD turbulent regime, they evolve with the inverse cascade where the energy at a small scale is transferred to the larger scales~\cite{Son:1998my,Muller:2000zz,Banerjee:2004df,Kahniashvili:2012uj,Brandenburg:2016odr}, which results in stronger magnetic fields with larger coherence length and is suitable for explaining the present IGMFs. (See, however, the discussion on the constraints from the baryogenesis in the next subsection.)
In the following we examine two representative mechanisms that can give rise to primordial helical magnetic fields and how chiral effects play roles there. 

\subsubsection{\it Pseudoscalar inflation}

The first example to generate helical magnetic fields is the pseudoscalar inflation or axion inflation~\cite{Turner:1987bw,Garretson:1992vt,Anber:2006xt}. 
Inflation~\cite{Starobinsky:1980te,Kazanas:1980tx,Guth:1980zm,Sato:1980yn}, which is now a part of modern cosmology, is a mechanism that expands the primordial Universe in an accelerating manner so that the present Universe does not suffer from the flatness and horizon problems as well as the monopole problem. 
A difficulty in constructing a concrete model of inflation lies in that a sufficiently flat potential of inflaton against quantum corrections is required. 
Inflation models~\cite{Freese:1990rb,Adams:1992bn} driven by the potential energy of a homogeneous and slow-rolling pseudo-Nambu-Goldstone boson or an axion-like particle (ALP)~\cite{Weinberg:1977ma,Wilczek:1977pj}, $\varphi(t)$, has an advantage from this point of view, since the flatness of the potential is guaranteed by an approximate shift symmetry, which is not spoiled by radiative corrections. They are now regarded as one of the most promising models of inflation. 

If the underlying global symmetry is anomalous under the U(1) (hyper)gauge symmetry of the SM, the inflaton couples to the hypergauge field via an effective Chern-Simons term, 
\begin{equation}
{\cal L}_\mathrm{eff} \ni \frac{\varphi}{4 f} Y_{\mu\nu} {\tilde Y}^{\mu\nu}, 
\end{equation}
which is supposed to be originated from the chiral anomaly of the underlying theory. 
Here $Y_{\mu\nu} \equiv \partial_\mu Y_\nu -\partial_\nu Y_\mu$ is the hypergauge field strength tensor and ${\tilde Y}^{\mu\nu} \equiv \epsilon^{\mu\nu\rho\sigma} Y_{\rho\sigma}/(2\sqrt{-g})$ is its dual with $g$ being the determinant of the metric. 
The parameter $f$ is related to the scale of the spontaneous breaking of the global symmetry, which appears in the axion potential.%
\footnote{Since the effective Chern-Simons term appears, e.g., by integrating out the fields that become massive through the spontaneous symmetry breaking of the underlying global symmetry, $f$ is not exactly the symmetry breaking scale but also depends on the detail of the matter contents in the UV theory.} 
With this effective Lagrangian, the gauge fields experience non-trivial dynamics during inflation. 
Note that in the presence of the homogeneous and monotonic dynamics of the ALP, it gives a similar effect to the CME with the correspondence, $\partial_t \varphi \leftrightarrow \mu_5$, so that the gauge fields exhibit a CPI-like instability. 
The advantage to consider the dynamics during inflation is that there are no thermal plasmas at that time and hence we do not have to take into account the friction for the gauge field dynamics due to the Ohmic current and the gauge field amplification is much more effective.%
\footnote{In principle one can also consider the dynamics of ALP in the radiation dominated Universe to amplify the gauge fields. One example is the study that takes advantage of the dark photon that kinetically mixes with the SM U(1) gauge fields~\cite{Choi:2018dqr} to weaken the thermal effect. But to the best of knowledge of the present authors, no comprehensive study on the gauge field amplification in the radiation dominated Universe due to the directly-coupled ALP dynamics has been done yet.
Note also that the effect of radiation during axion inflation is discussed in, e.g., Refs.~\cite{Ferreira:2015omg,Berghaus:2019whh}.}
As we will see, we still need to take into account the backreaction from the Schwinger effect.

Now we examine how (hyper)magnetic fields are generated during pseudoscalar inflation. 
We work with the Friedmann-Lema\^{i}tre-Robertson-Walker (FLRW) metric,
\begin{equation} 
{\rm d}s^2 = {\rm d}t^2 - a^2(t) {\rm d} {\bm x}^2 = a^2(\tau)({\rm d} \tau^2 - {\rm d}{\bm x}^2), \label{FLRW_metric}
\end{equation} 
with $a$ being the scale factor and $\tau$ being the conformal time.  
To investigate the gauge field in a model-independent way, we fix the metric with the de Sitter one, $\partial_t a/a=H_\mathrm{inf} (= \mathrm{const}.)$,%
\footnote{Generally, we define the Hubble parameter as $H\equiv \partial_t a/a$.  The subscript ``inf'' here denotes that it is evaluated during inflation.} or $a(\tau) = - 1/(H_\mathrm{inf} \tau)$. 
The ALP is a slow-rolling inflaton and has a non-vanishing time derivative, ${\partial_t \varphi}$, which is taken to be a constant. 
Then with this background dynamics and omitting the backreaction from the matter fields, the mode equation for the quantized hypergauge fields is given by 
\begin{equation}
\left[\partial_\tau^2 + k^2 \left( 1+\frac{2 \lambda \xi}{k\tau}\right)\right] Y_\lambda^k  (\tau) =0\,, \quad \xi \equiv \frac{{\partial_t \varphi}}{2f H_\mathrm{inf}},   \quad k = |{\bm k}|, \quad -\infty < \tau < 0,  \label{eom_hypergauge_inf}
\end{equation}
where $\lambda = \pm$ indicates the circular polarization mode of the gauge field and we have quantized the hypergauge fields as
\begin{equation}
Y_i ({\bm x},\tau) = \sum_{\lambda = \pm} \int \frac{{\rm d}^3 {\bm k}}{(2\pi)^{3/2}} \left[ Y_\lambda^k (\tau) e_i^\lambda({\bm k}) {\hat a}_\lambda({\bm k}) {\rm e}^{{\rm i} {\bm k} \cdot {\bm x}} +\mathrm{h.c.}\right]
\end{equation}
with adopting the radiation gauge $\partial^i Y_i =0$, $Y_0=0$. 
Here ${{\hat a}_\lambda}^\dagger({\bm k})$ and ${\hat a}_\lambda({\bm k})$ are the creation and annihilation operators satisfying usual commutation relations, and $e_i^\lambda({\bm k})$ is the circular polarization vector that satisfies
\begin{equation}
\label{polarization}
\sum_i k_i e_i^\lambda({\bm k})=0, \quad \sum_{j,l} {\rm i} \epsilon_{ijl} k_j e_l^\lambda({\bm k}) = \lambda k e_i^\lambda({\bm k}), \quad \sum_i e_i^\lambda ({\bm k}) {e_i^{\lambda'}}^* ({\bm k}) = \delta^{\lambda \lambda'}, \quad {e_i^{\lambda}}^* ({\bm k})  = e_i^{\lambda} (-{\bm k}) . 
\end{equation}
One can easily see that for $\xi>0$ and $\xi<0$, the $+$ and $-$ modes exhibit instabilities at small $k$, respectively, which guarantees the amplification of helical hypergauge fields. 
Requiring that the mode functions reduce to the ones in the Bunch-Davies vacuum in the asymptotic past,
\begin{equation}
\lim_{k\tau \rightarrow -\infty} Y_\lambda^k(\tau) = \frac{{\rm e}^{-{\rm i} k \tau}}{\sqrt{2 k}}\,, 
\end{equation}
we obtain the solution 
\begin{equation}
Y_\lambda(\tau, {\bm k})  = \frac{{\rm e}^{\lambda \pi \xi/2}}{\sqrt{2 k}} W_{-{\rm i} \lambda \xi, 1/2} (2 {\rm i} k \tau)\,,  \label{gen_mode_sol}
\end{equation}
where $W_{\kappa,\mu}(z)$ is the Whittaker function. 
To be concrete, hereafter we focus on the case with $\xi>0$, but  the case with $\xi<0$ can be analyzed in the same way. 

The resultant properties of the gauge fields are evaluated as follows. 
The typical hypermagnetic and electric field strength as well as the coherence length in the physical frame is defined as
\begin{align}
B_\mathrm{p} &= \langle B_i^2\rangle^{1/2} = \left( \frac{1}{a^4} \int \frac{{\rm d}^3{\bm k}}{(2 \pi)^3}  \sum_\lambda k^2 |Y_\lambda^k|^2 \right)^{1/2}, \label{infBp}\\
E_\mathrm{p} &= \langle E_i^2 \rangle^{1/2} = \left( \frac{1}{a^4} \int \frac{{\rm d}^3{\bm k}}{(2 \pi)^3}  \sum_\lambda   \left| \partial_\tau Y_\lambda^k \right|^2 \right)^{1/2}, \label{infEp} \\
\lambda_\mathrm{p} &=\frac{2 \pi}{a^3 B_\mathrm{p}^2} \int \frac{{\rm d}^3 \bm{k}}{(2\pi)^3} \sum_\lambda k |Y_\lambda^k|^2. \label{infcohp}
 \end{align}
With the asymptotic form of the Whittaker function at $- k \tau \ll 1$, we see that only $+$ modes are exponentially amplified and obtain the physical hypermagnetic and electric field at the end of inflation as
\begin{align}
B_\mathrm{p} & \simeq 1.2 \times 10^{-2} \frac{{\rm e}^{\pi \xi}}{\xi^{5/2}}H_\mathrm{inf}^2 \simeq 6 \times 10^{43}\,\mathrm{G} \ \frac{{\rm e}^{\pi \xi}}{\xi^{5/2}} \left(\frac{H_\mathrm{inf}}{10^{13}\,\mathrm{GeV}}\right)^2,  \label{Bp_inf_num}\\
E_\mathrm{p} & \simeq 1.1 \times 10^{-2}\frac{{\rm e}^{\pi \xi}}{\xi^{3/2}}H_\mathrm{inf}^2,  \label{Ep_inf_num}\\
\lambda_\mathrm{p} & \simeq 0.6 \xi \frac{2 \pi}{H_\mathrm{inf}}  \simeq 2 \times 10^{-51}\,\mathrm{Mpc} \ \xi \left(\frac{H_\mathrm{inf}}{10^{13}\,\mathrm{GeV}}\right)^{-1}, \label{lambdap_inf_num}
\end{align}
for large enough $\xi>4$ with the electric and magnetic fields being almost parallel. 
Note that the integrals in Eqs.~\eqref{infBp}, \eqref{infEp}, and \eqref{infcohp} formally suffer from the UV divergences 
and require appropriate renormalization~\cite{Ballardini:2019rqh}. 
Here we simply introduce a physically-motivated UV cutoff, $k_\mathrm{UV} = 2 a_\mathrm{end} \xi H_\mathrm{inf}$, with $a_\mathrm{end}$ being the scale factor at the end of inflation, which corresponds to the highest wavenumber that experienced the tachyonic instability. 
This estimate is found to be consistent with the one studied in Refs.~\cite{Anber:2006xt,Jimenez:2017cdr}. 
While we have taken $H_\mathrm{inf}$ and $\xi$ as constants in the present estimate, in the realistic inflation models they are time-dependent. 
However their time-dependence is suppressed by the slow-roll dynamics of inflation and gives only slight deviation from the estimate in the above. 

Let us suppose that the Universe is instantaneously reheated to become radiation dominated so that the electric fields are quickly screened (see also Refs.~\cite{Fujita:2015iga,Adshead:2016iae,Hashiba:2021gmn} that tried to address the gauge field amplification during the inflaton oscillation dominated era).  
We expect that the magnetic fields evolve according to the MHD where the velocity fields $v$ are induced by the magnetic fields. 
First they evolve adiabatically until the eddy turnover scale $\sim v t$ catches up the magnetic field coherence length and then evolve according to the inverse cascade for the maximally helical magnetic fields $B_\mathrm{p} \propto a^{-7/3}$ and $\lambda_\mathrm{p} \propto a^{5/3}$ until recombination~\cite{Banerjee:2004df,Kahniashvili:2012uj}. 
After recombination until today, magnetic fields are expected to evolve adiabatically again. 
Now suppose that the equipartition between the magnetic fields and velocity fields are established, 
$v \sim B_\mathrm{p}/\sqrt{\epsilon_\mathrm{rad}}$, where $\epsilon_\mathrm{rad} = (\pi^2 g_*/30) T^4$ with $g_*$ being 
the number of relativistic degrees of freedom and $T$ the (unperturbed) temperature of the Universe.    
Then the properties of the magnetic fields today are evaluated as~\cite{Jimenez:2017cdr} 
\begin{align}
\label{B_p}
B_\mathrm{p}  & \simeq 2.5 \times 10^{-19}\,\mathrm{G} \left(\frac{{\rm e}^{2 \pi \xi}}{\xi^4}\right)^{1/3} \left(\frac{H_\mathrm{inf}}{10^{13}\,\mathrm{GeV}}\right)^{1/2}, \\
\lambda_\mathrm{p} & \simeq 0.28\,\mathrm{pc} \left(\frac{B_\mathrm{p}}{10^{-14}\,\mathrm{G}}\right). 
\end{align}
If $\xi$ is sufficiently large, the magnetic fields remain as the IGMFs so that we may expect to detect them by observations of astrophysical objects such as blazars. 

\

Thus far we have taken into account only gauge fields and background axion dynamics. But if we identify the gauge fields as the SM hypergauge fields, we need to incorporate  the matter fields in the SM, which are produced by the amplified gauge fields via the Schwinger effect~\cite{Schwinger:1951nm}. 
Since the produced particles screen the electric field, it is expected that the gauge field amplification becomes less effective. 

Although in order to give a precise estimate for the dynamics of the system we need to solve the dynamics of the gauge fields and matter fields simultaneously, it is practically difficult. 
There are several trials to estimate the effect of the Schwinger pair-production on the gauge field production from the axion inflation~\cite{Domcke:2018eki,Domcke:2019mnd,Sobol:2019xls,Domcke:2019qmm,Domcke:2021fee,Gorbar:2021rlt,Gorbar:2021zlr,Gorbar:2021ajq,Fujita:2022fwc}. 
Here we introduce some of them, which would give the best estimate at the moment. 

One of the most difficult problems is that we need to study the particle production from the dynamical spacetime-dependent gauge fields. 
However, if the energy scale of the produced particle is much larger than the inverse of the time scale as well as the coherence length of the gauge fields, we may adopt the Schwinger induced current for the one with the electric and magnetic fields in parallel and constant in both space and time~\cite{Domcke:2018eki},
\begin{equation}
\label{j_Y}
\langle j_Y\rangle = \frac{(g' |Q|)^3}{6 \pi^2}  \coth\left(\frac{\pi B_\mathrm{p}}{E_\mathrm{p}} \right)\frac{E_\mathrm{p} B_\mathrm{p}}{H_\mathrm{inf}}\,, 
\end{equation}
which runs the same direction to the electric and magnetic fields. 
Here $g'$ is the hypergauge coupling constant and $Q$ is the hypercharge of the fermion. To take into account multiple fermion species, we can just sum up each contribution.
The induced current is summation of the contribution from the LLL, which reflects the chiral anomaly described in the previous section, and from the HLL. 
Note that we have assumed that all the fermions are massless, which is the case when the SM Higgs is stabilized by a sufficiently large Hubble induced mass. 
With these simplifications, the evolution equation of the energy density of the hypergauge fields, $\epsilon_Y\equiv  (E_\mathrm{p}^2+B_\mathrm{p}^2)/2$, is written as
\begin{equation}
{\partial_t \epsilon_Y} + 4 H \epsilon_Y = 2 \xi H_\mathrm{inf} E_\mathrm{p} B_\mathrm{p} - E_\mathrm{p} \langle j_Y\rangle, 
\end{equation}
where the first term of the right-hand side is the source term from the axion dynamics while the second term is the backreaction due to the Schwinger effect, which makes the first term less efficient. 

The basic approximation of the first method to give the estimate for the gauge field production including the backreaction from the Schwinger effect~\cite{Domcke:2018eki} is to identify the Schwinger current as the modification of the instability parameter $\xi$, 
\begin{equation}
\xi \rightarrow \xi_\mathrm{eff} \equiv \xi - \frac{(g' |Q|)^3}{12 \pi^2} \coth \left(\frac{\pi B_\mathrm{p}}{E_\mathrm{p}} \right) \frac{E_\mathrm{p}}{H_\mathrm{inf}^2} \label{xieff_def}\,.
\end{equation}
Since $\xi_\mathrm{eff}$ depends on the hypermagnetic and electric fields, we cannot simply solve the equation of motion~\eqref{eom_hypergauge_inf} by replacing $\xi$ with $\xi_\mathrm{eff}$. 
The ``equilibrium estimate''\footnote{The ``maximal estimate'' was also proposed in Ref.~\cite{Domcke:2018eki}. This estimate is now thought to be overestimated. The gradient expansion method~\cite{Sobol:2019xls} likely gives the real maximal estimate~\cite{Gorbar:2021rlt,Gorbar:2021zlr}.} proposed in Ref.~\cite{Domcke:2018eki} determines $\xi_\mathrm{eff}$ by solving Eq.~\eqref{xieff_def} with taking the physical hypermagnetic and electric fields Eqs.~\eqref{infBp} and \eqref{infEp} as the function of $\xi_\mathrm{eff}$, with $\xi$ in the mode function~\eqref{gen_mode_sol} being replaced by $\xi_\mathrm{eff}$. 
The field strength and coherence length of the physical hypermagnetic and electric fields are given by substituting $\xi_\mathrm{eff}$ to $\xi$ in Eqs.~\eqref{Bp_inf_num}, \eqref{Ep_inf_num}, and \eqref{lambdap_inf_num}. 
The following fitting formula with the SM particle contents was found in Ref.~\cite{Gorbar:2021zlr}:
\begin{equation}
\xi_\mathrm{eff} \simeq a_\mathrm{eq} \ln (|\xi| + b_\mathrm{eq}) + c_\mathrm{eq},  
\end{equation}
with 
\begin{align}
a_\mathrm{eq} &\simeq 0.3679-0.0004 \log_{10} \left(\frac{H_\mathrm{inf}}{10^{12}\,\mathrm{GeV}}\right), \\
b_\mathrm{eq} &\simeq -3.3668+0.0099 \log_{10} \left(\frac{H_\mathrm{inf}}{10^{12}\,\mathrm{GeV}}\right), \\
c_\mathrm{eq} &\simeq 3.7012-0.0152 \log_{10} \left(\frac{H_\mathrm{inf}}{10^{12}\,\mathrm{GeV}}\right). 
\end{align}

Another possible approximation is to identify the Schwinger current as the Ohmic current with the hyperelectric conductivity as
\begin{equation}
\sigma = \frac{(g' |Q|)^3}{6 \pi^2} \frac{B_\mathrm{p}}{H_\mathrm{inf}} \coth \left(\frac{\pi B_\mathrm{p}}{E_\mathrm{p}} \right).  
\end{equation}
With this assumption, Refs.~\cite{Sobol:2019xls,Gorbar:2021rlt,Gorbar:2021zlr} introduced the ``gradient expansion method,'' where the evolution of gauge fields are described by the evolution equation for the space-independent expectation values of 
${\cal E}^{(n)} (t) \equiv \langle {\bm E} \cdot ({\bm \nabla} \times)^n {\bm E} \rangle/(a^n H_\mathrm{inf}^4)$,  
${\cal G}^{(n)} (t) \equiv -\langle {\bm E} \cdot  ({\bm \nabla} \times)^n {\bm B} \rangle/(a^n H_\mathrm{inf}^4)$, and
${\cal B}^{(n)} (t) \equiv \langle {\bm B} \cdot  ({\bm \nabla} \times)^n  {\bm B} \rangle/(a^n H_\mathrm{inf}^4)$. 
Then the Schwinger conductivity is now the function of ${\cal E}^{(0)}$ and ${\cal B}^{(0)}$.  
The equation of motion for ${\cal E}^{(n)}$, ${\cal G}^{(n)}$, and ${\cal B}^{(n)}$ includes the $n+1$-th terms and the boundary terms originated from the horizon crossing of the physical modes, which depends on the damping factor $\Delta (\leq1)$ due to the induced conductivity. 
The number of equations of motion is here infinite and hence we need to truncate. 
The truncation method introduced in Refs.~\cite{Gorbar:2021rlt,Gorbar:2021zlr} is that for a given truncation order $n_\mathrm{max}$, the next-order quantities are set, e.g., by ${\cal E}^{(n_\mathrm{max}+1)} = r(\xi, \sigma)^2  {\cal E}^{(n_\mathrm{max}-1)}$ with $r(\xi, \sigma) \equiv |\xi| +\sqrt{\xi^2+ (\sigma/2H_\mathrm{inf})+ 
(\sigma/2H_\mathrm{inf})^2}$.  
With these criteria, expectation values of the hypergauge fields are numerically calculated. 
See Fig.~1 in Ref.~\cite{Gorbar:2021zlr} for the properties of the hypergauge fields in each method. 
In both methods, for $\xi\gtrsim 4$ the amplification of the hypergauge fields is found to be power-law in $\xi$, unlike the exponential amplification without the Schwinger effect suggested in Eqs.~\eqref{Bp_inf_num} and \eqref{Ep_inf_num}. 

Moreover, recently another approach that uses the mean field approximation is proposed~\cite{Fujita:2022fwc}.
In this approach, the electromagnetic fields are divided into the constant mean part and dynamical perturbation part. 
The Schwinger current in Eq.~\eqref{j_Y} is then divided into the mean part, perturbed part in the direction of the magnetic field fluctuation, and that in the direction of the electric field fluctuation. 
In this way, one can decompose the Schwinger current into the $\xi$-correction part and Ohmic current part so that perturbed equation of motion for the gauge fields that takes into account the backreaction from the Schwinger effect is uniquely determined.   
Solving the equation of motion numerically in a consistent manner, a stationary solution for the electromagnetic fields is obtained. 
The efficiency of the amplification of the electromagnetic fields is found to be suppressed, as expected, and the resultant electromagnetic field strength seems to come between the estimates of the ``equilibrium estimate'' and the ``gradient expansion method.''

Thus far we have discussed a model-independent way to evaluate the backreaction from the Schwinger effect by parameterizing inflation with the Hubble parameter during inflation $H_\mathrm{inf}$ and the instability parameter $\xi$. 
If we specify the inflation model, one can solve the inflaton dynamics as well as the gauge field amplification.  
For example, in the gradient expansion method the damping factor $\Delta$ can consistently be calculated.

The evolution of the hypergauge fields after inflation strongly depends on the dynamics of reheating of the Universe, when the inflation is connected to the hot Big Bang Universe. 
There can be further amplification of the hypergauge fields at that stage~\cite{Fujita:2015iga,Adshead:2016iae,Hashiba:2021gmn}, but at present no consistent study that takes into account the Schwinger effect has been done. 
In literature, the instant reheating assumption is often adopted, where the inflationary Universe is instantaneously turned to the radiation dominated Universe. 
Once the reheating is completed and the Universe becomes radiation dominated, the electric fields are supposed to be screened by thermal plasma but the magnetic fields evolve according to the MHD. 

An important consequence of the axion inflation is the chiral asymmetry (and also baryon and lepton asymmetry) is generated simultaneously, which is guaranteed by the Schwinger effect for the LLL, so that the anomaly equation~\eqref{helicity} holds. 
There can be late-time annihilation of the hypermagnetic helicity and chirality, which we will mention in the next subsection. 

Before closing this subsection, we note some remarks on the observational constraints. 
The amplified gauge fields during inflation can generate the cosmological perturbations as the second-order effect. 
Since the gauge field amplification is relatively large, the strength of the amplified gauge fields can be in principle constrained by the observations of the cosmic microwave background (CMB)~\cite{Barnaby:2011vw,Barnaby:2011qe,Meerburg:2012id,Pajer:2013fsa}, in which the constraints on $\xi$ are given in the absence of the Schwinger effect, as $\xi\lesssim 1.92 - 2.45$~\cite{Pajer:2013fsa}. 
These constraints given at the CMB scales are strong enough for the Schwinger effect not to be so relevant. 
However, the value of $\xi$ grows (slowly) as the inflation end approaches and can become much larger at a later time. The Schwinger effect can indeed be relevant at the last stage of inflation.
We can also expect that the magnetic fields generated by axion inflation can explain the IGMFs suggested by blazar observations, but this possibility is strongly constrained by the baryon overproduction, which will be discussed in the next subsection.

\subsubsection{\it Chiral plasma instability}
\label{sec:cosmo_CPI}

Hypermagnetic field can be amplified with the CPI discussed in Section~\ref{sec:CPI} in the early Universe during the radiation dominated era~\cite{Joyce:1997uy,Tashiro:2012mf}. 
In order for the CPI to occur efficiently, we need to have a large chiral asymmetry and the chirality-flipping interactions to be sufficiently weak. 
While the chirality producing process would not exist in the SM, once we explore the physics beyond the SM, one can imagine, for example, a heavy particle decay that can generate chiral asymmetry. 
Indeed, such a process is well-motivated to produce the baryon asymmetry of the Universe (BAU), as the Sakharov's condition~\cite{Sakharov:1967dj}
requires the CP-violation. 
For example, in the SU(5) GUT baryogenesis~\cite{Yoshimura:1978ex,Dimopoulos:1978kv,Toussaint:1978br,Weinberg:1979bt,Barr:1979ye}, the chiral asymmetry and the baryon asymmetry are tightly connected and produced simultaneously~\cite{Kamada:2018tcs,Domcke:2020quw}. 
Moreover, the chirality-flipping rate for the one carried by the right-handed electron is suppressed by the electron Yukawa coupling  $h_{\rm e} \simeq 2.9\times 10^{-6}$, and hence the CPI can grow before the chirality-flipping interaction through the electron Yukawa coupling becomes relevant~\cite{Campbell:1992jd,Bodeker:2019ajh} (see also Refs.~\cite{Boyarsky:2020cyk,Boyarsky:2020ani}), if the chiral asymmetry is sufficiently large. 
In the following, we adjust the chiral MHD to the radiation dominated Universe and study how the CPI grows to investigate its consequences. 

In the radiation dominated Universe, the CPI induced by a large chiral asymmetry is studied in terms of the chiral MHD. 
Since the Universe expands, we need to take into account the effects of the space-time dynamics. 
Omitting the contributions from the gauge fields and bulk velocity fields, the cosmic expansion is described by the Friedmann equation for the metric of the form in Eq.~\eqref{FLRW_metric}, 
\begin{equation}
3 H^2 M_\mathrm{pl}^2 = \epsilon_\mathrm{rad}, 
\quad {\partial_t  \epsilon_\mathrm{rad} } + 3 (1+w) H  \epsilon_\mathrm{rad} =0, \label{FriedmannEq_rad}
\end{equation}
where $M_\mathrm{pl}\simeq 2.43 \times 10^{18}\,\mathrm{GeV}$ is the reduced Planck mass and $w=1/3$ is the equation of state of radiation ($P = w \epsilon$).

By solving Eq.~\eqref{FriedmannEq_rad}, we find $a  \propto t^{1/2} \propto \tau $ and $T \propto a^{-1}$ (neglecting the change of the number of effective relativistic degree of freedom). Fortunately the system is conformally invariant at high temperature and we can hide the cosmic expansion in the MHD equations by moving to the ``conformal frame,'' where the variables are replaced by the conformal or comoving quantities as~\cite{Brandenburg:1996fc}
\begin{align}
&{\bm B} \rightarrow {\hat {\bm B}} \equiv a^2 {\bm B}, \quad {\bm E} \rightarrow {\hat {\bm E}} \equiv a^2 {\bm E}, \quad \mu_5 \rightarrow {\hat \mu_5} = a \mu_5, \\ 
&T \rightarrow {\hat T} = a T, \quad \epsilon \rightarrow {\hat \epsilon} = a^4 \epsilon, \quad P \rightarrow {\hat P} = a^4 P, \quad \sigma \rightarrow {\hat \sigma} = a \sigma, \quad \kappa \rightarrow {\hat \kappa} = a^3 \kappa, 
\end{align}
while the bulk velocity field ${\bm v}$ is unchanged. 
Note that at high temperature where the transport phenomena are induced by thermalized charged relativistic particles, we have the hyperelectric conductivity and shear viscosity as~\cite{Arnold:2000dr} 
\begin{equation}
\sigma \simeq \frac{T}{g'^2 \ln g'^{-1}}\,, \quad 
\kappa \simeq \frac{T^3}{g'^4 \ln g'^{-1}}\,. 
\end{equation}
Remember that the kinematic viscosity and shear viscosity are related by $\nu = \kappa/\epsilon$ (and hence $\nu \rightarrow {\hat \nu} = a^{-1} \nu$). 
Then the MHD equations for the relativistic particles with non-relativistic bulk motion adopted in Refs.~\cite{Rogachevskii:2017uyc,Schober:2017cdw} are given by
\begin{align}
\label{energy_bran}
{\cal D}_\tau {\hat \epsilon}&= - {\hat \epsilon} {\bm \nabla} \cdot {\bm v},  \\
{\hat \epsilon} {\cal D}_\tau {\bm v} &= \left({\bm \nabla} \times {\hat {\bm B}}\right) \times {\hat {\bm B}} - {\bm \nabla} {\hat P} +{\bm \nabla} \cdot (2 {\hat \nu} {\hat \epsilon} {\cal S} ) + {\hat \epsilon} {\bm f}, \label{NSbran} \\
\partial_\tau {\hat {\bm B}}& = {\bm \nabla} \times \left[{\bm v} \times {\hat {\bm B}} - \frac{1}{\hat \sigma}\left({\bm \nabla} \times {\hat {\bm B}} - C {\hat \mu_5} {\hat {\bm B}}\right)\right],   \\
{\cal D}_\tau {\hat \mu_5}&= D_5{\bm \nabla}^2 {\hat \mu}_5 + \frac{\lambda_C}{\hat \sigma} \left[ {\hat {\bm B}} \cdot \left({\bm \nabla} \times {\hat {\bm B}} \right) - C {\hat \mu_5}{\hat {\bm B}}^2 \right] -\Gamma_\mathrm{f} {\hat {\mu_5}}, 
\end{align}
where the isothermal fluid is assumed so that the temperature $T$ is constant and the pressure is determined by the constant equation of state ${\hat P} = w {\hat \epsilon}$ with $w=1/3$.%
\footnote{A factor $(1+w)$ in the right-hand side of Eq.~\eqref{energy_bran} and left-hand side of Eq.~\eqref{NSbran} looks missing in the studies of Refs.~\cite{Rogachevskii:2017uyc,Schober:2017cdw}.}
Here ${\cal D}_\tau \equiv \partial_\tau + {\bm v} \cdot {\bm \nabla}$ is the advective derivative, ${\cal S}_{ij} = (\partial_j v_i+\partial_i v_j)/2-\delta_{ij} {\bm \nabla} \cdot {\bm v}/3$ is the trace-free strain tensor, $D_5$ is a phenomenologically introduced diffusion constant for the chiral chemical potential $\mu_5$, and $\Gamma_\mathrm{f}$ is the chirality-flipping rate. 
Here the CME induced current is expressed as ${\hat {\bm j} }= C {\hat \mu_5} {\hat {\bm B}}$. 
Compared to the MHD equations for the relativistic electrons and non-relativistic protons in Section~\ref{sec:CCSN}, the energy and pressure get related with the constant equation of state so that the mass and energy conservation is unified into Eq.~\eqref{energy_bran}.
Also, diffusion of $\mu_5$ is considered with introducing a diffusion coefficient phenomenologically. The chirality-flipping term is explicitly included, though its effect is not so explored in the numerical simulations.

The factors $C$ and $\lambda_C$ depend on the particle contents of the fluid. 
For the QED with massless Dirac fermions with charge $e$, with the discussions with the arguments in the previous sections one finds $C=e^2/(2 \pi^2)$ and $\lambda_C=6C/T^2 = 3e^2/(\pi T)^2$. 
For the SM thermal plasma, which is of our interest for the early Universe, the situation is complicated since there are several particle species with different U(1) hyper charges and complicated interactions. 
Fortunately, the system can be safely reduced to the chiral MHD. 
This is because that we are interested in the dynamics of a long-wavelength magnetic fields and hence we can assume the equilibration of fast interactions and focus on the slow variables. 
It has been noticed that it depends on the temperature of the Universe if interactions between the charged particles in the SM, such as the Yukawa interactions and sphalerons~\cite{Kuzmin:1985mm,Manton:1983nd,Klinkhamer:1984di,Arnold:1987mh,Khlebnikov:1988sr}, are fast enough and how many 
(approximate) global charges $q_\bullet$ among 15 species of the chiral fermions in the SM%
\footnote{A convenient choice of the global charges is given as, for example, 
\begin{equation}
\left\{q_u, q_B, q_{d-b}, q_\tau, q_{u-c}, q_\mu, q_{B_1-B_2}, q_{d-s}, q_{u-d}, q_{2 B_1-B_2-B_3}, q_e, q_{B/3-L_e}, q_{B/3-L_\mu}, q_{B/3-L_\tau}, q_\Phi\right\}, \nonumber
\end{equation}
where small characters such as $u, d,\cdots$ represent the right-handed fermion numbers, $B_i$ represent the baryon number for the $i$-th generation ($B=\sum_i B_i$), $L_f$ represents lepton number for the flavor $f$, and $\Phi$ represents the Higgs particle number. Here the hypercharge neutrality is assumed and $q_Y$ is omitted.}
can be treated as slow variables~\cite{Joyce:1997uy,Domcke:2020quw,Garbrecht:2014kda,Domcke:2020kcp}. 
For example, around the temperature right above $10^{5}\,{\rm GeV}$, electroweak and strong sphalerons as well as all the Yukawa interactions other than the electron Yukawa interactions are fast enough to be in equilibrium so that the only the right-handed electron number remains as a slow variable (together with four exact conserved charges of the SM, hypercharge and the flavored $B/3-L_f$). 
Defining the chiral chemical potential associated with the hyper U(1) gauge interaction, $\mu_5^Y \equiv \sum_i \chi_i y_i^2 \mu_i$, where $i$ runs through all the fermions in the SM, $y_i$ is the hypercharge for the fermion $i$, and $\chi_i=\pm1$ is assigned for the right- and left-handed fermions, respectively, $\mu_5^Y$ is expressed in terms of $\mu_{{\rm e}_{\rm R}}$~\cite{Joyce:1997uy,Kamada:2018tcs,Domcke:2020quw}, 
\begin{equation}
\label{mu5^Y}
\mu_5^Y = \frac{711}{481} \mu_{{\rm e}_{\rm R}\,}. 
\end{equation}
Then one finds the factors are given as $C=g'^2/(4\pi^2)$ and
\begin{equation}
\lambda_C=\frac{711}{481}\frac{6}{T^2}C = \frac{2133}{962 \pi^2} \frac{g'^2}{T^2}\,. 
\end{equation}
Just with changing these factors, the system can be described by the chiral MHD. 

As has been discussed in the previous section, the CPI amplifies the magnetic fields. 
Since the MHD equations are highly nonlinear when the magnetic fields are sufficiently strong, numerical studies are needed to see how the system evolves until later times and if the system settles on to  a scaling regime. 
Recently the chiral MHD simulations have been performed by several groups, and it is still under the debate on the late time evolution of the system, including the study explained in Section~\ref{sec:ChMHD}. 
Here we review the results of recent direct numerical simulations by another group~\cite{Brandenburg:2017rcb,Schober:2017cdw,Schober:2018ojn} and follow the arguments. 

The magnetic fields in the chiral MHD are now treated as stochastic fields, whose properties are characterized by the energy spectrum, $E(k)$,  which satisfies
\begin{equation}
\mathcal{E}_{\rm M} = \int {\rm d}k E_{\rm M}(k),
\end{equation}
where $\mathcal{E}_{\rm M} \equiv ({1}/{2}) \langle {\bm B}^2 \rangle$ stands for the energy density of the magnetic fields. 
The coherence length of magnetic fields are evaluated with the spectrum as
\begin{equation}
\lambda = \frac{1}{\mathcal{E}_{\rm M}}\int {\rm d}k \frac{E_{\rm M}(k)}{k}\,. 
\end{equation}
By starting from the initial condition with large spatially-homogeneous chiral asymmetry ${\hat \mu}_{5,0}$ (and vanishing $\Gamma_{\rm f}$) and without velocity fields, they have identified three following phases that characterize the growth and saturation of the magnetic fields that are affected by the fluid turbulence produced by the magnetic fields themselves. 

\

\noindent
{\bf Phase 1: ``Laminar phase'' or a linear instability}

Due to the CPI, weak initial magnetic fields are amplified (Gaussian noise is applied in the simulations). 
When the velocity fields are weak enough, the magnetic fields are amplified as has been discussed in Section~\ref{sec:CPI}. 
Namely, 
\begin{equation}
E_{\rm M}(k = k_\mathrm{CPI}) \propto 
\exp \left( \frac{k_\mathrm{CPI}^2}{ {\hat \sigma}} \tau \right), \quad k_\mathrm{CPI}=\frac{C {\hat \mu}_{5,0}}{2} = \frac{g'^2 {\hat \mu}_{5,0}}{4 \pi^2}\,, 
\end{equation}
while ${\hat \mu}_5$ remains almost constant. 
The velocity fields are simultaneously amplified with twice the growth rate due to the Lorentz force term in the Navier-Stokes equation. 

\

\noindent
{\bf Phase 2: Chiral turbulence driven dynamo instability}

Once the velocity fields are amplified enough, the velocity fields are no longer negligible. 
The growth rate becomes smaller and the magnetic field amplification scale is smaller. 
The peak magnetic energy spectrum when the system enters the Phase 2 is found to be
\begin{equation}
E_{\rm M}(k_\mathrm{CPI}) = C_5 {\bar {\hat \epsilon}} \frac{{\hat \mu}_{5,0}}{ {\hat \sigma}^2}\,, 
\end{equation}
where $C_5$ is a numerical factor which are numerically found to be around 16. 
This can be understood when the eddy turnover scale $\sim {\bar  v} (k_\mathrm{CPI}^2/{\hat \sigma})^{-1}$ reaches the CPI scale $k_\mathrm{CPI}^{-1}$ with supposing the equipartition between the magnetic fields and velocity fields are quickly established, ${\cal E}_{\rm M} (\simeq k_\mathrm{CPI} E_{\rm M}(k_\mathrm{CPI})) \simeq {\cal E}_{\rm K} (\simeq {\bar {\hat \epsilon}} {\bar v}^2/2)$. 
After the onset of the Phase 2, the magnetic fields show a spectrum proportional to $k^{-2}$, which is expected for the weak turbulence~\cite{Brandenburg:2017rcb,Galtier:2000ce}, with keeping $E_{\rm M}(k=k_\mathrm{CPI}) \simeq C_5 {\bar {\hat \epsilon}} {\hat \mu}_{5,0}/ {\hat \sigma}^2$ as the UV cutoff. 
The peak scale is redshifted to the IR, and the strength of the magnetic energy spectrum is given as
\begin{equation}
E_{\rm M}(k_\mathrm{peak} ) \simeq C_5 {\bar {\hat \epsilon}} \frac{{\hat \mu}_{5,0}^3}{{\hat \sigma}^2 k_\mathrm{peak}^2}\,. 
\end{equation}
Since the CPI amplifies one helicity mode, the magnetic fields are maximally helical. 
The magnetic helicity is roughly estimated as
\begin{equation}
{\hat h}_{\rm M} \simeq E_{\rm M}(k_\mathrm{peak} ) \simeq C_5 {\bar {\hat \epsilon}} \frac{{\hat \mu}_{5,0}^3}{{\hat \sigma}^2 k_\mathrm{peak}^2}\,. 
\end{equation}
$k_\mathrm{peak}$ evolves smaller with time, and hence the magnetic helicity grows with time. 

\

\noindent
{\bf Phase 3: Inverse cascade after the saturation of the chirality transfer}

\begin{figure}[htbp]
\begin{center}
    \includegraphics[width=0.45 \textwidth]{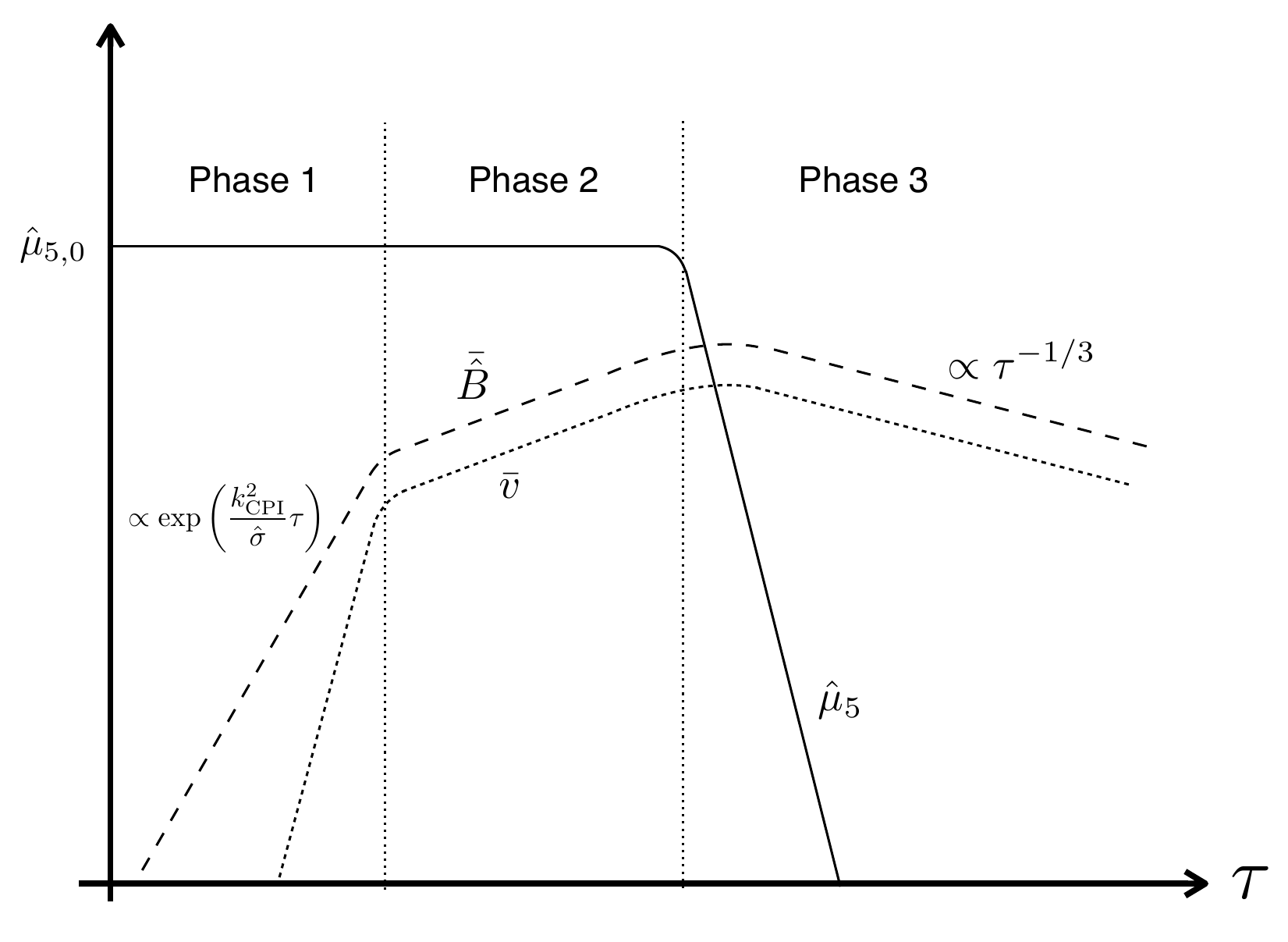}
    \includegraphics[width=0.5 \textwidth]{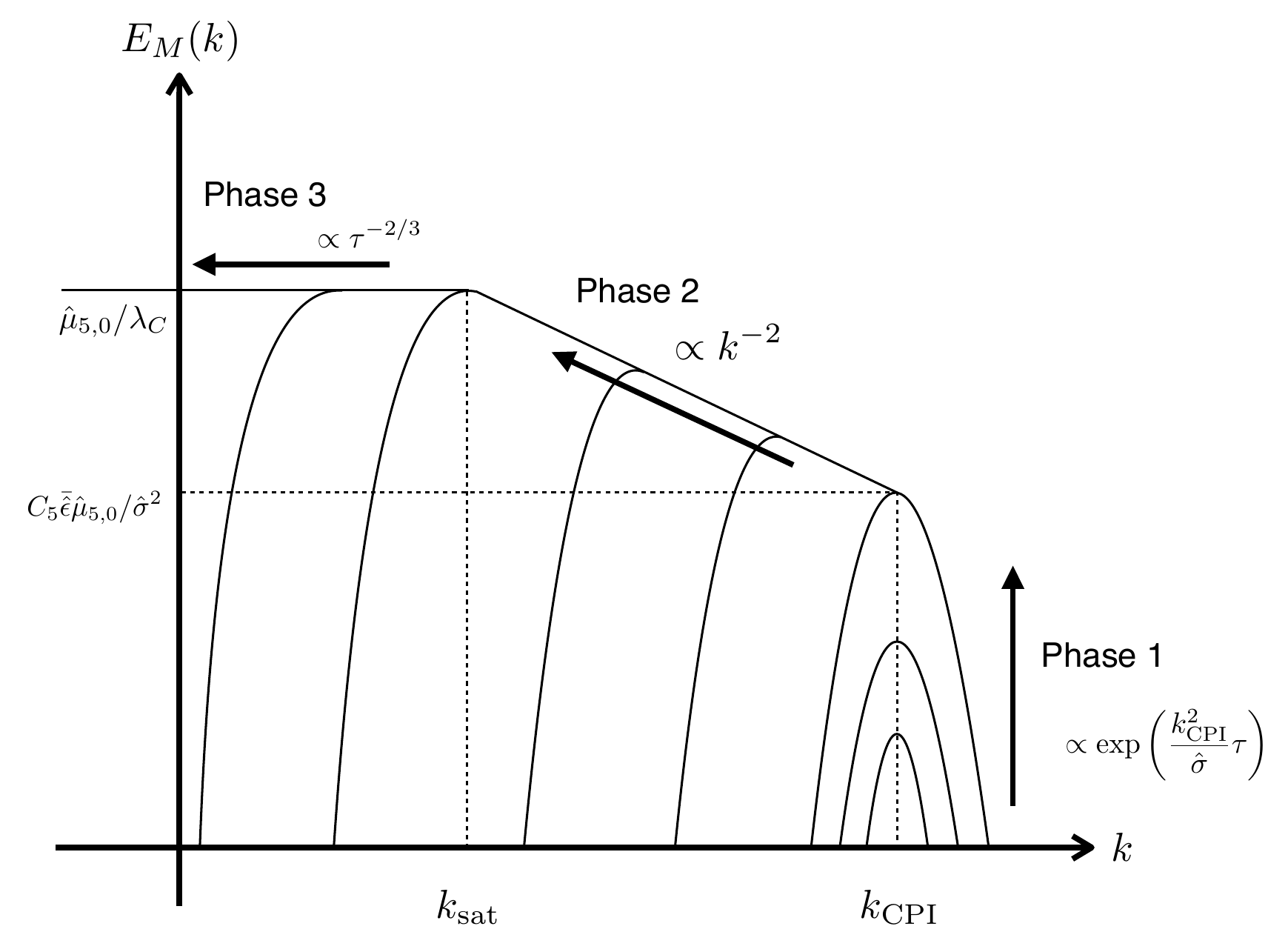}
\caption{\label{fig:CPI-cosmo}
    {\it Left.} Schematic picture of the time evolution of the chiral chemical potential ${\hat \mu}_5$ (solid line), typical 
   strength of magnetic fields ${\bar {\hat B}}$ (dashed line), typical strength of velocity fields ${\bar v}$ 
   (dotted line).  In the numerical simulation in Ref.~\cite{Schober:2017cdw}, 
   these quantities are all found to reach constant values, but it is understood as the finite box size effect and a numerical artifact.
    {\it Right.} Schematic picture of the time evolution of the magnetic field spectrum $E_{\rm M}(k)$. 
}
\end{center}
\end{figure}

Since the anomaly equation holds during the evolution of the system, the following equation is always satisfied,
\begin{equation}
{\hat \mu}_5+\lambda_C {\hat h}_{\rm M}=\mathrm{const} = {\hat \mu}_{5,0}.
\end{equation}
This means that the process in Phases 1 and 2 can be understood as the transfer of the chirality from the fermions to magnetic fields, and the chiral chemical potential decreases with time during these phases. 
The transfer saturates when all the initial chirality carried by fermions is transferred to the magnetic fields so that
\begin{equation}
\lambda_C {\hat h_{\rm M}} \simeq {\hat \mu}_{5,0}. \label{helsat}
\end{equation}
At that time the peak momentum is given as
\begin{equation}
k_\mathrm{peak} \simeq  k_\mathrm{sat} \equiv \sqrt{C_5 \lambda_C {\bar {\hat \epsilon}}}\frac{{\hat \mu}_{5,0}}{{\hat \sigma}}\,. 
\end{equation}
In the case of our interest, we have $\lambda_C {\bar {\hat \epsilon}} \ll {\hat \sigma}^2 $, and hence we have $k_\mathrm{sat} \ll k_\mathrm{CPI}$. 
After the saturation, it is found that the chiral chemical potential ${\hat \mu}_5$ starts to decay. It is argued that the chiral effect is no longer relevant and the magnetic field spectra evolve with the usual inverse cascade law for the typical scale and field strength of the maximally helical magnetic fields, 
\begin{equation}
k_\mathrm{peak} \propto \tau^{-2/3}, \quad {\bar {\hat B}} \equiv \sqrt{2 {\cal E}_{\rm M}} \simeq \sqrt{2 k_\mathrm{peak} E_\mathrm{peak} } \propto \tau^{-1/3}, \label{inversecascade}
\end{equation}
and ${\hat \mu}_5$ decays quickly.
(see, however, the discussion in Section~\ref{sec:CPI}). 
With this treatment, one can estimate the properties of the magnetic fields in later Universe. Figure~\ref{fig:CPI-cosmo} shows the schematic overview of the time evolution of the chiral chemical potential, typical strength of magnetic fields and of velocity fields as well as that of magnetic field spectrum. 

\

Results in other groups agree with the discussion above that magnetic fields are amplified until the saturation in Eq.~\eqref{helsat}, but the late time evolution is still under debate. 
Note that in the numerical simulations of Ref.~\cite{Schober:2017cdw}, the decay of ${\hat \mu}_5$ stops relatively quickly due to the finite box size and the scaling evolution in Eq.~\eqref{inversecascade} is not clearly confirmed. 
The estimate presented in the above gives the strongest magnetic fields with longest coherence length. 
Thus we here explain its late-time cosmological consequence as the most optimistic one. 

Going back to physical variables, one can estimate that the temperature when the CPI strongly amplifies is 
\begin{equation}
T\simeq 6.8 \times 10^6\,\mathrm{GeV} \left(\frac{\alpha_Y}{10^{-2}}\right)^2 \left(\frac{\sigma}{10^2 T} \right)^{-1} \left(\frac{g_*}{106.75}\right)^{-1/2} \left(\frac{\mu_{5,0}/{T_0}}{10^{-2}}\right)^2, 
\end{equation}
where $\alpha_Y \equiv g'^2/(4\pi)$ is the hyperfine structure constant, and $T_0$ is the initial temperature. 
Note that $\sigma/T$ is temperature independent. 
For the CPI to be effective, the chirality-flipping interaction should be negligible. 
The weakest chirality-flipping interaction that determines the chirality-flipping rate is the electron Yukawa interaction for the SM plasma in the early Universe. Note that $\mu_{{\rm e}_{\rm R}}$ as well as $\mu_5$ can have finite value when the electron Yukawa interaction is out of equilibrium, $T>T_{h_\mathrm{e}} \simeq 10^2\,{\rm TeV}$. 
Thus, we have a critical value for the initial chiral chemical potential, $\mu_{5,0}^\mathrm{c}/T_0 \simeq 10^{-3}$, for the magnetic fields to be strongly amplified with
\begin{equation}
\frac{\mu_{5,0}}{T_0} > \frac{\mu_{5,0}^\mathrm{c}}{T_0}\,.
\end{equation}
Supposing that magnetic fields evolve according to the inverse cascade until recombination and then evolve adiabatically, the properties of magnetic fields today are evaluated as
\begin{align}
B_\mathrm{p} & \simeq 9.9 \times 10^{-16}\,\mathrm{G} \left(\frac{\lambda_C}{C T^2/6} \right)^{-1/3}\left(\frac{\gamma}{10^{-2}}\right)^{-1/3} \left(\frac{\alpha_Y}{10^{-2}}\right)^{-1/3} \left(\frac{\mu_{5,0}/{T_0}}{10^{-2}}\right)^{1/3}\left(\frac{g_*}{106.75}\right)^{1/3}, \\
\lambda_\mathrm{p} & \simeq 6.9 \times 10^{-3}\,\mathrm{pc} \left(\frac{\lambda_C}{C T^2/6} \right)^{-1/3}\left(\frac{\gamma}{10^{-2}}\right)^{2/3} \left(\frac{\alpha_Y}{10^{-2}}\right)^{-1/3} \left(\frac{\mu_{5,0}/{T_0}}{10^{-2}}\right)^{1/3}\left(\frac{g_*}{106.75}\right)^{-2/3}, 
\end{align}
where $\gamma$ is introduced to determine the ratio between the magnetic fields and velocity fields, $v = \gamma B_\mathrm{p} \sqrt{\bar \epsilon}$ and it is supposed that the coherence length coincides with the eddy turnover scale, $\lambda_\mathrm{p} \simeq v t$. 
One would expect that for sufficiently large $\mu_{5,0}/T_0$ the IGMFs suggested by blazar observations can be explained, but once more, this possibility is strongly constrained by baryon overproduction. 
Instead, it can be the origin of the BAU, which is the topic of the next section.

Before closing this subsection, let us mention related studies in Refs.~\cite{Schober:2021yav,Schober:2021iws}. 
In these studies it is found that small-scale spatial fluctuations of $\mu_5$ can amplify the magnetic fields through the CPI, even if the mean value of $\mu_5$ is zero.

\subsection{\it Baryogenesis \label{sec:baryogenesis}}

An advantage to consider the generation of the helical (hyper)magnetic fields in the early Universe is that they can generate the BAU. 
The BAU is one of the most important mysteries for the particle physics and cosmology. 
In the modern cosmology, which includes primordial inflation, there should be a mechanism to generate the BAU after inflation before the Big Bang Nucleosynthesis (BBN) at when the temperature of the Universe is around $T \sim 1\,{\rm MeV}$, otherwise the matter and antimatter has been annihilated each other and the resultant abundance of the light elements such as helium does not reach the one in the present Universe. 
It is well known that the Sakharov's conditions, that is, i) baryon number violation, ii) C and CP violation, and iii) deviation from thermal equilibrium, are needed to be satisfied~\cite{Sakharov:1967dj}, but it is difficult to satisfy them within the SM and General Relativity. 
There have been many studies to explain the BAU by introducing new physics~\cite{Kuzmin:1985mm,Fukugita:1986hr,Affleck:1984fy,Riotto:1999yt}, but none of them is conclusive. 
Once we assume that helical hypermagnetic fields are generated before the electroweak symmetry breaking at $T\gtrsim 100\,{\rm GeV}$, it is found that the baryon asymmetry is generated automatically through the hypermagnetic helicity decay without introducing any new physics~\cite{Giovannini:1997gp,Giovannini:1997eg,Bamba:2006km,Anber:2015yca,Fujita:2016igl,Kamada:2016eeb,Kamada:2016cnb} (see also Ref.~\cite{Giovannini:2021xbi}). 
Here the Sakharov's conditions are satisfied as follows, 
i) chiral anomaly in the SM which involves the U(1) hypercharge interaction breaks baryon number, 
ii) the existence of the long-range helical hypermagnetic fields violates C and CP, 
iii) the long-range hypermagnetic fields are not in thermal equilibrium. 
This mechanism is noteworthy since the magnetic fields would remain until today as the IGMFs and in principle testable. 
In this subsection, we examine how baryon asymmetry is generated 
with the helical hypermagnetic fields. 

The chiral anomaly that involves the SM gauge group leads to the violation of the conservation of baryon and lepton currents ($j_\mu^B$ and $j_\mu^L$) and numbers ($Q_B$ and $Q_L$) as~\cite{tHooft:1976rip}
\begin{equation} \label{anomaly_b+l}
\partial_\mu j^\mu_B = \partial_\mu j^\mu_L = N_g \left( \frac{g^2}{16 \pi^2} \mathrm{Tr} [W_{\mu\nu} {\tilde W}^{\mu\nu}] - \frac{g'^2}{32 \pi^2} Y_{\mu\nu} {\tilde Y}^{\mu\nu}\right)
\end{equation}
or
\begin{equation}
\Delta Q_B = \Delta Q_L = N_g \left(\Delta N_\mathrm{CS}^L - \frac{g'^2}{16 \pi^2 } \Delta {\cal H}_Y\right), 
\end{equation}
where $N_g=3$ is the number of fermion generation, $W_{\mu\nu}$ and ${\tilde W}^{\mu\nu}$ are the gauge field strength tensor and its dual for the weak SU(2)$_L$ interaction, $g$ is the gauge coupling for the SU(2)$_L$ gauge interaction, $N_\mathrm{CS}^L$ is the SU(2)$_L$ Chern-Simons number, and ${\cal H}_Y$ is the hypermagnetic helicity. 
From this anomaly equation, it is clear that if the hypermagnetic helicity in the system changes, the baryon asymmetry is induced. 
Note that the SU(2)$_L$ part causes the washout of preexisting $B+L$ asymmetry through the electroweak sphaleron process~\cite{Kuzmin:1985mm,Manton:1983nd,Klinkhamer:1984di}, which will be also taken into account in formulating the kinetic equation for the baryon asymmetry~\cite{Arnold:1987mh,Khlebnikov:1988sr}.  

Let us consider the case where the long-range helical hypermagnetic fields exist before the electroweak symmetry breaking and are well described by the hydrodynamic approximation. 
The hypermagnetic helicity is a good approximate conserved quantity in this regime, but decays slightly in two ways. 
The first is the magnetic diffusion due to the finite electric conductivity and the other is the electroweak symmetry breaking when the long-range magnetic fields change from the hypermagnetic one to the electromagnetic one. 
It depends on the details of the electroweak symmetry breaking which contribution is dominant for baryogenesis, and for the electroweak crossover with $125\,{\rm GeV}$ Higgs mass it turned out the latter would be dominant~\cite{Kamada:2016cnb}. 
By taking into account the washout by the electroweak sphaleron and the fast ``spectator'' processes which determine the relation between the chemical potentials of each particle in the thermal bath one finds that the number of approximately conserved quantities is reduced, as we have also discussed in the previous subsection. 
The relevant quasi-conserved quantities are found to be the baryon (and lepton) number, and the kinetic equation for the volume average of baryon asymmetry, or specifically baryon-to-entropy ratio, $\eta_B$, around the electroweak symmetry breaking is given as
\begin{equation}
\frac{{\rm d} \overline{\eta}_B}{{\rm d} x} = (\text{source}) - (\text{washout})
\end{equation}
with
\begin{equation} 
(\text{source}) = \frac{3}{2} (g^2+g'^2) \frac{{\rm d} \theta_\mathrm{w}}{{\rm d} \ln x} \sin 2 \theta_\mathrm{w} \mathcal{S}_\mathrm{AB}\,, \quad  (\text{washout}) = \frac{111}{34} \gamma_{\mathrm{w, sph}} \overline{\eta_B}\,,  \label{KineticEqn}
\end{equation}
where $x = M_0/T$  with $M_0 \equiv M_\mathrm{pl}/\sqrt{\pi^2  g_*/90}$. During the electroweak crossover, a linear combination of the U(1) hypergauge field and the third component of the SU(2)$_L$ gauge field remains massless~\cite{Kajantie:1996qd,DOnofrio:2015gop}, which is denoted as ${\cal A}$.  It is parameterized by the temperature-dependent effective weak mixing angle,  $\theta_\mathrm{w}$, which gradually changes from 0 to $\theta_{\mathrm{w}0} \equiv \tan^{-1} (g'/g)$. 
The function in the source term, $\mathcal{S}_{\mathrm{AB}}$, is defined as 
\begin{equation}
    \mathcal{S}_{\mathrm{AB}}
    =\frac{H}{8\pi^2a^3sT}{\overline{\boldsymbol{A}_{\mathcal{A}}\cdot\boldsymbol{B}_{\mathcal{A}}}},  \label{ansource}
\end{equation}
where $s=(2 \pi^2/45)g_{*s}T^3$ is the entropy density with $g_{*s}$ being the relativistic degrees of freedom for entropy, and ${\bm A}_{\cal A}$ and ${\bm B}_{\cal A}$ are the vector potential and comoving magnetic fields for ${\cal A}$, respectively. The overline represents the volume average. 
The parameter in the washout term, $\gamma_{\mathrm{w, sph}}$, is the transport coefficient for the electroweak sphaleron process, which is evaluated by the recent lattice simulations as 
$\gamma_{\mathrm{w, sph}}\simeq \exp[-(146.6\pm 1.0)+(0.83\pm0.01) T/\mathrm{GeV}]$~\cite{DOnofrio:2014rug}.
The factor 111/34 is determined by the equilibrium condition of the fast spectator processes. 
It turned out that the right-hand side of the anomaly equation~\eqref{anomaly_b+l} reduces to the source term in Eq.~\eqref{ansource}. 

When the sphaleron process is faster than the cosmic expansion, $\gamma_{\mathrm{w, sph}} x \gtrsim 1$, the system with the kinetic equation~\eqref{KineticEqn} enters the equilibrium state, 
${\rm d}\overline{\eta}_B/{\rm d}x \simeq 0$, 
which determines the mean baryon asymmetry as~\cite{Kamada:2016cnb} 
\begin{equation}
    \overline{\eta}_B \simeq \frac{17}{37} \frac{g^2+g'^2}{\gamma_{\mathrm{w, sph}}} \frac{{\rm d}\theta_{\mathrm{w}}}{{\rm d}\ln x} \sin 2 \theta_\mathrm{w} \mathcal{S}_{\mathrm{AB}}\,, \label{equil}
\end{equation}
before the freezeout of the sphaleron process. 
As the temperature of the Universe decreases, the electroweak sphaleron freezes out around $T=T_\mathrm{fo} \simeq 135\,{\rm GeV}$. 
(Hereafter the subscript ``fo" represents the quantity evaluated at the sphaleron freeze out, $T=T_\mathrm{fo}$.)
Just after the sphaleron freezeout, the electroweak symmetry breaking 
completes, and the baryon asymmetry is fixed. 
The resultant baryon asymmetry is evaluated in a simplified form as~\cite{Kamada:2016cnb}  
\begin{equation}
    \overline{\eta}_B
    =\left.\frac{17}{296\pi^2}(g^2+g'^2)\frac{{\rm d} \theta_\mathrm{w}}{{\rm d} \ln x} \sin 2 \theta_\mathrm{w}\frac{\overline{\boldsymbol{A}_{\mathcal{A}}\cdot\boldsymbol{B}_{\mathcal{A}}}}{a^3s}\right|_{T=T_\mathrm{fo}}.
\end{equation}
Therefore, if we have non-vanishing $\overline{\boldsymbol{A}_{\mathcal{A}}\cdot\boldsymbol{B}_{\mathcal{A}}}$ 
on average, we will obtain non-vanishing net baryon asymmetry. 
For a peaky spectrum of the hypermagnetic fields with the characteristic comoving field strength $B_\mathrm{c}$ (or the physical strength $B_\mathrm{p} = a^{-2} B_\mathrm{c}$) and the comoving coherence length $\lambda_\mathrm{c}$ (or the physical coherence length $\lambda_\mathrm{p} = a \lambda_\mathrm{c}$) with the helicity fraction $\varepsilon_\mathrm{hf}$, we can write  $\overline{\boldsymbol{A}_{\mathcal{A}}\cdot\boldsymbol{B}_{\mathcal{A}}} \simeq \varepsilon_\mathrm{hf} \lambda_\mathrm{c} B_\mathrm{c}^2 = a^3 \varepsilon_\mathrm{hf} \lambda_\mathrm{p} B_\mathrm{p}^2$ and evaluate the net baryon asymmetry as~\cite{Kamada:2016cnb}  
\begin{align}\label{eq:ProducedBAUBlambda}
    \overline{\eta}_B
    &=\left.\frac{17}{296\pi^2}(g^2+g'^2)\frac{{\rm d} \theta_\mathrm{w}}{{\rm d} \ln x} \sin 2 \theta_\mathrm{w}\frac{\varepsilon_\mathrm{hf} \lambda_\mathrm{p} B_\mathrm{p}^2}{s}\right|_{T=T_\mathrm{fo}} \notag \\
    & \sim 10^{-10} \varepsilon_\mathrm{hf} \left.\sin 2 \theta_\mathrm{w} \frac{{\rm d} \theta_\mathrm{w}}{{\rm d} \ln x} \left(\frac{\lambda_\mathrm{p} }{10^6\,\mathrm{GeV}^{-1}}\right) \left(\frac{B_\mathrm{p}}{10^{-3}\, \mathrm{GeV}^2}\right)^2\right|_{T=T_\mathrm{fo}}. 
\end{align}
Note that from the one-loop analytic estimate for the time dependence of the weak mixing angle~\cite{Kajantie:1996qd}, 
\begin{equation}
    \cos^2\theta_{\rm w}=\cos^2\theta_{{\rm w}0}\left(1+\frac{11}{12}\frac{g_3^2\sin^2\theta_{{\rm w}0}}{\pi m_W(T)}\right),
\end{equation}
where $g_3$ is the three dimensional SU($2$) gauge coupling, and $m_W(T)$ is the perturbative $W$ boson mass, we roughly obtain~\cite{Kamada:2020bmb}
\begin{equation}\label{eq:1-loop}
    \left.\frac{{\rm d} \theta_\mathrm{w}}{{\rm d} \ln x} \sin 2 \theta_\mathrm{w}\right|_{T=T_\mathrm{fo}}=
    \left.-T\frac{{\rm d} \theta_\mathrm{w}}{{\rm d} T} \sin 2 \theta_\mathrm{w}\right|_{T=T_\mathrm{fo}}
    \simeq 0.14. 
\end{equation}
However, since the lattice calculation~\cite{DOnofrio:2015gop} shows slightly different values of the effective weak mixing angle with the analytic estimate for $140\,{\rm GeV} < T < 160\,{\rm GeV}$ and does not give the information of the effective weak mixing angle around $T_\mathrm{fo}$, we do not rule out the possibility that the value $ \left.\frac{{\rm d} \theta_\mathrm{w}}{{\rm d} \ln x} \sin 2 \theta_\mathrm{w}\right|_{T=T_\mathrm{fo}}$ is relatively smaller within the SM electroweak crossover. 

If the magnetic fields are maximally helical, $\varepsilon_\mathrm{hf}=1$, and already in the turbulent regime at the electroweak symmetry breaking, 
they are expected to evolve until recombination according to the inverse cascade, 
$B_p \propto a^{-7/2}$, and $\lambda_\mathrm{p} \propto a^{5/3}$~\cite{Banerjee:2004df,Kahniashvili:2012uj}, 
and then evolve adiabatically. 
With this consideration, the present baryon asymmetry can be written in terms of the properties of the IGMFs today as~\cite{Kamada:2016cnb,Kamada:2020bmb}
\begin{equation}
\overline{\eta}_B^0 \simeq 10^{-10} \left(\frac{   \left.\frac{{\rm d} \theta_\mathrm{w}}{{\rm d} \ln x} \sin 2 \theta_\mathrm{w}\right|_{T=T_\mathrm{fo}}}{0.14}\right) \left(\frac{\lambda_0}{10^{-9}\,\mathrm{Mpc}}\right) \left(\frac{B_0}{10^{-17}\,\mathrm{G}}\right)^2. 
\end{equation}
This suggests that the maximally helical hypermagnetic fields generated by, e.g., CPI or axion inflation can be responsible for the BAU but cannot explain the intergalactic fields suggested by blazar observations unless the efficiency factor of the baryogenesis, 
$\left.\frac{{\rm d} \theta_\mathrm{w}}{{\rm d} \ln x} \sin 2 \theta_\mathrm{w}\right|_{T=T_\mathrm{fo}}$ is 
significantly suppressed, much smaller than $10^{-9}$. 
In the case of axion inflation, we need to investigate it more carefully, because the baryon asymmetry is already generated at the end of inflation in the opposite sign to the hypermagnetic helicity so that they can completely annihilate away~\cite{Domcke:2018eki,Domcke:2019mnd}. 
The annihilation can be proceeded by magnetic diffusion or CPI. Fortunately there seems to be a parameter space where the sphaleron process washes out only baryon asymmetry so that the cancellation becomes incomplete before both the magnetic helicity and baryon asymmetry are annihilated away, and the present baryon asymmetry is explained by the 
hypermagnetic helicity decay at the electroweak crossover~\cite{Domcke:2019mnd}. 
Moreover, once the system involves relatively light right-handed neutrinos, whose mass $m_N$ is, at the lightest, as light as the electron Yukawa equilibration temperature, 
$m_N>T_{h_\mathrm{e}} \simeq10^2\,{\rm TeV}$, 
a recently proposed mechanism, the ``wash-in'' leptogenesis, can take place so that the net baryon asymmetry is the summation of these two contributions and non-zero baryon asymmetry remains~\cite{Domcke:2020quw}.

Thus far we focused on the volume average of the BAU. Since the baryon asymmetry is generated by the hypermagnetic fields with a finite coherence length, it also has a spatial fluctuation at the same scale.  
Such baryon isocurvature fluctuations are constrained by the observations of the CMB~\cite{Planck:2018jri} as well as the modification of the prediction of the BBN~\cite{Applegate:1987hm,Alcock:1987tx,Inomata:2018htm}. 
From Eq.~\eqref{eq:ProducedBAUBlambda} one might think that for smaller helicity ratio $\varepsilon_\mathrm{hf}$ stronger hypermagnetic fields with larger coherence length can be allowed even for the fiducial value of  the efficiency factor of the baryogenesis, 
$\left.\frac{{\rm d} \theta_\mathrm{w}}{{\rm d} \ln x} \sin 2 \theta_\mathrm{w}\right|_{T=T_\mathrm{fo}}\simeq 0.14$. 
Unfortunately, it turned out stronger hypermagnetic fields with longer coherence length is severely constrained by the constraints from the success of the BBN, independent of the helicity ratio~\cite{Kamada:2020bmb}. 
As a result, any hypermagnetic fields generated before the electroweak symmetry breaking can be responsible for the BAU but not the IGMFs suggested by the blazar observations, unless the efficiency factor of the baryogenesis, 
$\left.\frac{{\rm d} \theta_\mathrm{w}}{{\rm d} \ln x} \sin 2 \theta_\mathrm{w}\right|_{T=T_\mathrm{fo}}$, is highly suppressed.

\subsection{\it Chiral gravitational waves \label{sec:GW}}

Another interesting consequence of the chiral phenomena in the early Universe cosmology is the production of primordial GWs. 
Once the primordial magnetic fields are generated, they have anisotropic stress in the energy momentum tensor and source the GWs. 
If the primordial magnetic fields carry non-zero helicity, the produced GWs are expected to be helical. 
Since the GWs propagate almost without being scattered to other particles or waves, they carry the information of the very early Universe, and are expected to be detected by gravitational detectors~\cite{maggiore2008gravitational} such as LIGO~\cite{ligo}-Virgo~\cite{virgo}-KAGRA~\cite{kagra} around the frequency of $10^2\,{\rm Hz}$ and also at other broad range of frequencies as a stochastic background. 
While there are many models of particles physics and cosmology that emit GWs such as from inflation, preheating, first-order phase transition, and cosmic strings (for a review, see Ref.~\cite{Caprini:2018mtu}), those from primordial magnetic fields (with turbulence) are identified as yet another interesting source of the primordial GWs to explore the early Universe cosmology. 
In particular, the non-zero helicity of the GWs are characteristic feature of the those from the helical magnetic fields and the possibility of its detection in the stochastic GW background is now being investigated~\cite{Seto:2007tn,Seto:2008sr,Crowder:2012ik,Smith:2016jqs,Domcke:2019zls,Seto:2020zxw,Papageorgiou:2017yup}. 
Helical primordial GWs can also induce parity-odd signatures in the CMB, which can be detected in future observations (see Ref.~\cite{Komatsu:2022nvu} and references therein). 
Note that helical GW production is associated with a lepton number generation~\cite{Alexander:2004us,Lyth:2005jf,Maleknejad:2012wqk,Maleknejad:2014wsa,Caldwell:2017chz,Papageorgiou:2017yup} through the gravitational chiral anomaly~\cite{Kimura:1969iwz,Alvarez-Gaume:1983ihn}, which can be another origin of the present BAU.%
\footnote{It is practically difficult to explain the present BAU if the helical GWs are produced through the direct coupling between inflaton and gravitational Chern-Pontryagin density~\cite{Fischler:2007tj,Kamada:2019ewe,Kamada:2020jaf}.}
In this subsection, we explain how (helical) GWs are emitted from primordial helical magnetic fields. 

GWs are described by the transverse and traceless mode of the metric perturbation in the FLRW background, $h_{ij}^\mathrm{TT}$, 
\begin{equation}
{\rm d}s^2 =a^2(\tau) ({\rm d}\tau^2 - (\delta_{ij} + h_{ij}^\mathrm{TT} ){\rm d}x^i {\rm d}x^j), \quad \text{with} \quad \delta^{ij} h_{ij}^\mathrm{TT} = \partial^i h_{ij}^\mathrm{TT} =0. 
\end{equation}
By perturbing the Einstein equation with taking the first order in $h_{ij}^\mathrm{TT}$, we obtain the equation of motion for the GWs as
\begin{equation}
\left(\partial_\tau ^2+2 \frac{\partial_\tau a}{a} \partial_\tau - {\bm \nabla}^2 \right) h_{ij}^\mathrm{TT} = \frac{2}{M_\mathrm{pl}^2}T_{ij}^\mathrm{TT}\,, 
\end{equation}
where $T_{ij}^\mathrm{TT}$ is the transverse and traceless part of the energy momentum tensor.  
Let us move to the Fourier space representation with the circular polarization tensor $p_{ij}^s$ ($s={\rm R,L}$), 
\begin{equation}
h_{ij}^\mathrm{TT} (\tau, {\bm x}) = \frac{1}{(2\pi)^{3/2}} \int {\rm d}^3 {\bm k} \sum_{s={\rm R,L}} \sqrt{2} p_{ij}^s({\bm k}) h_{\bm k}^s (\tau) {\rm e}^{{\rm i} {\bm k} \cdot {\bm x}}\,,
\end{equation} 
where the circular polarization tensors are defined as\footnote{We adopt the convention for the circular polarization that $+$ mode for vector corresponds to R mode for tensor, and $-$ mode for vector corresponds to L mode for tensor.}
\begin{equation}
p_{ij}^s ({\bm k}) \equiv e_i^s ({\bm k}) e_j^s({\bm k}), 
\end{equation}
and satisfy the following equations,
\begin{align}
&\sum_i k^i p_{ij}^s({\bm k})  =\sum_j k^j p_{ij}^s({\bm k}) =0, \\
&\sum_{i,j}p_{ij}^s({\bm k}) {{p^{ij}}^{s'}}^* ({\bm k}) = \delta_{s s'} , \\
&\sum_{j,p}k_p \epsilon^{mpj} p_{ij}^s ({\bm k}) = -{\rm i} \lambda_{\bm k}^s k {p^m_i}^s ({\bm k}),  
\end{align}
with $\lambda_{\bm k}^{\rm R}=+1$, $\lambda_{\bm k}^{\rm L}=-1$. 
Then the equation of motion is rewritten as
\begin{equation}
\label{h_k}
\left(\partial_\tau^2+2 \frac{\partial_\tau a}{a} \partial_\tau +k^2 \right) h_{\bm k}^s= \frac{2}{M_\mathrm{pl}^2}T_{\bm k}^s\,, 
\end{equation}
where $T_{\bm k}^s$ is also the Fourier transform of the transverse and traceless part of the energy momentum tensor with the circular polarization basis. In terms of the Fourier transform, $T_{ij} (\tau, {\bm k}) \equiv(1/(2 \pi)^{3/2}) \int {\rm d}^3 {\bm k} T_{ij}(\tau, {\bm x}) {\rm e}^{{\rm i} {\bm k}\cdot {\bm x}}$, it is given as $T_{\bm k}^s (\tau) =\sum_{i,j} {{p^{ij}}^s}^* ({\bm k})T_{ij}(\tau, {\bm k})/2$. 
Thus a non-zero $T_{\bm k}^s$ sources the GWs and the difference between $T_{\bm k}^{\rm L}$ and $T_{\bm k}^{\rm R}$ makes it chiral.  

The spatial part of the energy momentum tensor of the electromagnetic fields is given as
\begin{equation}
T_{ij} = a^2 \left(\frac{1}{2} \delta_{ij}\sum_{k,l} \delta^{kl} (E_k E_l+B_kB_l) - E_i E_j -B_i B_j\right). 
\end{equation}
After going to the momentum space, the transverse and traceless part is obtained as
\begin{equation}
T_{ij}^\mathrm{TT} (\tau, {\bm k}) = \sum_{k.l} \Lambda_{ij,kl} ({\bm k}) T_{kl}(\tau, {\bm k}), 
\end{equation}
where the projection tensor $\Lambda_{ij,kl} ({\bm k})$ is defined as
\begin{equation}
\Lambda_{ij,kl} ({\bm k})  \equiv P_{ik} ({\bm k}) P_{jl}({\bm k}) - \frac{1}{2} P_{ij} ({\bm k}) P_{kl}({\bm k}) \,,\quad P_{ij} ({\bm k})\equiv \delta_{ij} - \frac{k_i k_j}{k^2}\,. 
\end{equation}
With the Green function for the operator $\partial_\tau^2+ 2 (\partial_\tau a /a)+k^2$ in Eq.~(\ref{h_k}), $G(\tau, \tau')$, the mode function of the GWs can be written as
\begin{align}
h_{\bm k}^s (\tau) = - &\frac{2}{M_\mathrm{pl}^2} \int^\tau {\rm d}\tau' a^2 (\tau') G(\tau, \tau') \int \frac{{\rm d}^3 {\bm q}}{(2 \pi)^{3/2}} \sum_{i,j} {p^{ij s}}^*({\bm k}) 
\notag \\
& \times \left[ \partial_\tau A_i ({\bm q}, \tau')  \partial_\tau A_j ({\bm k}-{\bm q}, \tau') - \sum_{k,l,m,n}\epsilon_{i kl} \epsilon_{jmn} q_k (k_m-q_m) A_l    ({\bm q}, \tau')  A_n ({\bm k}-{\bm q}, \tau') \right].
\end{align}
Let us decompose the vector potential in the integral, $A_i$, into the mode function in the circular polarization basis, $A_+$ and $A_-$, with
\begin{equation}
A_i ({\bm x}, \tau) = \int \frac{\mathrm{d}^3 {\bm k}}{(2\pi)^{3/2}} \left[A_i ({\bm k},\tau) {\rm e}^{{\rm i} {\bm k} \cdot {\bm x}} + \mathrm{h.c.} \right] = \sum_{\lambda = \pm} \int \frac{\mathrm{d}^3 {\bm k}}{(2\pi)^{3/2}}  \left[A_\lambda({\bm k}, \tau) e^\lambda_i ({\bm k})  {\rm e}^{{\rm i} {\bm k} \cdot {\bm x}} + \mathrm{h.c.} \right], 
\end{equation} 
where $e_i^\lambda ({\bm k})$ is the circular polarization vector introduced in Section~\ref{sec:primordial}. 
If the electromagnetic fields are maximally helical (without loss of generality, one can take $A_-=0$), and they have a Gaussian distribution, $\langle A_+({\bm k}) A_+ ({\bm k}') \rangle \propto \delta ({\bm k}+{\bm k}')$ and $\langle \partial_\tau A_+ ({\bm k}) \partial_\tau A_+ ({\bm k}') \rangle \propto \delta ({\bm k}+{\bm k}')$, one can show that the two-point function of the helicity-$s$ graviton is given by
\begin{align}
\langle h^s_{\bm k}(\tau) h^s_{{\bm k}'}(\tau) \rangle = \frac{4}{M_\mathrm{pl}^4} \delta({\bm k}+{\bm k}') \int &{\rm d}\tau' {\rm d} \tau'' a^2 (\tau') a^2 (\tau '') G(\tau, \tau') G(\tau, \tau'') \notag \\
\times &\int  \frac{{\rm d}^3 {\bm q}}{(2\pi)^3} \sum_{i,j} |{e^s_i}^*({\bm k}) e_i^+ ({\bm q})|^2  |{e^s_j}^*({\bm k}) e_j^+ ({\bm k}-{\bm q})|^2 F_A({\bm k}, {\bm q}, \tau', \tau''), 
\end{align} 
with $F_A({\bm k}, {\bm q}, \tau', \tau'')$ being a function determined by $A_+({\bm k},\tau)$. 
With the property of the polarization vector, 
\begin{equation}
\sum_{i,j}|{e^s_i}^*({\bm k}) e_i^+ ({\bm q})|^2 = \frac{1}{4} \left(1+\lambda_s \frac{{\bm k} \cdot {\bm q}}{k q}\right), 
\end{equation}
one can see that the helical GWs are generated from maximally helical electromagnetic fields. 
Note that the GWs are not maximally helical in this case.  

For example, in the case with U(1) gauge field production during axion inflation without Schwinger backreaction (see Eq.~\eqref{gen_mode_sol}), the correlation function of the each helicity mode of the graviton is evaluated as~\cite{Sorbo:2011rz}
\begin{align}
\langle h_{\rm R}({\bm k}) h_{\rm R} ({\bm k}) \rangle &\simeq 8.6 \times 10^{-7} \frac{H_\mathrm{inf}^4}{M_\mathrm{pl}^4} \frac{{\rm e}^{4 \pi \xi}}{\xi^6} \frac{\delta ({\bm k}+{\bm k}')}{k^3} \,, \\
\langle h_{\rm L}({\bm k}) h_{\rm L} ({\bm k}) \rangle &\simeq 1.8 \times 10^{-9} \frac{H_\mathrm{inf}^4}{M_\mathrm{pl}^4} \frac{{\rm e}^{4 \pi \xi}}{\xi^6} \frac{\delta ({\bm k}+{\bm k}')}{k^3} \,. 
\end{align}
We can see that the GWs are not maximally helical but strongly helical. 
Compared with the stochastic GWs from inflation, $\langle h({\bm k}) h({\bm k}')\rangle \sim (H_\mathrm{inf}/M_\mathrm{pl})^4 \delta ({\bm k}+{\bm k}')/k^3$, it is a strong enhancement for $\xi \gg 1$, and one may expect the detection of such a stochastic GW background at the GW detectors such as LIGO or LISA~\cite{Cook:2011hg,Anber:2012du,Domcke:2016bkh,Garcia-Bellido:2016dkw}. 
However, it turned out such a detectable enhancement suffers from baryon overproduction~\cite{Jimenez:2017cdr} unless the procedure of the electroweak symmetry breaking is modified so that the efficiency factor of the baryogenesis, $\left.\frac{{\rm d} \theta_\mathrm{w}}{{\rm d} \ln x} \sin 2 \theta_\mathrm{w}\right|_{T=T_\mathrm{fo}}$, is highly suppressed if the U(1) field is the SM hyper U(1) gauge field.
See the discussion in the previous subsection. 
Note that the impact of the Schwinger effect is discussed in Ref.~\cite{Domcke:2019qmm}, where the spectrum of GWs is computed with the equilibrium estimate. 
In this case, the conclusion does not seem to be unchanged. 
In order to give a more conclusive statement, however, we need a more precise estimate of the Schwinger backreaction with taking into account the time-dependent $\xi$, which has not been obtained yet.
If the U(1) field is a dark photon that is not related to the SM, we are free from the baryon overproduction problem. 
In other words, the detection of such chiral GW background can be a smoking gun of the existence of the gauge field beyond the SM.%
\footnote{In the case where the SU(2) gauge field are coupled with axion as the inflaton, with the solid-like field configuration $W^a_i \propto \delta^a_i$, one circular polarization mode shows the tachyonic instability due to the direct coupling to the SU(2) gauge field configuration and maximally helical GWs are produced~\cite{Adshead:2013qp}.}

The evaluation of the GW background from the helical magnetic fields generated by the CPI is more involved since the MHD effect needs to be taken into account for the evolution of the magnetic fields. 
Recently a semi-analytic~\cite{Anand:2018mgf} and numerical MHD simulation~\cite{Brandenburg:2021aln} are performed to investigate the chiral GW production from the CPI. It is suggested that the generated GWs are helical. 

One may wonder if the direct coupling between the chiral chemical potential $\mu_5$ and gravitational Chern-Pontryagin density, $R_{\mu\nu\alpha\beta} {\tilde R}^{\mu\nu\alpha\beta}$, is induced by the gravitational chiral anomaly as we have seen in the U(1) gauge field. 
Indeed, in the presence of $\mu_5$, there can be an induced energy momentum tensor proportional to the tensor perturbation~\cite{Manes:2012hf,Jensen:2012kj,Sadofyev:2017zqc,Barrie:2017mmr},
\begin{equation}
T^i_j \simeq 4 \sum_{m,n} \epsilon^{jmn} \left(\frac{\mu_5}{192 \pi^2} \partial_m (\partial_t^2 - {\bm \nabla}^2 ) + \frac{\partial_t \mu_5}{192 \pi^2} \partial_m \partial_t\right)h_{in}^\mathrm{TT}, 
\end{equation}
in the Minkowski background. This can be understood as the gravitational counter part of the CME and may be called the chiral gravitational effect.
This corresponds to the effective action which couples to $\mu_5$ and  gravitational Chern-Pontryagin density, $R_{\mu\nu\alpha\beta} {\tilde R}^{\mu\nu\alpha\beta}$. 
One may expect that an instability in GWs can arise in a similar way to CPI. 
However, this induced energy momentum tensor does not cause the tachyonic instability since it acts as the modification of the kinetic term of the GWs. 
In particular, for constant $\mu_5$ it does not change the propagation of the GWs at all. (Too large $\mu_5$ causes an apparent catastrophic ghost-like instability, which would not be physical but suggests at least breakdown of the EFT.) 
For time-dependent $\mu_5$, a difference in the dispersion relation between the right- and left-handed polarization modes causes an oscillatory memory effect in the time dependence of the $\mu_5$ due to the change of $\mu_5$ induced by the chirality-flipping interaction or the dilution due to the cosmic expansion~\cite{Kamada:2021kxi}. 
Note that at non-zero temperature, it has been indicated that non-local contribution to the induced energy momentum tensor can arise in a non-covariant form~\cite{Manes:2012hf,Jensen:2012kj}. 
Such a contribution may result in a different behavior of the GWs, but a covariant description is desired to describe it in a consistent way. 

There are also other mechanisms for the generation of chiral GWs due to the presence of chiral fermions during inflation~\cite{Anber:2016yqr,Adshead:2019aac} and thermalized chiral fermions (like neutrinos) in the radiation dominated era~\cite{Gubler:2022zmf}.

\section{Summary and outlook \label{sec:summary}}
In this review, we summarized recent developments of the theoretical formulations for the chiral effects as well as the following possible novel phenomena and possible solutions to the unsolved problems (raised in Section~\ref{sec:introduction}) due to the chiral effects in high-energy astrophysics and early Universe:
\begin{itemize}
    \item{Pulsar kicks: large momentum of a neutron star may be generated by anisotropic neutrino emission along a magnetic field (or rotation) at the time of the supernova explosion. This macroscopic  chiral transport of neutrinos stems from the parity-violating effects of the weak interaction in non-equilibrium many-body systems of leptons.} 
    \item{Magnetars: the strong magnetic field of a magnetar may be induced by the CPI that also originates from the parity-violating effects of the weak interaction. The stability of its macroscopic magnetic field may be explained by the magnetic helicity resulting from the CPI.}
    \item{Core-collapse supernova explosions: the chiral effects can reverse the cascade direction of turbulence from direct cascade to inverse cascade in 3D. This inverse cascading feature, which has not been taken into account in the previous numerical simulations, could lead to a coherent fluid motion and potentially help the supernova explosion.}
    \item{Primordial magnetic field: in the early Universe, helical magnetic fields can be generated by a mechanism related to the chiral anomaly. One such mechanism is the axion inflation, where the Chern-Simons term is responsible for the tachyonic instability of U(1) gauge fields. The numerical study of the CPI as another possible mechanism is also recently developed. The resultant helical magnetic field shows the inverse cascade, which may explain the large-scale magnetic field in the present Universe.}
    \item{Baryogenesis: baryon asymmetry can be generated by the decay of the hypermagnetic helicity in the early Universe, especially around the electroweak phase transition. On the other hand, too large baryon fluctuation generated in this mechanism results in the failure of the BBN, which strongly constraints the abundance of the primordial magnetic fields.}
    \item{Chiral gravitational waves: circularly polarized stochastic gravitational wave background is generated if helical magnetic fields existed in the early Universe. It can be a possible probe of the chiral phenomena in the early Universe.}
\end{itemize}
These are still possible hypotheses at this stage, and it would be necessary to verify the relevance of the chiral effects more quantitatively. To investigate the first three questions on neutron stars and supernovae, it would be indispensable to perform numerical simulations of the chiral radiation transport theory for neutrinos. 
For the last three issues on the early Universe, implementing realistic particle physics models is necessary to give precise predictions. For example, (asymmetric) particle production associated with helical magnetogenesis and the detail of the electroweak phase transition in the SM and well-motivated models beyond SM should be quantitatively determined. We hope this review motivates these directions in the future.

\section*{Acknowledgments}
The authors would like to thank V.~Domcke and J.~Schober for comments on the manuscript.
K.~K. is supported by JSPS KAKENHI Grant Number JP19K03842. 
N.~Y. is supported by the Keio Institute of Pure and Applied Sciences (KiPAS) project at Keio University and JSPS KAKENHI Grant Number JP19K03852.
D.-L.Y. is supported by National Science and Technology Council (Taiwan) under Grant Number MOST 110-2112-M-001-070-MY3.

\bibliographystyle{elsarticle-num.bst}
\bibliography{chiral_ref.bib}
\end{document}